\documentclass[12pt,oneside,a4]{report}

\usepackage{amsmath}
\usepackage{amssymb}
\usepackage{graphicx}
\usepackage{chemsym}
\usepackage[rflt]{floatflt}

\input{Qcircuit.tex}

\newcommand{\ket}[1]{\ensuremath{|#1\rangle}}

\newcommand{\bra}[1]{\ensuremath{\langle#1|}}

\newcommand{\para}{\textit{para}}

\newcommand{\ortho}{\textit{ortho}}

\newcommand{\pHH}{\para-hydrogen}

\newcommand{\dihyd}{dihydrogen}

\newcommand{\ie}{\textit{i.e.},}

\newcommand{\eg}{\textit{e.g.},}

\newcommand{\braket}[2]{\ensuremath{\langle #1|#2\rangle}}

\newcommand{\expectation}[3]{\ensuremath{\langle #1|#2|#3\rangle}}

\newcommand{\average}[1]{\ensuremath{\langle #1\rangle}}

\newcommand{\dihydride}{\Ru(\H)_2(\C\O)_2(dppe)}

\newcommand{\dppestable}{\Ru(\C\O)_3(dppe)}

\newcommand{\dppeunstable}{\Ru(\C\O)_2(dppe)}

\newcommand{\diagonal}[1]{\ensuremath\{\{#1\}\}}

\newcommand{\etal}{\textit{et al}.}

\newcommand{\ashydride}{\mbox{$\text{Ru(H)}_2\text{(CO)}_2\text{(dpae)}$}}

\newcommand{\AX}{\textbf{AX}}

\newcommand{\AB}{\textbf{AB}}

\begin{document}

\begin{titlepage}
\thispagestyle{empty} \mbox{} \vspace{1in}
\begin{center}
{\LARGE\bf NMR Quantum Information Processing\\
with \textit{Para}-Hydrogen}
\end{center}
\vspace{1in}
\begin{center}

\begin{figure}[h]
\begin{center}
\includegraphics[scale=0.8]{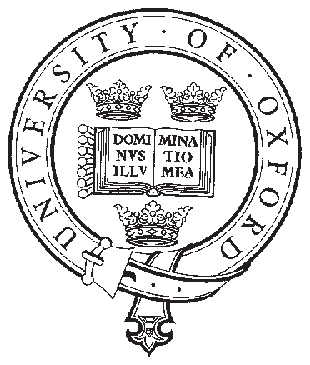}
\end{center}
\end{figure}

\end{center}
\vspace{1in}
\begin{center}
{\large
A thesis submitted for the degree of\\
\textit{Doctor of Philosophy}}
\end{center}
\vspace{0.25in}
\begin{center}
{\large
Muhammad Sabieh Anwar\\
University College, Oxford\\
Trinity Term 2004}
\end{center}
\end{titlepage}

\begin{abstract}

\begin{center}
{\bf NMR Quantum Information Processing with \textit{Para}-hydrogen}
\end{center}

\begin{center}
A thesis submitted for the degree of Doctor of Philosophy
\end{center}

\begin{center}
Muhammad Sabieh Anwar\\
University College, Oxford\\
Trinity Term 2004
\end{center}

Liquid state NMR quantum information processors are dominated by the
problem of preparing pure initial states. Traditionally these
devices employ pseudopure states instead of truly pure states, but
this approach is not scalable. Furthermore, almost all previous
demonstrations of information processing in liquid state NMR used
separable states which can in principle be described by classical
models. This thesis addresses the two-fold problem of initialization
and separability by preparing pure quantum states lying above the
entanglement threshold. Our pure state quantum computer derives its
purity from the highly polarized nuclear spin states in the \pHH\
molecule.

The thesis begins with a critique of conventional NMR based quantum
information processing in Chapter \ref{critique}, outlining the
major strengths and weaknesses of the technology. Chapter \ref{phip}
describes the enhanced magnetic ordering of the nuclear spin states
in \pHH\ and Chapter \ref{pure} describes an initialization
experiment exploiting this effect to achieve pure, entangled states.
These states can indeed be used as initial states in implementing
quantum algorithms. This is shown in Chapter \ref{QCimplement},
which describes an implementation of the Deutsch and the Grover
quantum algorithms.

The ``twirl'' operation converts a completely arbitrary input state
to a Werner singlet. The NMR implementation of this operation is
taken up in Chapter \ref{twirl}. Chapter \ref{sharing} analyzes the
possibility of sharing the purity of some highly polarized qubits in
a quantum computer onto quantum subspaces of arbitrary dimensions,
and whether these sharing operations increase or decrease the
likelihood of entanglement.

Finally, this dissertation has six Appendices which contain notes on
singlet-triplet mixtures, isotropic mixing sequences, systematic
errors in singlet detection, data processing and decoherence
modelling.

\end{abstract}


%
%
%
%


\begin{titlepage}
\pagestyle{empty}

\begin{center}
\bf Acknowledgements
\end{center}
\vspace{12pt}

My D.Phil. work was a part of a cross-university research
initiative. As such, I had the opportunity to participate in a truly
wonderful mix of specialist skills. It is my great pleasure to thank
Simon Duckett and Damir Blazina, without whose involvement, this
project would have been impossible to realize. From the preparation
of ultra-cold \pHH\ to implementing our pulse sequences, they showed
great skill (and patience) in deciphering my comments and
incorporating them into their experiments. Thanks indeed! Similarly,
Hilary Carteret in Montreal always kept me on my toes in her
scintillating email conversations regarding the many theoretical
aspects of our project. Many thanks to her.

Many thanks also to my post-doctoral colleague Li Xiao for
implementing my twirl sequences and promising to read this thesis. I
would also like to thank my doctoral colleague Tony Short for acting
as an instant first source of reference for my many mathematical
curiosities, and for adding a new impetus to my work on the twirl
operations. Another colleague who deserves my utmost thanks is Iain
Day in the Chemistry Department, for assisting and guiding me
through my experiments on the NMR spectrometer. I would also like to
extend my thanks to Lieven Vandersypen and to Thomas
Schulte-Herbr{\"{u}}ggen for kindly answering my questions on
decoherence modelling and unitary control.

My financial and, to a large extent, my social life were taken care
of by the Rhodes House at Oxford. Mary Eaton, Sheila Partridge and
Catherine King have helped me in numerous ways. My sincere thanks to
them as well.

A great number of friends have contributed to my Oxford life in
various ways. I would specially like to mention Adeel Malik, Ali
Abbas, Omer Suleman, Zunnoor Tarique, Sajid Javed, Samina Yasmin,
Usman Latif, Nadir Cheema, Niharika Gupta, Edward Dewey and Eleni
Hatzivassiliou for their love and constant encouragement; Mark
Bowdrey, Garry Bowen, Jon Walgate, Salvador Venegas, Yasser Omar,
Lucien Hardy, Mike Mosca, Raymond Laflamme and Alex Pines for many
useful discussions; the Oxford University Pakistan Discussion Forum
for accomplishing my literary missions; the Islamic Society for the
Friday prayers in Wadham College and Gilian Evisson for giving me
the master key to unlock the many treasures inside the Bodleian
Libraries. All of them make my three years in Oxford look so
precious.

There is however a debt of gratitude I \textit{cannot} repay; to my
supervisor Jonathan Jones who has helped me at every single step of
my research. Knowing Jonathan as a teacher and as a friend has been
the most satisfying experience in Oxford. His encouraging help and
wisest counsel in matters of both physics and life beyond, are
definitely the most valuable memories I shall be taking back home.

There is many a time when words fall short of feelings. It is
difficult to express my thoughts for my family. It is fair to admit
that all my accomplishments, large or small, are possible because of
my parents' unparalleled support and their constant prayer for me.
My fondest wishes also go to Rafay, Abu Bakar and Mehwesh who have
devoted much time in their joyous youth to thinking about their
brother here in Oxford. I have missed them as well.

Last but not the least, I am still thinking of a way to possibly
thank my grandmother who could not outlive the three very long years
she had spent, longing to see her grandson back in her loving
presence.

\vspace{0.2in}
%

{\bf The Wayfarer's Supplication}
\begin{verse}
``Submitting myself to the test of patience,\\
Like the rose's scent, I bid farewell to my garden.

In my search of the wine of knowledge,\\
I have left my homeland, hunting new taverns.

A tree in the desert, my eyes cling onto the clouds\\
Of \textit{Your} Magnanimity, and not on the gardener.

O Lord, keep me far ahead of my caravan,\\
So that they consider me as their destination.

The nest I had assembled straw by straw,\\
O God, May I see that same abode of mine.

And may I lay my forehead at my parents' feet,\\
by whose blessing, I possess the secret of Love.''

\end{verse}

\end{titlepage}

\thispagestyle{plain}
\renewcommand{\thepage}{}
\tableofcontents

\renewcommand{\thepage}{\arabic{page}}
\setcounter{page}{1}

\chapter{Liquid state NMR quantum information processing---a critique}\label{critique}

Liquid state nuclear magnetic resonance (NMR)
\cite{Abragam,ErnstNMR,LevittSpinDynamics,FreemanChoreo,HandbookNMRFreeman,Cavanagh,Toolkit}
is a vital spectroscopic technique in analytical chemistry. It also
plays a crucial role in present day medicine as a valuable imaging
technology. Recently, NMR has also sought new relationships:
interfacing with myriad fields in the biological and physical
sciences, probing living and non-living specimens of all kinds, and
now, in the form of \textit{functional} imaging, it is taking the
first few strides towards an understanding of the way humans think
and act! One such exciting partnership is with quantum information
processing (QIP) \cite{NielsenChuang,BDiV1}, now blossoming into an
established discipline in its own right. This thesis explores some
facets of this `NMR-QIP' interface.

NMR-QIP capitalizes on fifty years of experience in probing nuclear
spins in liquid solution with radio-frequency (RF) pulses. In fact,
several essential features of experimental QIP, such as Rabi
oscillations \cite{RabiOscillations} in single qubits, coherent
manipulations of multi-qubit systems and decoherence control have
long been developed and implemented in traditional NMR, decades
before QIP was investigated as an independent discipline. Today,
liquid state NMR is not only a leading technology for demonstrating
quantum computing (QC) in small physical systems, but also enriches
QIP, by inspiring new ideas for theoretical and experimental
investigation.

Like all the different ideas being pursued for the physical
realization of quantum computers, liquid state NMR comes with its
strengths and weaknesses. A useful set of criteria was laid out by
DiVincenzo \cite{DiVincenzoCrit} and since then, it has become
customary to evaluate practical QC schemes
\cite{Roadmap,QCPhysicalRSoc,JonesPracticeNature} on the basis of
these rules, which are listed below: 
\begin{enumerate}
\item there must exist well-defined, physical qubits and the system should be scalable
to higher dimensions;
\item the qubits must be initializable into a well-defined quantum
state;
\item the decoherence times must be long compared to the times for gate
operation;
\item there must also exist a set of quantum gates capable of implementing universal quantum
dynamics; and
\item it should be possible to perform single-qubit measurements.
\end{enumerate}
Most of my D.Phil. research revolves around one of the major
problems with liquid state NMR\footnote{From here onwards, NMR means
`liquid state NMR', unless the solid state is explicitly
mentioned.}, namely initialization (criterion 2). Consequently, in
this chapter, I choose to present a detailed description of the
intrinsic weaknesses in conventional NMR, and only touch upon the
strengths. Furthermore, this chapter is not intended to become
another review of NMR-QIP; there are already excellent tutorials and
monographs on the subject, such as
\cite{NielsenChuang,JonesRev,CoryNMRQCReview,CoryProcNatl,VandersypenNMRQCCtrl,VandersypenThesis}.
Instead, the specific aim of this chapter is to introduce the
basics, to guide the reader through the material in subsequent
chapters and also, to come up with a motivation for the forthcoming
chapters.

Liquid state NMR quantum computers, reaching seven qubits
\cite{VandersypenShor}, are known to be far ahead of competing
technologies, at least in terms of the implementation of quantum
algorithms
\cite{VandersypenShor,JonesDeutsch,ChuangDeutsch,ChuangGrover,Jonesapproxcount,JonesGrover,
CollinsDeutschEnt,KumarGroverTomography,KumarDeutschEntangled,KumarDeutsch2D}.
(Recently, the Deutsch-Jozsa algorithm has been implemented on a two
qubit ion trap quantum computer \cite{DeutschIonTraps}.) This
relative success has been made possible by the ease as well as
robustness with which unitary operations can be implemented in NMR.
These strengths are mentioned in Section \ref{section-unitary},
which describes the NMR states, coherences and universal unitary
dynamics (criteria 1, 3 and 4) and also introduces the necessary
notation. The problems of initialization and single qubit read-out
(criteria 2 and 5) are discussed in Sections
\ref{section-initialization-problem-nmr} and
\ref{section-measurement}. Another problem, not directly encompassed
in DiVincenzo's criteria, is the issue of separability and
entanglement. Traditional NMR is critically analyzed in view of this
requirement in Section \ref{section-entanglement-NMR} and finally,
general methods for addressing the problem of low polarizations are
addressed in Section \ref{section-hyper-polarization}.

\section{NMR States, Coherence and Dynamics}\label{section-unitary}

A qubit is a two level quantum system. A spin $s$ nucleus has $2s+1$
energy levels; the qubits in NMR are (generally) spin $1/2$ nuclei,
and so I often use the words `qubit' and `spin' interchangeably. The
two energy levels are labelled $\ket{\alpha}$ and $\ket{\beta}$ (or
$\ket{0}$ and $\ket{1}$), and these states also serve as the
computational eigenstates \cite{JonesRev}. The most readily
available two level systems are ^1\H, ^{19}\F, ^{31}\P, ^{13}\C\ and
^{15}\N, and what characterizes these qubits are their
\textit{Larmor frequencies} \cite{LevittSpinDynamics}, which in the
presence of a magnetic field, $B_0$, are,
\begin{equation}\label{Larmor-frequency-def}
\omega=\,-\gamma B_0\
\end{equation}
$\gamma$ being the gyromagnetic ratio of the nucleus. The qubits are
therefore ``identified'' in frequency space, and can be perturbed by
applying resonant electromagnetic radiation. For the nuclear spin
levels, the Larmor frequencies (and therefore the resonant
radiation) are in the RF region of the spectrum. An NMR spectrometer
is a highly sophisticated, computer controlled radio receiver and
transmitter, capable of delivering electromagnetic radiation to
nuclei in a liquid solution placed in a strong superconducting
magnet, and of detecting a time-resolved RF signal generated by the
nuclei. The signal is Fourier transformed and a frequency spectrum
is obtained, with the qubits identifying themselves by their
individual frequencies. In short NMR furnishes well characterized
qubits, partially satisfying the first DiVincenzo criterion. About
the other half, that is scalability, I will say more in Section
\ref{section-initialization-problem-nmr}.

\subsection{Ensemble quantum computing}\label{section-ensemble}

In the NMR setting, each qubit resides in a molecule, along with
other nuclei, with arbitrary spins. In the sample, there will be a
large number of these molecules; for example, in $1~$ml of a $1~$mM
solution, there will be roughly $10^{17}$, rapidly tumbling in the
liquid. As there is one $n$ qubit quantum computer per molecule, so
the sample will contain an ensemble of quantum computers. All these
quantum computers are identical, act in parallel and collectively
participate in producing a time-dependent macroscopic magnetization,
which is manipulated and detected in the NMR experiment. A
convenient language to describe such ensembles is the density
operator formalism \cite{NielsenChuang,CohenTannoudji1}. An $n$
qubit state, for example, is represented as an $2^n\times 2^n$
density matrix, expanded in some operator basis. In this thesis, I
shall use the notation $N=2^n$ for the Hilbert dimensions of an $n$
qubit quantum system. Different basis sets, each comprising $4^n$
basis elements, are in use \cite{ErnstNMR}; however, one set which
is the most common in the NMR community is the \textit{product
operator}
\cite{ErnstNMR,SorensenProdOp,Podkorytov,ProductOperatorABX} set,
which I now describe.

\subsection{Product operator
basis}\label{section-product-operator-define}

The product operator basis elements are represented as $N\times N$
matrices and are simply re-scaled versions of products of the Pauli
operators \cite{NielsenChuang}. For a single qubit, the
correspondence\footnote{An operator basis is
representation-independent whereas ``writing'' the basis elements in
matrix form requires a representation basis. All matrices in this
thesis are written in the so-called Zeeman basis
\cite{ErnstNMR,Cavanagh,JonesRev}, which is formed by taking the
direct products of the Zeeman eigenstates, which are also the
computational eigenstates. The ordering that I follow in writing
these matrices is the traditional
$\{\ket{\alpha\alpha},\ket{\alpha\beta},\ket{\beta\alpha,\ket{\beta\beta}}\}$
order.} is,
\begin{eqnarray}\label{po-Pauli-one-qubit-E}
\frac{\mathbf{1}_2}{2}=&\,\begin{pmatrix}1/2 & 0 \\ 0 &
1/2\end{pmatrix}=\,\frac{1}{2}\sigma_0,\\
I_x=&\,\begin{pmatrix}0 & 1/2 \\ 1/2 &
0\end{pmatrix}=\,\frac{1}{2}\sigma_x,\label{po-Pauli-one-qubit-Ix}\\
I_y=&\,\begin{pmatrix}0 & -i/2 \\ i/2 &
0\end{pmatrix}=\,\frac{1}{2}\sigma_y,\quad\text{and}\label{po-Pauli-one-qubit-Iy}\\
I_z=&\,\begin{pmatrix}1/2 & 0 \\ 0 &
-1/2\end{pmatrix}=\,\frac{1}{2}\sigma_z.\label{po-Pauli-one-qubit-Iz}
\end{eqnarray}
The symbol $\mathbf{1}_N$ is an example of a general notation I
shall use for an $N\times N$ identity matrix. The matrix,
$\mathbf{1}_2/2$, is commonly denoted as $\tfrac{1}{2}E$ in the NMR
literature, but I shall keep to the notation introduced above. A
single qubit density matrix, $\rho$ can always be uniquely expanded
in terms of these four basis elements,
\begin{equation}\label{po-basis-single-qubit-expand}
\rho=\,\frac{\mathbf{1}_2}{2}+\sum_{i=x,y,z}\,c_iI_i,
\end{equation}
which has unit trace matrix as it should \cite{NielsenChuang}, and
the coefficients $c_i$ are given by,
\begin{equation}\label{coefficient}
c_i=\,Tr(\rho I_i)/Tr(I_i^2)=\,2\,Tr(\rho I_i),
\end{equation}
$Tr(\ldots )$ representing the trace operation. It is traditional to
represent operators of the qubits with labels $I$, $S$, $R$, and so
on\footnote{Sometimes, the notation $I_1$, $I_2\ldots$ is used,
especially for homonuclear systems.}. For a two qubit system, the
$4^2=16$ product operators are constructed by taking the tensor
products of single spin operators according to,
\begin{eqnarray}
&2\,\times& I_i\otimes S_j,\quad i,j=x,y,z\label{PO-2-qubit-1}\\
&2\,\times& \frac{\mathbf{1}_2}{2}\otimes S_j,\quad j=x,y,z\label{PO-2-qubit-2}\\
&2\,\times& I_i\otimes \frac{\mathbf{1}_2}{2},\quad i=x,y,z\label{PO-2-qubit-3}\quad\text{and}\\
&2\,\times& \frac{\mathbf{1}_2}{2}\otimes
\frac{\mathbf{1}_2}{2}=\frac{\mathbf{1}_4}{2}\label{PO-2-qubit-4}.
\end{eqnarray}
It is also common practice to drop the multiplication signs, and by
slight abuse of notation, represent the two qubit operators with the
same symbols as used for the single qubit $(\eg\ 2\,\times
{\mathbf{1}_2}/{2}\otimes S_j=\mathbf{1}_2\otimes S_j\equiv S_j)$,
the exact meaning being apparent from the context. For example, the
sixteen product operators for a two qubit system are given in Table
\ref{Table-product-operator-2-qubit}. Any two qubit state can be
expanded in terms of these product operators,
\begin{equation}\label{po-basis-two-qubits-expand}
\rho=\,\frac{1}{2}(\frac{\mathbf{1}_4}{2})+\sum_{i=x,y,z}c_iI_i+\sum_{i=x,y,z}d_iS_i+\sum_{i=x,y,z}e_{ij}2\,I_iS_j.
\end{equation}
The factor of $1/2$ in front of the scalar element $\mathbf{1}_4/2$
ensures that $\rho$ has unit trace and the coefficients $c$, $d$ and
$e$ are calculated in a manner similar to \eqref{coefficient}.
Furthermore, the $\mathbf{1}_4/4$ term does not evolve under unitary
operations \cite{CohenTannoudji1} and from the point of view of the
state dynamics, is totally ``dormant''. The state is called the
\textit{maximally mixed state} and is often dropped out from the
state expansion, \eqref{po-basis-two-qubits-expand}. As a result,
only the \textit{traceless} operators are written in the expansion
and a unit trace scalar operator is \textit{assumed}. The traceless
part is called the \textit{deviation} matrix
\cite{Toolkit,JonesRev}. Although the distinction between the
``real'' and the deviation state is immaterial for traditional NMR
and signal detection, for QIP calculations it is better to carefully
appreciate this difference: therefore, I strictly use the symbol
`$\sim$' whenever I write only the deviation part of the state
expansion. Moreover, I represent diagonal matrices with the notation
$\diagonal{\ldots}$, with the diagonal elements inside the braces
and reserve single braces $\{\ldots\}$ for writing general lists of
numbers or vectors. For the physical meaning (or lack thereof,) of
the product operators, one may refer to \cite{LevittSpinDynamics}.

\begin{table}
\begin{center}
\begin{tabular}{ll}
\hline \\
Scalar element & ${\mathbf{1}_4}/{2}$ \\
Single quantum coherences & $I_x$, $I_y$, $S_x$,
$S_y$ \\
& $2I_xS_z$, $2I_yS_z$, $2I_zS_x$, $2I_zS_y$\\
Populations & $I_z$, $S_z$, $2I_zS_z$ \\
Multiple quantum coherences & $2I_xS_x$, $2I_yS_y$, $2I_xS_y$,
$2I_yS_x$ \\
  \hline
\end{tabular}
\end{center}
\caption{The $16$ product operator basis elements for a two qubit
($IS$) system. The basis elements are grouped according to their
coherence orders.\label{Table-product-operator-2-qubit}}
\end{table}

\subsection{Coherences}\label{section-coherences}

In Table \ref{Table-product-operator-2-qubit}, I have grouped
together the product operators in terms of their \textit{coherence
order}. The concept can be defined in terms of the positions of
non-zero elements in their matrix representations. The
\textit{single quantum coherences} are located in the positions,
\begin{equation}\label{single-quantum-coherence}
\begin{pmatrix}
& \Box & \Box & \\
\Box & & & \Box \\
\Box & & & \Box \\
& \Box & \Box & \\
\end{pmatrix},
\end{equation}
and the operators
$\{I_x,I_y,S_x,S_y,2I_xS_z,2I_yS_z,2I_zS_x,2I_zS_y\}$ comprise only
these coherences. The zero and double quantum coherences are in the
positions,
\begin{equation}\label{multiple-quantum-coherence}
\begin{pmatrix}
&  &  & \\
&  & \Box & \\
& \Box &  & \\
&  &  & \\
\end{pmatrix}\quad\text{and}\quad
\begin{pmatrix}
&  &  & \Box \\
&  &  & \\
&  & & \\
\Box &  &  & \\
\end{pmatrix},
\end{equation}
and are collectively designated \textit{multiple quantum
coherences}. Similarly, the diagonal terms are called the
\textit{populations} and in certain respects behave like zero
quantum terms. (The matrix representations of the $16$ product
operators are tabulated in \cite{Cavanagh,Toolkit}.)

The spin operators involving multiple coherence terms are also
expressed in terms of their linear combinations,
\begin{eqnarray}
&ZQ_x=&\,\frac{1}{2}\bigl(2I_xS_x+2I_yS_y\bigr),\label{ZQx-define}\\
&ZQ_y=&\,\frac{1}{2}\bigl(2I_yS_x-2I_xS_y\bigr),\label{ZQy-define}\\
&DQ_x=&\,\frac{1}{2}\bigl(2I_xS_x-2I_yS_y\bigr)\quad\text{and}\label{DQx-define}\\
&DQ_y=&\,\frac{1}{2}\bigl(2I_yS_x+2I_xS_y\bigr).\label{DQy-define}
\end{eqnarray}
This allows us to define a new basis, which I call the
\textit{product operator-multiple quantum} basis, spanned by the
product operators for single quantum terms and populations and
$\{ZQ_x,ZQ_y,DQ_x,DQ_y\}$ for the multiple quantum terms.

Two other bases will be occasionally used in parts of this thesis.
One is the \textit{spherical basis} \cite{ErnstNMR}, in which the
single qubit operators are related to the product operators through,
\begin{equation}
I^{\pm}=\,I_x\pm iI_y,\label{spherical-I}
\end{equation}
and $I_z$ and $\mathbf{1}_2/2$ are left unchanged; the multiple
qubit operators are formed analogously to
\eqref{PO-2-qubit-1}--\eqref{PO-2-qubit-4}, by taking tensor
products. The other basis I shall occasionally refer to, is the
\textit{polarization operator} \cite{ErnstNMR} basis, in which $I_x$
and $I_y$ are unchanged, and $I_z$ and $\mathbf{1}_2/2$ are replaced
by,
\begin{eqnarray}
I_\alpha&=&\,\frac{\mathbf{1}_2}{2}+I_z\quad\text{and}\label{polarization-operator-Ialpha}\\
I_\beta&=&\,\frac{\mathbf{1}_2}{2}-I_z,\label{polarization-operator-Ibeta}
\end{eqnarray}
which are in fact equal to the projection operators
$\ket{\alpha}\bra{\alpha}$ and $\ket{\beta}\bra{\beta}$.

\subsection{Detection in NMR}\label{section-detection-NMR}

All NMR experiments can be described in terms of transfers of
coherence orders \cite{LevittSpinDynamics,Cavanagh}. Even the
detection process can be clarified using the language of coherences.
For example, in a standard NMR experiment, only the single quantum
coherence terms can be detected. As such, they are called the
\textit{direct observables}. Suppose we wish to detect a state
$\rho$: then only the single quantum terms in the expansion
\eqref{po-basis-two-qubits-expand} will contribute to the signal,
while all other terms will remain unobservable,
\begin{equation}\label{direct-observables}
\rho\xrightarrow{\text{observe}}\,\{I_x,I_y,S_x,S_y,2I_xS_z,2I_yS_z,2I_zS_x,2I_zS_y\}.
\end{equation}
The terms $\{I_x,I_y,S_x,S_y\}$ represent \textit{in-phase} and
$\{2I_xS_z,2I_yS_z,2I_zS_x,2I_zS_y\}$ indicate \textit{anti-phase}
magnetization. The single quantum coherences are also assigned
\cite{LevittSpinDynamics} a positive or negative sign, depending on
whether they are positioned above or below the diagonal. The way
quadrature detection \cite{LevittSpinDynamics,Claridge} is
conceptually employed in modern spectrometers, measures only one
kind of, (say, the negative) coherence terms; the mechanism is
carefully spelled out in \cite{LevittSpinDynamics}.

The spherical and polarization operators are used in predicting the
NMR spectrum acquired from a state $\rho$. I define a \textit{signal
vector}, \textbf{Sg}, with the terms,
\begin{equation}\label{signal-vector}
\,\{Tr(\rho\, I_+S_{\alpha}),\,Tr(\rho\, I_+S_{\beta}),\,Tr(\rho\,
I_{\alpha}S_+),\,Tr(\rho\, I_{\beta}S_+)\}.
\end{equation}
The two leftmost entries indicate resonances of the $I$ spin, and
the two rightmost entries indicate resonances of the $S$ spin.
Mathematically, for an exponentially decaying sinusoidal time domain
signal, called the \textit{free induction decay} (FID), these
resonances appear as Lorentzian lineshapes
\cite{LevittSpinDynamics,Cavanagh,Toolkit} in the spectrum; real
values in $\textbf{Sg}$ indicate absorptive and imaginary values
represent dispersive character.

\subsection{Chemical
shift}\label{section-chemical-shift}

The $I$ and $S$ spin resonances, appearing in the NMR spectrum are
distinct because the $I$ and $S$ qubits are different nuclei having
different Larmor frequencies, \eqref{Larmor-frequency-def}. This is
most obviously true for a \textit{heteronuclear} system, involving
spins with different chemical identities, such as ^1\H\ and ^{13}\C
. However, NMR spectra reveal that even if the $I$ and $S$ spins
have the same chemical identity, \ie\ they form a
\textit{homonuclear} system, their frequencies can be different.
Each spin is surrounded by a cloud of electrons, which
\textit{shields} the externally applied field, $B_0$, changing the
effective field\footnote{Some authors use the `deshielding
convention', in which $\B_0$ is replaced by $B_0(1+\delta)$.} seen
by the nucleus to $B_0(1-\sigma)$. The Larmor frequency,
\eqref{Larmor-frequency-def}, of each spin is, therefore, shifted,
\begin{equation}
\omega=\,-\gamma
B_0(1-\sigma).\label{Larmor-frequency-chemical-shifted}
\end{equation}
If different spins belonging to a homonuclear system, have different
electronic environments and therefore, different \textit{chemical
shifts}, they can still act as distinct qubits. A useful parameter
is the difference in frequencies, $2\pi\delta$ ($\delta$ is assumed
to be in Hz), signifying how far apart the qubit resonances are: for
a pair of homonuclear spins such as two protons, $\delta$ is of the
order of a kHz, whereas for heteronuclear systems, it can easily be
a few hundred MHz.

\subsection{Spin-spin coupling}\label{section-spin-spin-coupling}

For purposes of QIP, it is important, that we have two or more
qubits per quantum computer, and that these qubits interact with one
another. The interaction between spins can assume different forms
\cite{ErnstNMR,SchmidtRohr,SolidNMR}, isotropic and anisotropic,
direct and indirect, depending on the molecular motions, and the
physical state of the sample. In solids and liquid crystals, the
interactions are dominated by direct, dipole-dipole magnetic
interactions between the nuclei. The NMR of such anisotropic media
is a separate realm altogether; a very good textbook on the subject
is \cite{SchmidtRohr} and examples of QIP with dipolar coupled
qubits are presented in
\cite{LiquidCrystalNMRQCVandersypen,SolidSingleCrystalNMRQC,KumarSolidNMRQC}.

In liquids, with which we are concerned, motional averaging
\cite{LevittSpinDynamics} leaves us with a non-zero isotropic value
of a kind of indirect coupling between the spins, which is called
the $J$ or \textit{scalar coupling}. This magnetic spin-spin
coupling is mediated by electrons bonding the nuclei, and is a kind
of Fermi-contact interaction. Typical values of $J$ are a few Hz for
^1\H--^1\H\ couplings and slightly larger, that is between
$100$--$200$~Hz for one bond ^{13}\C--^{1}\H\ couplings. The signal
vector, \eqref{signal-vector} suggests the excitation of four lines
in the spectrum of a two qubit system: in fact, the scalar coupling
causes each peak to split into two halves. The first two elements,
$Tr(\rho\,I_+S_{\alpha})$ and $Tr(\rho\, I_+S_{\beta})$, for
example, represent the two $I$ spin resonances, when the
neighbouring $S$ spin is in the $\ket{\alpha}\bra{\alpha}$ or
$\ket{\beta}\bra{\beta}$ state.

In the NMR tradition, a pair of spins is classified as
\textit{weakly coupled}, if $\delta\gg J$; the system is said to be
of the type \textbf{AX}. This will be the case for heteronuclear
systems as well as for many homonuclear systems in the presence of
strong magnetic fields, $B_0$. The system is classified as
\textit{strongly coupled} \cite{Cavanagh,Toolkit}, and of the type
$\textbf{AB}$, if $\delta \lesssim J$. Such systems are the
prevalent case in weak or near zero fields. The extreme case is an
\textbf{A$_2$} system, in which $\delta=0$, where the spins become
\textit{equivalent} and have identical frequencies. Even two
heteronuclear spins, in zero field conditions, would form an
\textbf{A$_2$} system.

\subsection{The NMR Hamiltonian}\label{section-Hamiltonian}

For a two qubit, \AB\ system, the NMR Hamiltonian
\cite{Abragam,ErnstNMR,LevittSpinDynamics,Cavanagh} is given by,
\begin{equation}\label{Hamiltonian-AB}
H=\,\hbar\bigl(\omega_II_z+\omega_sS_z+\pi
J\,2\,\mathbf{I}\cdot\mathbf{S}\bigr),
\end{equation}
where $\mathbf{I}\cdot\mathbf{S}$ is short-hand for
\begin{equation}\label{isotropic-define}
\frac{1}{2}\bigl(2I_xS_x+2I_yS_y+2I_zS_z\bigr),
\end{equation}
and in the weak coupling limit, \ie\ for an \AX\ system, it becomes,
\begin{equation}\label{Hamiltonian-AX}
\hbar\bigl(\omega_II_z+\omega_sS_z+\pi J\,2\,I_zS_z\bigr).
\end{equation}
I call these Hamiltonians \AB\ and \AX\ Hamiltonians. In these
expressions, $\omega_I$ and $\omega_S$ represent the chemically
shifted Larmor frequencies,
\eqref{Larmor-frequency-chemical-shifted}, of the two qubits, and I
subsequently follow the convention of setting $\hbar =1$. It is also
customary, to write the rotating frame \cite{LevittSpinDynamics}
versions of these Hamiltonians, replacing each $\omega$ by
$\Omega=\omega-\omega_{\textit{rf}}$, $\omega_{\textit{rf}}$ being
the frequency  of the rotating frame\footnote{The symbol for
rotating frame `rf'' should not be confused with `RF' for radio
frequency.}, generally taken to be equal to the transmitter
frequency \cite{SignsLevitt}. In this dissertation, I shall mostly
deal with homonuclear systems in the weak coupling limit.

The Hamiltonian discussed so far is the background or the internal
Hamiltonian, under which the system evolves at all times. A resonant
magnetic field is applied to perturb the spins. This control field,
$B_1$ is perpendicular to the static field, $B_0$ and oscillates in
the perpendicular plane (the \textit{transverse} plane) with a
frequency $\omega_{RF}$, expected to be close to the Larmor
frequencies of the spins in order to satisfy the resonance
condition. The oscillatory field, in addition to a frequency and
amplitude parameter, also possesses a phase factor $\phi$ in the
transverse plane; the control Hamiltonian for a perturbation on a
single qubit is given by,
\begin{equation}\label{hamiltonian-RF-control-single-qubit}
H_{RF}=\,\omega_1\bigl(I_x\cos{(\omega_{RF}t+\phi)}+I_y\sin{(\omega_{RF}t+\phi)}\bigr),
\end{equation}
where $\omega_1=-\gamma B_1/2$ is called the \textit{nutation
frequency}. The nutation frequency (amplitude), phase, radio
frequency and the time for which this control field is applied, can
all be modulated in time to achieve the desired state dynamics.

\subsection{Dynamics}\label{section-dynamics}

Corresponding to a Hamiltonian that is piecewise constant in time,
$H_1,H_2,\ldots H_k$, acting for time intervals $t_1,t_2,\ldots
t_k$, we can construct a propagator,
\begin{equation}\label{propagator}
U=\,\mathcal{T}_k\,\exp{(-iH_kt_k)},
\end{equation}
where $\mathcal{T}$ is the Dyson time-ordering operator
\cite{ErnstNMR}, and a state $\rho$ evolves according to the
Liouville-von Neumann equation \cite{ErnstNMR},
\begin{equation}
\rho\xrightarrow{}\,U\rho U^{\dag};
\end{equation}
the correspondence between this equation and the Schrodinger's
equation is analogous to the relationship between the the density
matrix and the pure state. 
What is referred to as a \textit{pulse} in this thesis, is the pulse
propagator,
\begin{equation}\label{pulse-propagator}
P=\,\exp{(-iH_{RF}t_p)}.
\end{equation}
Selecting suitable times $t_p$, it is possible to achieve desired
flip angles; for example, if $\omega_1t_p=\pi/2$, the pulse flips
the magnetization through $90^\circ$. I shall represent a pulse with
a flip angle of $\theta^\circ$, with a phase angle $\phi$ by the
symbol $\theta_\phi$, and for RF fields along the cardinal axes, I
shall simply represent the phases by $\{x,y,-x,-y\}$. For example,
in my notation, a pulse with flip angle $180^\circ$ and phase of
$180^\circ$, would be written as $180_{-x}$. So far, I have assumed
all axes of rotation to lie in the transverse plane. This is not a
strict assumption, as rotations around arbitrary axes in the Bloch
sphere \cite{NielsenChuang} can also be implemented using composite
rotations
\cite{FreemanChoreo,HandbookNMRFreeman,CompositeLevitt,CompositeCounsell,
TyckoBroadBandPopInversion,WimperisComposite}. This includes
rotations around the $\pm z$ axes \cite{CompositeZFreeman}.

The product operators also facilitate predicting the dynamics of the
spin operators. We make use of a theorem \cite{Cavanagh,Toolkit}:
for a set of three operators satisfying $[A,B]=iC$ and its cyclic
variations, $A$ evolves under the propagator $\exp{(-i\theta C)}$
as,
\begin{equation}\label{evolution-ABC}
A\xrightarrow{\exp{(-i\theta C)}}\,A\cos{(\theta)}+B\sin{(\theta)}.
\end{equation}

\subsection{Quantum logic gates}\label{quantum-logic}

The standard network model of quantum computation
\cite{NielsenChuang} assumes a network built up of logic gates
\cite{LogicGatesDiVincenzo}. In NMR, these gates are generally
implemented
\cite{JonesNMRLogicGates,LogicGatesCory,LogicGatesCoryPRA,UniversalSWAP,MultiQubitGatesNMRCory}
through a sequence of pulses and intervals of free precession (under
the internal Hamiltonian). Stated simply, single qubit gates can be
implemented by pulses exciting only one qubit in the system, whereas
two qubit gates normally exploit the $J$ coupling between the spins,
and often make use of the spin-echo pulse sequence
\cite{LevittSpinDynamics,FreemanChoreo,Cavanagh}. Logic gate design
often involves ``sculpting'' the Hamiltonian, making use of the
concepts learned from average Hamiltonian theory \cite{ErnstNMR}.
Sometimes, use is made of frequency selective pulses, so called
\textit{soft pulses} \cite{FreemanChoreo}, exciting peaks from a
single qubit, or ``finer tuned'', transition selective pulses
\cite{KumarTransitionSelective,CoryPhysicaD}, exciting only one peak
in a group. These selective pulses, also come with their concomitant
problems of cross-talk with other qubits and coupling evolutions
during finite pulse lengths; different methods have been devised
\cite{TransientBlochSiegertShift,VandersypenSimultaneousSoft,FreemanFreezingCoupling}
to ameliorate these effects. In our work
\cite{AnwarPH,AnwarDeutsch}, we mostly implemented selective pulses
using hard pulses and periods of free precession, a technique based
on the Jump-and-Return \cite{JumpReturnOriginal} sequences. I shall
return to this method in Chapter \ref{pure}. Furthermore, in Chapter
\ref{twirl}, we achieve selective excitation, using a method devised
by Fortunato \etal\ \cite{CoryHamiltonianModulate}, which
numerically searches for a shaped pulse profile, that maximizes some
pre-defined fidelity measure. Experimentally, all pulses, be they
soft or frequency non-selective (\textit{hard})
\cite{FreemanChoreo}, always have instrumental imperfections leading
to \textit{systematic} errors \cite{VandersypenNMRQCCtrl,Claridge},
and these can also be tackled
\cite{CumminsNJPComposite,JonesOffsetTailored,JonesIsing,JonesRobustPhilTrans}
very effectively. (These errors are also considered in Appendix
\ref{app-errors}.) In this dissertation, I represent soft pulses
achieving a flip angle of $\theta$, as $\theta I_{\phi}$ or $\theta
S_\phi$, exciting the $I$ or $S$ spin, and hard pulses as
$\theta_\phi$ without a spin label.

It was shown early on \cite{BarencoQuantumGates} that a
controlled-NOT (c-NOT) gate together with single qubit rotation,
suffices for implementing universal quantum computation
\cite{UniversalQC}. NMR readily provides c-NOT and single qubit
rotations, and can therefore, \textit{simulate} universal dynamics,
to some required accuracy, satisfying the fourth DiVincenzo
criterion. In fact, a major factor contributing to the early success
of liquid state NMR, was the reliability and flexibility of applying
arbitrary unitary propagators to the system, making it a convenient
test-bed for many quantum experiments.

\subsection{Decoherence}\label{section-decoherence}

Decoherence \cite{NielsenChuang} is the decay of coherences, leading
to the gradual decay of off-diagonal terms in the density matrix.
The time-scale over which this decay takes place is important,
because quantum computations can only last for times shorter than
this characteristic length of time, putting an upper limit on the
length of computations, before the quantum computer ``collapses''
into a set of so-called \textit{pointer} states
\cite{ZurekDecoherence}. In NMR, decoherence is captured by the term
\textit{relaxation}
\cite{Abragam,ErnstNMR,LevittSpinDynamics,Cavanagh} and two
characteristic time scales are defined: $T_1$ which determines the
return of the state to the equilibrium condition and $T_2$, which
accounts for phase randomization\footnote{Strictly speaking, the
$T_1$ relaxation also contributes to the phase randomization.}. (A
detailed treatise on the theory of relaxation in NMR is
\cite{RelaxationMcConnell}.) For all spin $1/2$ nuclei, $T_2\le
T_1$, and therefore, in a simple decoherence model, $T_2$ is assumed
to be the limiting coherence time scale. Typically, a two qubit gate
can be completed in a time proportional to $1/J$~s, so a rough
measure of the number of (two qubit) operations that can be
completed in the coherence time is $T_2/(1/J)$, a parameter we
desire to maximize. I shall return to this argument in the
introduction to Chapter \ref{QCimplement}. It is difficult, in
general, to estimate the number of two qubit operations that can be
performed before decoherence takes its toll, but a rough estimate
would suggest about a thousand operations \cite{JonesNMRcrit}.

There have been several attempts
\cite{DecoherenceModelCory,DecoherenceModelLindbaldCory} to
\textit{model} the effects of decoherence in small NMR quantum
information processors. We also used a simple model
\cite{VandersypenShor} to describe the decrease in signal intensity
from our two qubit implementation of the Deutsch algorithm,
attributable to relaxation. The experiment is explained in Chapter
\ref{QCimplement} and the model itself is outlined in Appendix
\ref{app-operator-sum-decoherence}. Interestingly, it is also
possible to \textit{reverse} the effects of unwanted
system-environment interactions
\cite{ViolaDynamicalCtrl,BangBangViolaKnillPRL} by perturbing the
system, a technique called \textit{dynamical decoupling}; or
\textit{avoiding} these interactions altogether, by working in
\textit{decoherence-free subspaces} \cite{DFSLidar}. These noiseless
sub-spaces have now been experimentally realized in NMR quantum
computers \cite{DFSLidarNMRImplement,DFSCoryScience}.

\subsection{Gradients}\label{section-gradients}

However, there are sometimes good reasons of intentionally
introducing decoherence into the ensemble: as the name suggests, the
\textit{gradient} magnetic field
\cite{Cavanagh,Toolkit,KeelerGradients} introduces a spatially
linear magnetic field in the sample. If the gradient
\textit{strength} is $G$, a point in the sample at the spatial
coordinate $z$, will experience a field of $B_0+Gz$, instead of the
otherwise homogenous field, $B_0$, resulting in a position-dependent
Larmor frequency,
\begin{equation}
\omega=\,-\gamma (B_0+Gz)(1-\sigma).
\end{equation}
If we consider only the spatially dependent part, the two-qubit
Hamiltonian during the gradient pulse, $H_g$ can be written as,
\begin{equation}
H_g(z)=\,-\gamma_I Gz(1-\sigma_I)I_z-\gamma_S Gz(1-\sigma_S)S_z;
\end{equation}
and furthermore, assuming a homonuclear system with
$\sigma_I\approx\sigma_S=\sigma\approx 0$,
\begin{equation}
H_g(z)=\,-\gamma Gz(I_z+S_z),
\end{equation}
and the propagator is given by,
\begin{equation}
U_g(z)=\,\exp{(-iH_g(z)\,t)}=\,\exp{\bigl(-i\phi(z)\,(I_z+S_z)\bigr)},
\end{equation}
where $\phi(z)=-\gamma Gzt$, is a space dependent phase, different
points in the sample incurring different phases. Now consider a
coherence term $\ket{r}\bra{s}$ (an element in the density matrix),
with coherence order $p=M_r-M_s$, where $M_r$ and $M_s$ are the
magnetic quantum numbers\footnote{For example, the quantum number
$M_r$ is defined by the operator equation $I_z\ket{r}=M_r\ket{r}$.} of the states $\ket{r}$ and $\ket{s}$. The propagator transforms this coherence to, 
\begin{eqnarray}
&&U_g\ket{r}\bra{s}U_g^{\dag}\nonumber\\
&&=\exp{\bigl(-i\phi(z)(I_z+S_z)\bigr)}\,\ket{r}\bra{s}\,\exp{\bigl(i\phi(z)(I_z+S_z)\bigr)}\nonumber\\
&&=\exp{(-i\phi(z)M_r)}\exp{(i\phi(z)M_s)}\,\ket{r}\bra{s}\nonumber\\
&&=\exp{(-i\phi(z)p)}\,\ket{r}\bra{s}\label{gradient-coherence},
\end{eqnarray}
indicating that the $z$-dependent phase also depends on the
coherence order: populations and homonuclear zero quantum
coherences, having $p=0$, acquire zero phase, irrespective of the
spatial position. In the post-gradient measurement, the detection
coil integrates the coherences originating from across the length of
the sample, and if $G$ is sufficiently strong, (or the duration of
the gradient is long enough,) the exponential terms in
\eqref{gradient-coherence} will add up to zero (if $p\ne 0$). The
action of an ideal, strong gradient field (also called a
\textit{crush}), is therefore to ``crush'' all the single and double
quantum terms in the homonuclear density matrix,
\begin{equation}\label{crush-effect}
\begin{pmatrix}
\Box & \Box & \Box & \Box \\
\Box & \Box & \Box & \Box \\
\Box & \Box & \Box & \Box \\
\Box & \Box & \Box & \Box \\
\end{pmatrix}\xrightarrow{\text{crush}}\,\begin{pmatrix}
\Box &  &  &  \\
 & \Box & \Box &  \\
 & \Box & \Box &  \\
 & & & \Box \\
\end{pmatrix}.
\end{equation}
Gradients used in this way, are often called \textit{purge}
gradients \cite{Cavanagh}. I will sometimes call them \textit{crush}
sequences as well. Note that in heteronuclear systems, even the zero
quantum terms will be suppressed. However, researchers have also
come up with techniques for suppressing homonuclear zero quantum
terms; for an example see \cite{GradientZQKeeler}. We shall use
gradient fields in different contexts through the course of this
work.

\subsection{Quantum state tomography}\label{section-tomography-QST}

\textit{Quantum state tomography} \cite{NielsenChuang} is the
process of determining a density operator $\rho$. A two qubit $\rho$
has sixteen elements, and is characterized by fifteen independent
parameters \cite{FanoDensityMatrixRMP}, for example, the fifteen
coefficients in \eqref{po-basis-two-qubits-expand}, and state
tomography entails determining these parameters. Direct detection of
$\rho$, gives access to only the eight direct observables,
\eqref{direct-observables} and so, additional experiments are
needed, converting the unobservable terms into observables. For
example, a two qubit $\rho$ can be characterized in four
experiments; possible sets are the experiments involving,
\begin{eqnarray}
&&\{\mathbf{1},90_x,90_y,90\,I_y\},\quad\text{and}\nonumber\\
&&\{90\,I_x,90\,I_y,90\,S_x,90\,S_y\},\textit{etc.},\ \nonumber
\end{eqnarray}
where $\mathbf{1}$ represents an identity operation, corresponding
to direct detection. (The four-step tomography also suggests that
the nine-step scheme used in \cite{ChuangDeutsch,ChuangGrover} is
excessive.) In some cases, when there is some prior knowledge about
the state, fewer tomography experiments may also be sufficient. One
obvious method is using a crush gradient, which forces all single
and double quantum coherences to the value zero. In Section
\ref{section-partial-twirl}, I present an example of a one-shot
tomography on a state which is constrained to have only two
independent parameters. Recent demonstrations of state tomography in
NMR can be found in
\cite{AnwarPH,EraserQuantum,MehringNuclearELectronEnt}. The
reconstruction of density matrices from read-outs in an NMR
experiment is exemplified in \cite{TomographyNMRLong01}.

\section{Initialization in
NMR}\label{section-initialization-problem-nmr}

As I have mentioned, NMR is very good at unitary operations, with
fidelities approaching unity. However, its main weakness lies in
non-unitary operations, namely initialization and single-spin
read-out. This section discusses the problem of initialization and
the following section touches upon the issue of measurement.

\subsection{Initial states in ``thermal''
NMR}\label{section-initial-state-nmr}

Consider a weakly coupled spin system. The equilibrium density
matrix of an ensemble of $n$ qubits is given by
\cite{LevittSpinDynamics,CoryNMRQCReview,CoryProcNatl},
\begin{equation}\label{equilibrium-rho}
\rho_{eq}=\,\frac{\exp{(-H/kT)}}{Z},
\end{equation}
where,
\begin{equation}\label{partition-function}
Z=\,\text{Tr}(\exp{(-H/kT)})
\end{equation}
is the partition function and serves for normalizing the state and
$H$ is the system Hamiltonian. The operator $\rho_{eq}$ will be
diagonal in the basis of the Hamiltonian $H$ \cite{CohenTannoudji1},
(the Zeeman basis). If $q$, $r$ and $s$ are indices for the energy
levels and $i$ is an index for the qubits, the individual matrix
elements in $\rho_{eq}$ will be given by,
\begin{eqnarray}
\expectation{r}{\rho_{eq}}{s}=&\,{\expectation{r}{\exp{(-H/kT)}}{s}}\bigg/{\sum_{q=1}^{2^n}\expectation{q}{\exp{(-H/kT)}}{q}}\nonumber\\
=&\,{\delta_{rs}\exp{(-E_{r}/kT)}}\bigg/\sum_{q=1}^{2^n}{\exp{(-E_{q}/kT)}}\label{equilibrium-rho-element},
\end{eqnarray}
where $E_r$ is the energy of the $r$th eigenlevel (Zeeman level),
\begin{eqnarray}\label{energy-rth-level-1}
E_r=&\,\hbar\omega_r\quad\text{and}\\
\omega_r=&\,\sum_{i=1}^{n}m_i\omega_i\label{energy-rth-level-2} \\
\stackrel{\text{homonuclear}}{\approx}&\,\omega_i\sum_{i=1}^{n}m_i=\,\omega_iM_r,\label{energy-rth-level-3}
\end{eqnarray}
$\omega_i=-\gamma_iB$ being the Larmor frequency of the $i$th qubit,
and the spin magnetic quantum number \cite{Sakurai}, $m_i$ being
$1/2$ or $-1/2$ depending\footnote{Without loss of generality, I
\textit{assume} that $m=1/2$ for the $\ket{\alpha}$ state and $-1/2$
for the $\ket{\beta}$ state. This translates into $\ket{\alpha}$
being a low energy state for positive $\gamma$ nuclei such as
hydrogen. In fact, understanding the signs of frequencies in NMR is
a convoluted undertaking; see, for example \cite{SignsLevitt}.} on
whether the corresponding qubit is $\ket{\alpha}$ or $\ket{\beta}$.
In writing \eqref{energy-rth-level-2}, I assume that the coupling
frequencies $J$ (hundreds of Hz) are very small as compared to the
Larmor frequencies (hundreds of MHz), and in writing
\eqref{energy-rth-level-3}, I also neglect effects of chemical shift
differences between the qubits.

\begin{figure}
\begin{center}
\includegraphics[scale=0.9]{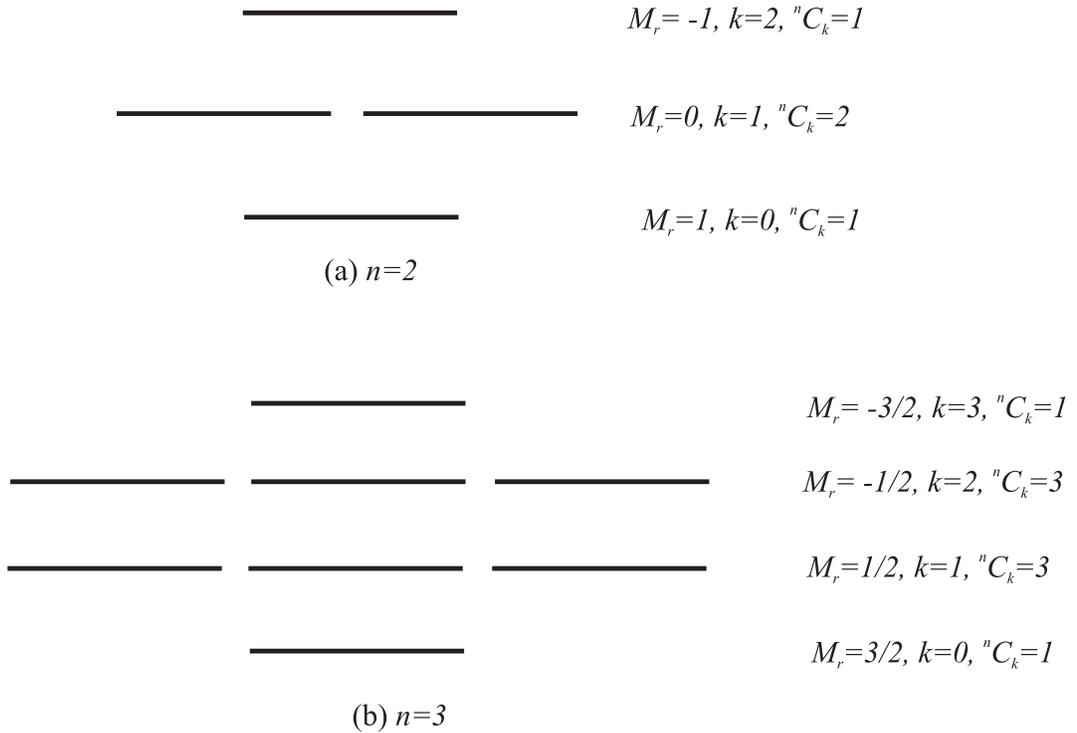}
\caption{Conceptual representation of energy levels in steps on a
Pascal's ladder, and correspondence with the total magnetic quantum
number $M_r$. Case (a) is for two qubits ($n=2$) and case (b)
represents three qubits ($n=3$). The $M_r$, step index $k$ and
number of levels in each step, $^nC_k$ are written alongside the
steps.}\label{Figure-energy-manifold}
\end{center}
\end{figure}

In \eqref{equilibrium-rho-element}, $\expectation{r}{\rho_{eq}}{r}$
represents the \textit{fractional population} in the $r$th level,
suggesting a Boltzmann distribution of qubits in the $2^n$ available
levels. For a homonuclear system, these levels are distributed in
$n+1$ ``steps'', in a Pascal's triangle fashion. The distribution of
levels and steps in the energy manifold is exemplified in Table
\ref{Table-Pascal} and illustrated for two and three qubits in
Figure \ref{Figure-energy-manifold}. The summary is that the
distributions of levels in steps of a ``Pascal's ladder'' is a
direct result of the addition of the magnetic quantum numbers $m_i$.
The sum of the magnetic quantum numbers, $M_r=\sum_im_i$, can take
up a value from the range, $\{n/2,n/2-1,n/2-2,\ldots
,-(n/2-1),-n/2\}$. The number of steps is $n+1$, and if we index the
steps $k=0,1,\ldots ,n$, then the number of levels in the $k$th step
is $^nC_k$, and the $M_r$ of the individual levels belonging to this
step is $n/2-k$.

\begin{table}
\begin{center}
\begin{tabular}{cccc}
\hline   \\
no. of qubits & no. of levels & no. of ``steps'' & levels per step
\\
 \hline \\
 $1$ & $2$ & $2$ & $1$ \\
 $2$ & $4$ & $3$ & $1,2,1$ \\
 $3$ & $8$ & $4$ & $1,3,3,1$ \\
 $\vdots$ &  &  &  \\
 $n$ & $2^n$ & $n+1$ & $1,^nC_1,^nC_2,\ldots ,^nC_{n-1},^nC_n$ \\
  \hline
\end{tabular}
\caption{Arrangement of energy levels in the nuclear spin energy
manifold for $n$ qubit homonuclear systems, neglecting coupling
interactions.} \label{Table-Pascal}
\end{center}
\end{table}

With this background, we can derive a product operator form for an
$n$ qubit initial state. This will also be a convenient entry into
understanding the problem of initialization in NMR. Consider a
single qubit $n=1$; there are two energy levels, each level in a
separate step, $\ket{\alpha}$ being lower in energy than
$\ket{\beta}$. Suppose the fractional populations in these states
are $n_\alpha$ and $n_\beta$; from \eqref{equilibrium-rho-element},
$n_\alpha$ is given by,
\begin{eqnarray}
n_{\alpha}&=\,&\exp{(-E_\alpha/kT)}\big/\bigl(\sum_{q=\alpha ,\beta}\exp{(-E_q/kT)}\bigr)\nonumber\\
&=\,&\exp{(-\hbar\omega_\alpha/kT)}\big/\bigl(\exp{(-\hbar\omega_\alpha/kT)}+\exp{(-\hbar\omega_\beta/kT)}\bigr),\quad\text{(from \eqref{energy-rth-level-1})}\nonumber\\
&=\,&\exp{(-\hbar\omega/2kT)}\big/\bigl(\exp{(-\hbar\omega/2kT)}+\exp{(\hbar\omega/2kT)}\bigr),\quad\text{(from
\eqref{energy-rth-level-3}),\quad}
\end{eqnarray}
where $\omega$ is the Larmor frequency of the first and the only
qubit
. In the high temperature limit, $kT\gg \hbar\omega$, $n_\alpha$ is
approximated as,
\begin{eqnarray}
n_\alpha&{\approx}&\,\bigl(1-\hbar\omega/2kT\bigr)\big/\bigl(1-\hbar\omega/2kT+1+\hbar\omega/2kT\bigr)\nonumber\\
&=&\,{\bigl(1-\hbar\omega/2kT\bigr)}\big/{2}\label{fractional-population-alpha},
\end{eqnarray}
and similarly,
\begin{equation}
n_\beta{\approx}\,{\bigl(1+\hbar\omega/2kT\bigr)}\big/{2}\label{fractional-population-beta}.
\end{equation}
With these fractional populations, the equilibrium state can be
written as,
\begin{eqnarray}
\rho_{eq}&=&\,\diagonal{n_\alpha,n_\beta}\nonumber\\
&=&\,\frac{1}{2}
\begin{pmatrix}
1-\hbar\omega/2kT & 0 \\
0 & 1+\hbar\omega/2kT
\end{pmatrix}\nonumber\\
&=&\frac{1}{2}\bigl(\mathbf{1}_2-\mathcal{B}I_z\bigr)\sim
I_z\label{equilibrium-rho-one-qubit},
\end{eqnarray}
where $\mathcal{B}=\hbar\omega/kT$ is what I refer to as the
\textit{Boltzmann factor}. Similarly, for two homonuclear qubits,
the equilibrium state is
\begin{equation}
\frac{1}{4}\big(\mathbf{1}_4-\mathcal{B}(I_z+S_z)\bigr)\sim
I_z+S_z,\label{equilibrium-rho-2-qubit}
\end{equation}
and for $n$ qubits,
\begin{equation}
\frac{1}{N}\big(\mathbf{1}_N-\mathcal{B}(\sum_{i=1}^nI_{iz})\bigr)\sim
\sum_{i=1}^nI_{iz}.\label{equilibrium-rho-n-qubits}
\end{equation}

For a proton frequency of $400$~MHz, the high temperature
approximation will be valid for $T\gg 0.02$~K, indicating that the
equilibrium states \eqref{equilibrium-rho-n-qubits} are an extremely
good description at temperatures of even a few hundred millikelvins,
and therefore, must also accurately describe our initial states at
room temperature. I shall refer to these high temperature
equilibrium states as \textit{thermal} states in the remainder of
this thesis.

Now some important comments on the energy gaps and population
differences between levels, which are significant because the NMR
signal arises from \textit{differences} in level populations. The
ball-park figures derived in the following discussion all assume a
proton frequency of $400$~MHz and a temperature of $295$~K. The
\textit{Zeeman splitting} defined as the energy gap between levels
in adjacent steps, is always $\hbar\omega$, resulting in the steps
being evenly spaced. This splitting is extremely small, a tiny
$2.6\times 10^{-25}$~J, and the available thermal energy $kT\approx
4\times 10^{-21}$~J is four orders of magnitude larger, resulting in
\textit{almost} equal populations between the levels. Another way of
stating this is that $\mathcal{B}$ is exceedingly small: about
$6.5\times 10^{-5}$. However, the populations in the adjacent levels
are not \textit{exactly} identical, otherwise thermal NMR would not
have been possible in the fist place. In fact, the fractional
population difference between adjacent levels is proportional to
$\mathcal{B}$. For example, in a single qubit, the difference is
$\mathcal{B}/2\approx 3.2\times 10^{-5}$: which means that for
roughly every $3\times 10^4$ spins, there will be one extra spin in
the lower $\ket{\alpha}$ level and if the sample contains $10^{15}$
spins, the net excess of spins will be of the order of $10^{10}$;
not a very large number, keeping in view the very small magnetic
moment of the nucleus. For all other nuclei (except radioactive
tritium, ^3\H_1), the Zeeman splitting is even smaller.

\subsection{Pseudopure initial states}\label{section-pseudopure}

The foregoing discussion explains the origins of the notoriously low
sensitivity of NMR arising from the tiny Zeeman splittings: small
gaps lead to small population differences which in turn lead to
small signals. The insensitivity is a problem in analytical NMR,
especially if the concentration of the sample is small. A lot of
effort, therefore, is devoted to attaining increased polarization or
hyperpolarization and some techniques for achieving this will be
touched upon in Section \ref{section-hyper-polarization}. However,
in the context of quantum information processing, the low magnetic
to thermal energy ratio is also a limiting factor in a slightly
different way and I now outline this limitation.

Every computational process must start off by placing the
computational register in a well-defined initial state; for a
quantum computer, we normally expect the (register of) qubits to be
placed in the energetically ground state
$\ket{\mathbf{0}}=\ket{00\ldots 0}=\ket{\alpha\alpha\ldots \alpha}$,
a \textit{pure} quantum state. For example, for two qubits, a
suitable initial state would be,
\begin{eqnarray}
\diagonal{1,0,0,0}=&\,\frac{\mathbf{1}_4}{4}+\frac{I_z+S_z+2I_zS_z}{2}\nonumber\\
\sim&\,I_z+S_z+2I_zS_z\label{initial-state-ideal},
\end{eqnarray}
having only one non-zero eigenvalue. What nature furnishes instead
is the thermal state, \eqref{equilibrium-rho-2-qubit} $\sim
I_z+S_z=\diagonal{1,0,0,-1}$, having two non-zero eigenvalues. We
immediately encounter two problems: \textit{first}, the thermal and
desired initial states are not identical and \textit{second}, the
amount of the (deviation) thermal state is proportional to a very
small number, $\mathcal{B}$. How do we then prepare a pure initial
state, suitable for quantum computing? The answer was provided by
Cory \etal , \cite{CoryProcNatl,CoryPhysComp}, who suggested that it
is \textit{not} necessary to prepare pure states, rather it is
sufficient to prepare only \textit{pseudopure} states, that are
given by the general form,
\begin{equation}\label{pseudopure-n-qubit}
\rho_{ps}=\,(1-\varepsilon)\frac{\mathbf{1}_N}{N}+\varepsilon\ket{\psi}_p\bra{\psi}_p,
\end{equation}
the subscript ``$ps$'' and ``$p$'' indicating pseudopure and pure
states. For a two qubit system, a conventional pseudopure state is,
\begin{equation}\label{pseudopure-2-qubits}
\rho_{ps}=\,(1-\varepsilon)\frac{\mathbf{1}_4}{4}+\varepsilon\ket{00}\bra{00},
\end{equation}
comprising $\varepsilon$ parts of the desired pure state mixed with
$1-\varepsilon$ parts of the maximally mixed state. This factor,
$\varepsilon$, is what I term the \textit{polarization}, signifying
the amount of ``polarized'' pure component in the overall state, the
maximally mixed (unpolarized) state being NMR silent. The dynamics
of $\rho_{ps}$ are exactly identical to $\ket{00}\bra{00}$, and so
the pseudopure state effectively behaves as a pure state, albeit
with smaller polarization and hence lower sensitivity.

How do we prepare these effective pure states? The eigenvalue
spectra for the thermal and pure states are different, and so some
kind of non-unitary operation is required
\cite{CohenTannoudji1,UnitaryBoundsSorensen}: as a result, the
different approaches for assembling pseudopure states all involve
some kind of non-unitary averaging of eigenvalues. At the end of the
averaging process, we are left with one \textit{non-degenerate}
eigenvalue. (A pure state has one \textit{non-zero} eigenvalue.)

The different strategies for pseudopure state generation are
reviewed in \cite{JonesRev,VandersypenThesis}. In summary, the
earliest proposals \cite{CoryProcNatl,CoryPhysComp}, called
\textit{spatial averaging}, employed field gradients, whereby the
averaging is achieved by entangling the spin degrees of freedom with
spatial dimensions and later taking a partial trace over space.
Another approach \cite{KnillChuangEffectivePure,TemporalMarx},
called \textit{temporal averaging}, performs the averaging by
carrying out multiple experiments, each experiment using a different
initial state unitarily derived from the thermal state. This second
approach, in a sense, entangles the spin degrees of freedom with
time and throws away the time information at the end. A third scheme
\cite{Gershenfeld,LogicalLabellingVandersypen}, called
\textit{logical labelling}, selects a subset of qubits embedded in a
higher dimensional quantum space, conditioned upon the state of the
unselected qubits. This last approach is elegant in principle, but
the complexity of the pulse sequences has limited its applicability
to small homonuclear systems. There is yet another efficient
strategy \cite{KnillCatBenchamrk} for ``extracting'' an $n-1$ qubit
pseudopure state from $n$-multiple quantum coherence terms (related
to ``cat'' states \cite{NielsenChuang}) which are selected using
filtration sequences based on the use of gradients or phase cycling
\cite{LevittSpinDynamics}.

\subsection{Climbing ``Mount Scalable''?}\label{section-scalable}

A large part of this thesis concerns the preparation of a two qubit
NMR based quantum computer, so what method do we use to prepare
pseudopure states? The answer is, simply, we do not prepare
pseudopure states, instead our system is directly initialized in an
almost pure state. A part of the very first DiVincenzo requirement
\cite{DiVincenzoCrit} is scalability and the pseudopure strategy
does not satisfy this requirement, as I now review. Most of the
following criticism was anticipated in Warren's paper \cite{Warren}
and more comprehensive reviews on the scalability issue can be found
in \cite{JonesNMRcrit,MountScalable}, the latter reference being the
source of the term ``Mount Scalable''.

Warren derived an upper bound on the polarization factor for a
pseudopure state in a homonuclear system, which I call the
\textit{Warren bound}. The derivation is based on the fact that the
theoretical maximum polarization achievable from a thermal state is
upper bounded by the maximum fractional population difference in an
$n$ qubit system: the maximum difference is between the
$\ket{\alpha\alpha\ldots \alpha}$ and $\ket{\beta\beta\ldots \beta}$
states, which are the furthest apart in energy. These ``all
$\alpha$'' and ``all $\beta$'' states have magnetic quantum numbers
$n/2$ and $-n/2$, and using
\eqref{equilibrium-rho-element}--\eqref{energy-rth-level-3}, the
corresponding difference\footnote{This difference enables excitation
of the multiple quantum coherence of order $n$ \cite{FreemanChoreo}.
In reference \cite{Warren}, the expression
\eqref{Warren-bound-derive-1} does not carry the negative sign,
however, this sign is consistent with our assumption that for
positive $\gamma$ (and negative $\omega$) nuclei, $\ket{\alpha}$ is
the lower energy state, resulting in $n_{\text{all\
}\alpha}>n_{\text{all\ }\beta}$.} can be derived as,
\begin{eqnarray}
n_{\text{all\ }\alpha}-n_{\text{all\
}\beta}&=&\,{\bigl(\exp{(-n\hbar\omega/2kT)}-\exp{(n\hbar\omega/2kT)}\bigr)}\big/{Z}\nonumber\\
&=&\,{-2\sinh{(n\hbar\omega/2kT)}}\big/{Z}.\label{Warren-bound-derive-1}
\end{eqnarray}
The partition function $Z$ can be written in compact form as the
Binomial expansion,
\begin{eqnarray}
Z&=&\,\sum_{k=0}^n\,^nC_k\bigl(\exp{(-\hbar\omega/2kT)}\bigr)^{n-k}\bigl(\exp{(\hbar\omega/2kT)}\bigr)^{k}\nonumber\label{Warren-bound-derive-2}\\
&=&\,\bigl(\exp{(-\hbar\omega/2kT)}+\exp{(\hbar\omega/2kT)}\bigr)^n\label{Warren-bound-derive-3}\nonumber\\
&=&\,2^n\cosh^n{(\hbar\omega/2kT)}\label{Warren-bound-derive-4},
\end{eqnarray}
and therefore, Warren's bound is given by,
\begin{equation}
(n_{\text{all\ }\alpha}-n_{\text{all\ }\beta})\big/Z
=\,{-2\sinh{(n\hbar\omega/2kT)}}\big/\bigl(2^n\cosh^n{(\hbar\omega/2kT)}\bigr),\label{Warren-bound-derive-5}
\end{equation}
which in the limit of high temperature and ignoring the negative
sign, is,
\begin{equation}
\frac{n}{2^n}(\hbar\omega/kT)=\,\frac{n}{2^n}\mathcal{B}\label{Warren-bound-high-T},
\end{equation}
where I have used the limiting values of the hyperbolic functions,
$\lim_{x\rightarrow 0}{\sinh{x}}=x$ and $\lim_{x\rightarrow
0}{\cosh{x}}=1$.

The exact polarization extracted from the the thermal state, depends
on the method used to prepare the pseudopure state, however in all
cases, the Warren bound \eqref{Warren-bound-high-T} sets an upper
limit. In the high temperature limit, the bound results in an
exponential decrease in polarization as we linearly increase the
number of qubits; and therefore to defeat the exponential loss in
sensitivity, exponentially more resources are required, such as
exponentially larger samples. This is not only impractical but also
defeats the fundamental gain expected (and desired) from a quantum
computer, that is exponentially bigger Hilbert dimensionality with
at the most, polynomial resources. It appears that thermal NMR based
on assembling pseudopure states is not scalable and this statement
is the gist of the initialization problem.

\subsection{Concentrating
polarization}\label{section-schulman-vazirani-algorithmic}

However, the literature suggests that the problem of low
polarization and scalability can be sidestepped, using a
computational (technology independent) approach outlined by Schulman
and Vazirani \cite{SchulmanVazirani}. The method centres around
polarization \textit{concentration}: given an ensemble of $n$ weakly
polarized qubits, it is possible to extract $k<n$ hyperpolarized
qubits, at the expense of maximally randomizing the remaining $n-k$
qubits. The scheme for extracting pure qubits can be represented as,
\begin{equation}\label{schulman-vazirani-scheme}
\bigotimes_{n}\bigl((1-\varepsilon)\frac{\mathbf{1}_2}{2}+\varepsilon\ket{0}\bra{0}\bigr)\xrightarrow{\text{S.
V.}}\,\bigl(\bigotimes_{k}\ket{0}\bra{0}\bigr)\otimes\bigl(\bigotimes_{n-k}\frac{\mathbf{1}_2}{2}\bigr),
\end{equation}
$\bigotimes_j$ denoting the tensor product between $j$ qubits.  The
algorithm operates on classical states and conserves the classical
entropy $S$ \cite{NielsenChuang} of the system. From this condition,
an upper limit on the number of pure qubits extracted can be
determined: the entropy of a single qubit with polarization
$\varepsilon$ (and fractional populations $(1\pm\varepsilon)/2$) is
\begin{equation}\label{entropy-def}
S(\varepsilon)=\,-(\frac{1+\varepsilon}{2})\lg_2{(\frac{1+\varepsilon}{2})}-(\frac{1-\varepsilon}{2})\lg_2{(\frac{1-\varepsilon}{2})},
\end{equation}
and as the entropy is conserved, we can write,
\begin{eqnarray}
nS(\varepsilon)&=&\,kS(1)+(n-k)S(0)\label{shannon-bound}\\
&=&\,k(0)+(n-k)(1)\nonumber\\
\implies k&=&\,n(1-S(\varepsilon))\nonumber\\
&=&n\bigl(1+(\frac{1+\varepsilon}{2})\lg_2{(\frac{1+\varepsilon}{2})}+(\frac{1-\varepsilon}{2})\lg_2{(\frac{1-\varepsilon}{2})}\bigr),\\
\end{eqnarray}
and using the expansion $\lg_2(1\pm x)/2=-1\pm
x/\ln{2}-x^2/(2\ln{2})\pm\mathcal{O}(x^3)$ for small $x$, 
\begin{equation}
k\approx\frac{n\varepsilon^2}{2\ln{2}}\approx
n\varepsilon^2\quad(\text{for large\
}n).\label{schulman-vazirani-scaling}
\end{equation}
The equation indicates that we can extract $k$ pure qubits from
$k/\varepsilon^2$ low-bias qubits; the scaling is only linear in
$k$. This may seem to be an exciting possibility for NMR but taking
a more practical point of view, the restrictively small values of
$\varepsilon\approx 10^{-5}$ in thermal NMR, require a prohibitively
large ($\approx 10^{10}$) number of initial qubits, even to extract
a single pure qubit. However, despite this limitation, an NMR
demonstration of the basic building block in the polarization
boosting has already been demonstrated \cite{VandersypenSchulman}.

Another variant of the above technique is the so-called
\textit{algorithmic cooling} method \cite{AlgCoolProcNatlAcad} which
is an entropy non-conserving algorithm for polarization
concentration. The method beats the Shannon bound
\eqref{shannon-bound} through the use of rapidly relaxing qubits in
an open quantum system and an initial implementation of this
non-adiabatic version has also been reported \cite{AlgoCoolingJose}.
However, it seems that both of these computational schemes will
become practically useful in small sized spin systems, only when
used in conjunction with higher initial polarizations.

\subsection{Re-initialization}\label{section-re-initialize}

Quantum error correction
\cite{ErrorCorrectionShor,ErrorCorrectionSteane} is an attempt to
counteract the effects of decoherence, reversing errors once they
have occured. Thermal NMR is not well suited for this purpose,
because of two major reasons: \textit{first}, most error detection
and correction schemes work by encoding a physical qubit in several
logical qubits, the increased number of qubits reducing the
sensitivity; and \textit{second}, repeated error correction requires
one to reuse the ancillary qubits and currently, no practical
schemes are known for resetting NMR qubits into pure states at will.
Thermal NMR cannot realize pure states \textit{initially}, let alone
provide a \textit{continuous} supply of pure qubits \textit{during}
the experiment and so, even though, simple error detection
\cite{ErrorCorrectionVandersypen} and correction
\cite{ErrorCorrectionNMRCory98} have now been implemented in small
NMR systems, they mostly remain proof of principle experiments.

\section{Measurement in NMR}\label{section-measurement}

Liquid state NMR does not implement projective measurements
\cite{NielsenChuang} on single qubits. A projective (``strong'')
measurement on a qubit in the state
$(c_0\ket{0}+c_1\ket{1})(c_0^*\bra{0}+c_1^*\bra{1})$, yields
$\ket{0}$ with probability $|c_0|^2$ and $\ket{1}$ with probability
$|c_1|^2$, leaving the state in $\ket{0}\bra{0}$ or
$\ket{1}\bra{1}$, also showing that such measurements constitute one
(though not the only) way of initializing the qubit. In liquid state
NMR, we do not have access to single qubits but rather ensembles of
qubits, and the so-called ``weak'' measurements performed in these
systems provide ensemble averages \cite{CoryProcNatl}: the final
result is a noisy expectation value of a spin operator, such as
$\average{I_x}$ or $\average{I_y}$, taken over \textit{all}
molecules in the sensitive region of the detection coil.
Furthermore, the wavefunction does not collapse into one
post-measurement state as the system is only feebly affected by the
detection process \cite{SignsLevitt,DetectionVirtualPhotons}. In
short, it appears that these weak measurements are statistically
equivalent to standard, single-qubit measurements, and will suffice,
until we make enough progress to start actively considering
resetting qubits for quantum error correction.

It is not out of place to comment upon a few research initiatives
aimed at single-spin or single-molecule detection. Single sub-atomic
particles have small magnetic moments \cite{CohenTannoudji1} and are
difficult to measure in normal circumstances. There has, however,
been recent progress in detection of single electron spins using the
spin-dependent magnetic force on a ferromagnetic tip
\cite{SpinDetectionMRFM} and spin-dependent charge transport
\cite{SpinDetectionVandersypen}. Direct spin detection of protons
and nuclear spins is even more difficult, as their gyromagnetic
ratios are much smaller. The detection of single nuclear spins,
therefore, requires some kind of ``amplification'' of nuclear Zeeman
splittings by correlating them with the wider energy gaps present in
another degree of freedom. For example, in the first
optically-detected magnetic resonance experiments
\cite{ODMR1,ODMR2}, the magnetic transitions (microwave) in the
electronic spectrum of a \textit{single molecule} were indirectly
detected as optical transitions, which are intense and can be
monitored with much higher quantum efficiencies. Another interesting
experiment in this regard is the optical detection of nuclear
magnetic transitions in a single quantum well
\cite{ODNMRQuantumWell}. It is widely felt that any envisaged
single-nuclear spin detection scheme is likely to involve detection
of optical photons.

I have now discussed liquid state NMR in terms of the DiVincenzo
criteria. There remains another important and more fundamental
aspect of thermal NMR that has even raised questions about its
acceptance as a truly ``quantum'' process. In the next section, I
take up the widely debated issue of entanglement in liquid state
NMR.

\section{Entanglement in NMR}\label{section-entanglement-NMR}

Entanglement \cite{NielsenChuang} is undoubtedly an important
resource in quantum information processing; some researchers would
argue that it is \textit{the} ``characteristic trait of quantum
mechanics'' \cite{Schrodinger}. Entanglement is interesting because
it enables quantum states to exhibit correlations that have no
classical analog. Of its many example applications in quantum
information science, entanglement enables quantum teleportation
\cite{TeleportBennettOriginal,TeleportBeowmeesterExp}, superdense
coding \cite{SuperdenseCodingBennettOriginal,BellInterConversion}
and secure communication \cite{BB84,BBB92}.

A bipartite pure state will be entangled if its Schmidt number is
greater than one \cite{NielsenChuang} and a mixed state of two or
more qubits will be entangled if it cannot be written as the convex
sum of the direct products of the component density operators
(Section \ref{section-separable-nmr}). Several operational criteria
for detecting and quantifying entanglement will be taken up in the
following subsections.

As an example of the consequences of entanglement, consider the
maximally entangled state,
\begin{equation}\label{phi-plus}
\ket{\phi^+}=\,\frac{1}{\sqrt{2}}(\ket{01}+\ket{10}).
\end{equation}
Tracing out the second qubit, the reduced density matrix
\cite{NielsenChuang} of the first qubit will be the maximally mixed
state, $\rho_A=\,1/2\,\diagonal{1,1}$, implying that measuring this
qubit along any axis would result in a random result, and thus
measuring one subsystem \textit{alone} provides no information about
the preparation of the composite state, \eqref{phi-plus}. In this
sense, the entangled state contains ``hidden information'' which
cannot be revealed by measurements on either qubit. However, if both
subsystems are measured and the results are compared, they will be
perfectly correlated.

Conventional NMR based QIP uses highly mixed, separable
(non-entangled) states. In this section we explore the relations
between entanglement and the purported power of quantum computers,
more specifically, NMR quantum computers. For example, can NMR
devices exhibit entanglement? And are current devices truly quantum?
In fact, most of the work in the current thesis addresses the
two-fold problem of initialization and entanglement. Subsequent
chapters will discuss in detail, our experimental implementations,
of \textit{initializing} a two qubit system in an \textit{entangled}
state. A gentle introduction to the role of entanglement in NMR can
also be found in \cite{EntanglementNMRLaflamme}.

\subsection{Separable and entangled states}\label{section-separable-nmr}

Traditional QIP experiments using liquid state NMR furnish and
employ separable states. A two qubit pure state will be separable if
it can be written as a tensor product
\cite{NielsenChuang,CohenTannoudji1} of the component
pure states, 
\begin{equation}\label{separable-pure}
\ket{\chi}_{AB}=\,\ket{\psi}_A\otimes\ket{\phi}_B,
\end{equation}
where $A$ and $B$ label the qubits. Examples of separable states
include $\ket{00}=\ket{0}\otimes\ket{0}$ and
$(\ket{01}+\ket{11})/\sqrt{2}=\bigl((\ket{0}+\ket{1})/\sqrt{2}\bigr)\otimes\ket{1}$.
A pure state is non-separable, or entangled, if it cannot be brought
into the form \eqref{separable-pure}; examples are the Bell states
\cite{NielsenChuang}, defined as, 
\begin{eqnarray}
\ket{\phi^\pm}=&\,\bigl(\ket{00}\pm\ket{11}\bigr)/{\sqrt{2}}\label{Bell-phi}\\
\ket{\psi^\pm}=&\,\bigl(\ket{01}\pm\ket{10}\bigr)/{\sqrt{2}}\label{Bell-psi}.
\end{eqnarray}

However, as pointed out earlier, states encountered in traditional
NMR are mixed and have exceedingly small polarizations. Separability
for such mixed states is, at least, well-\textit{defined}: a two
qubit mixed state $\rho$ will be separable if it can be
decomposed as a weighted sum of \textit{product} states, 
\begin{equation}\label{separable-mixed}
\rho=\,\sum_i\,p_i\,\bigl(\rho_A\otimes\rho_B\bigr)_i,\quad\text{with\
} p_i\ge0{\text{\ and\ }\sum_ip_i=1}.
\end{equation}
The product state $\bigl(\rho_A\otimes\rho_B\bigr)_i$, prepended
with a non-negative classical probability $p_i$, has the property
that the $A$ and $B$ quantum sub-systems are independent: knowledge
of one sub-system does not depend on or affect the other sub-system.
Product states are not entangled; and neither are their convex
combinations, such as \eqref{separable-mixed}. All states in
low-polarization NMR can be written in the form
\eqref{separable-mixed}, and are therefore, always separable.

A state we shall frequently encounter in this work is the Werner
state \cite{WernerState}. For two qubits the state is given by,
\begin{equation}\label{Werner-def}
\rho_\varepsilon=\,(1-\varepsilon)\frac{\mathbf{1}_4}{4}+\varepsilon\ket{\psi_{\textit{ent}}}\bra{\psi_{\textit{ent}}},
\end{equation}
where $\ket{\psi_{\textit{ent}}}$ is one of the four maximally
entangled Bell states, \eqref{Bell-phi} and \eqref{Bell-psi}.
Whether $\rho_{\varepsilon}$ is entangled or not, depends on the
precise value of $\varepsilon$, and is investigated in Section
\ref{section-entanglement-2-qubit}. The headline result is that the
Werner state \eqref{Werner-def} will be entangled when
$\varepsilon>1/3$.

\subsection{Separability bounds for mixed states in
NMR}\label{label-bounds-separability}

Until now, I have defined separability and given examples of two
qubit entangled states. Can we also come up with general
separability tests for $n$ qubit states, where $n\ge 2$? This
question is important in understanding the scalability
characteristics of quantum computers, and in context of NMR, was
first addressed by Braunstein \etal\ \cite{Braunstein}. The major
results of their work and some interesting observations are
summarized in this section\footnote{It should be noted that since
Braunstein's original work, tighter bounds have been proposed
\cite{Pittenger,Gurvits}. These new bounds modify the finer details,
but do not affect the overall broad results.}. A general
method\footnote{The method is based on decomposing a density
operator in an over-complete basis, and based on the constraints on
the operator eigenvalues, looking for conditions that guarantee
non-negative probabilities in \eqref{separable-mixed}.} for deriving
the lower bound and constructing explicit decompositions of
separable states can also be found in
\cite{CavesSeparableLowerBound}.

\begin{figure}
\begin{center}
\includegraphics[scale=]{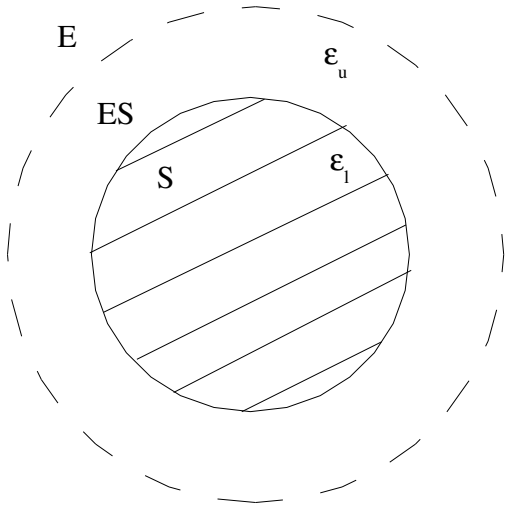}
\caption{Lower $\varepsilon_l$ and upper $\varepsilon_u$ bounds as
constructed by Braunstein and co-workers \cite{Braunstein}. The
regions \textbf{S}, \textbf{E} and \textbf{ES} indicate the
existence of definitely separable, definitely entangled and
``maybe-entangled'' states. The maximally mixed $M_n$ lives in the
middle of \textbf{S}. The continuous \textbf{S}-\textbf{ES} boundary
belongs to \textbf{S}, and the dashed \textbf{ES}-\textbf{E}
boundary belongs to \textbf{ES}.}\label{Figure-ESregions}
\end{center}
\end{figure}

\begin{figure}
\begin{center}
\includegraphics[scale=0.8]{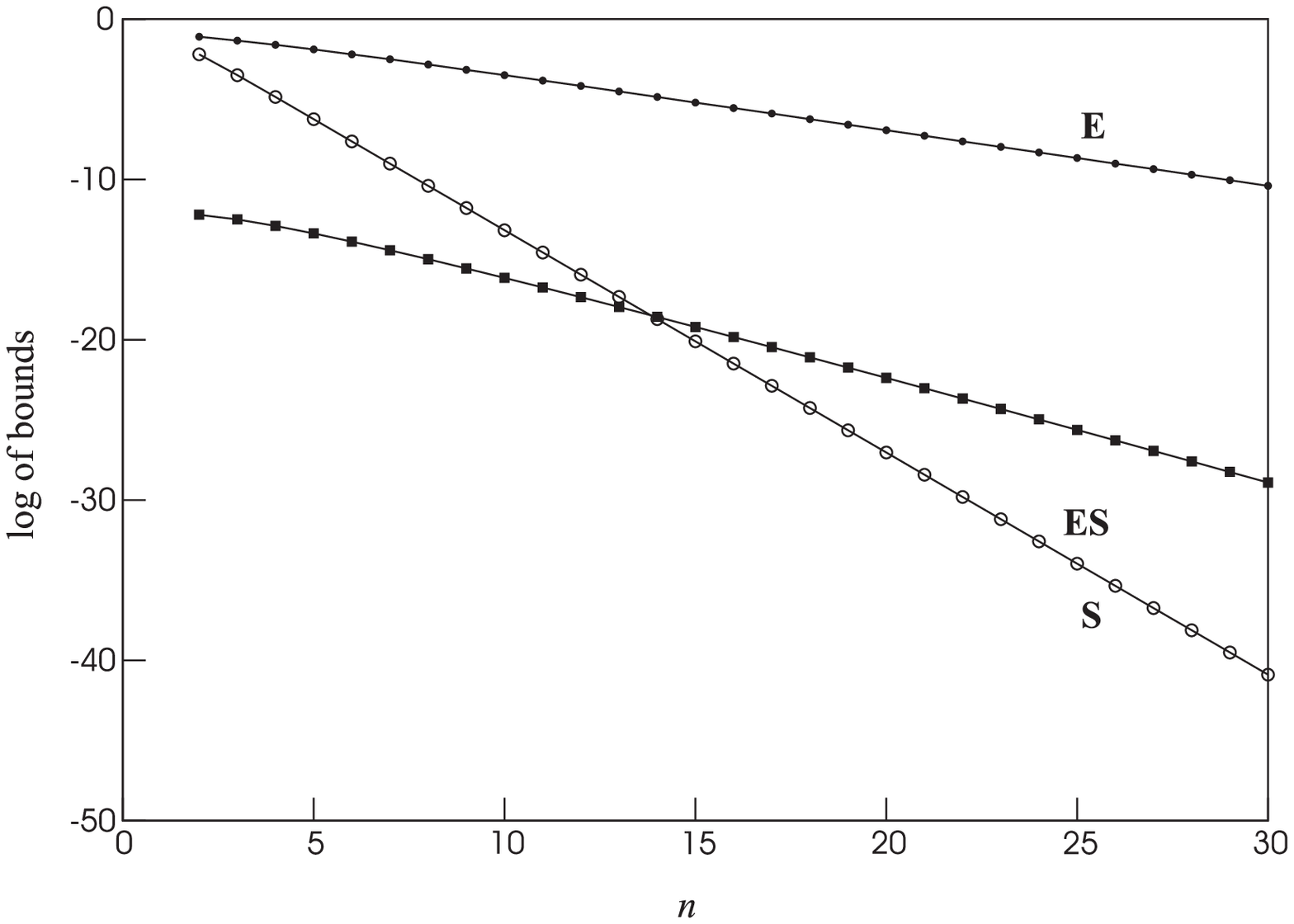}
\caption{Lower ($\varepsilon_l$) and upper ($\varepsilon_u$)
separability and Warren bounds as a function of the number of qubits
$n$. The ordinate scale is logarithmic. The line with the open
circles indicates $\varepsilon_l$, the \textbf{S}-\textbf{ES}
boundary and the one with filled circles represents $\varepsilon_u$,
separating \textbf{ES} and \textbf{E}. The line identified with
squares is the Warren bound, plotted assuming
$\mathcal{B}=10^{-5}$.}\label{Figure-bounds}
\end{center}
\end{figure}

NMR deals with highly mixed states. The maximally mixed state
$M_n=\mathbf{1}/2^n$ is clearly separable, as are states lying
sufficiently close to $M_n$. Braunstein and co-workers determined
\cite{Braunstein} lower and upper bounds on the size of the
separable neighbourhoods of $M_n$, for arbitrary $n$. The concept is
illustrated in Figure \ref{Figure-ESregions}, indicating that the
bounds result in three distinct regions: \textbf{S}, where all
states are provably separable, \textbf{E}, where entanglement can
definitely exist, or in other words, where the states are unitarily
equivalent \cite{GlaserScience98} to entangled states, and a region
in between, \textbf{ES}, where it is not clear whether entangled
states can or cannot exist. Consider the $n$ qubit state, with a
``distance'' $\varepsilon$ from $M_n$,
\begin{equation}
\rho=\,(1-\varepsilon)M_n+\varepsilon
\rho_1,\label{separable-neighbourhood}
\end{equation}
$\rho_1$ being an arbitrary density matrix. For sufficiently small
$\varepsilon$,
\begin{equation}\label{lower-bound-Braunstein}
\varepsilon\le\varepsilon_l=\frac{1}{1+2^{2n-1}}\stackrel{\text{large\
} n}{\approx}\frac{2}{4^n},
\end{equation}
$\rho$ lies in the region $S$. Similarly, for sufficiently large
values\footnote{The upper bound is determined by projecting an
arbitrary $2^d$ dimensional state of two \textit{qudits} onto a four
dimensional Werner state of two \textit{qubits}, and using Werner's
separability criterion \cite{PeresWernerSep} to estimate the
threshold.} of $\varepsilon$,
\begin{equation}\label{upper-bound-Braunstein}
\varepsilon>\varepsilon_u=\frac{1}{1+2^{n/2}}\stackrel{\text{large\
} n}{\approx}\frac{1}{2^{n/2}},
\end{equation}
entangled states \textit{can} exist. These bounds are shown in
Figure \ref{Figure-bounds}.

Some important observations emerging from the Braunstein and Warren
bounds are the following.
\begin{enumerate}
\item As the number of qubits increases, the size of \textbf{S}
shrinks, increasing the \textit{likelihood} of entanglement, even
for reasonably small values of $\varepsilon$. 
\item For a typical value of $\mathcal{B}\approx 10^{-5}$, and the
modest number of qubits presently employed in NMR quantum computers,
(the current record is at seven qubits \cite{VandersypenShor},) all
states lie in the region \textbf{S}. This has lead to interesting
discussions raising suspicions about the very nature of liquid state
NMR as a truly ``quantum'', computing paradigm.
\item Fixing $\mathcal{B}$, we can estimate the number of qubits at
which the \textbf{S} to \textbf{ES} cross-over will take place. For
$\mathcal{B}=10^{-5}$, this transition would occur with fourteen
qubits.
\item If instead of fixing $\mathcal{B}$, we choose to fix $n$ at seven qubits,
then $\mathcal{B}$ should be at least $0.002$ for the transition
into \textbf{ES}. On modern high-field spectrometers, this
translates into a maximum temperature of about $10$~K; most liquid
samples would be frozen at this temperature.
\item Figure \ref{Figure-bounds} also suggests that with low
polarization NMR, it will never be possible to cross over into the
provably entangled region \textbf{E}: a radically different approach
is therefore needed to realize entanglement in liquid state NMR.
\end{enumerate}

\subsection{(Quantum) nature of liquid state NMR
(quantum) information processing}\label{section-quantum-nature-NMR}

The above findings have led to considerable concern about the
\textit{nature} of liquid state NMR: the statistics of experimental
outcomes from separable states can be described perfectly well using
classical probability distributions, and therefore, in its very
essence, liquid state NMR may not be quantum at all
\cite{SchackClassicalNMR}! This might seem to be the end of the
story for liquid state NMR \textit{quantum} computing, but luckily,
the details are much subtler.

To start with, there are differing ``definitional'' stances on the
``quantumness'' of liquid state NMR, views revolving around the
question of whether entangled states are a \textit{prima facie}
signature of quantum behaviour, or if the real quantum character
lies in quantum mechanical dynamics, or possibly somewhere else. No
doubt, the quantum states are separable, but what about the
dynamics? Can quantum information also ``reside'' in states that are
not entangled, as discussed in the context of quantum discord
\cite{DiscordOllivier}? Finally, can we come up with efficient
classical models to simulate the \textit{dynamics} of separable
states?

In short, the role of entanglement in quantum computing \textit{per
se}, remains an open question
\cite{EntanglementNMRLaflamme,EntanglementNMRBrassard}. At the one
end, entanglement has been viewed as the source of the enhanced
information processing capabilities of quantum computers, a view
resonating in \cite{EkertJozsa98}. But at the same time, liquid
state NMR (quantum) information processors routinely implement
quantum dynamics governed by the Liouville-von Neumann Equation
\cite{ErnstNMR}, even though they do not exhibit entangled states.
Having said this, we must also remember that efficiency arguments
must be analyzed in light of the fact that the enhanced power of
quantum computers is only an assumption, although a very tempting
one, and except for the black box or oracle case, there is no proof
as yet of the increased power of quantum processors. Second, quantum
algorithms outperform only the best \textit{known} classical
algorithms. For example, in the case of the factoring problem, one
day somebody might also come up with a classical algorithm for
efficient factoring \cite{ShorAlgo}. In this case, most of our
fundamental perspectives on quantum information processing would be
subject to major change anyway. In light of this discussion, we can
safely conclude, that the essential nature of high temperature
liquid state NMR is an open-ended question, and is intertwined with
the \textit{supposed} role of entanglement in conferring
\textit{arguably} better information processing capabilities to
quantum computers.

The present work, however, by-passes this interesting discussion by
preparing \textit{de facto} entangled states; the first
demonstration of entanglement in liquid state NMR. As a result, most
of the charges made against the very nature of liquid state NMR
implementations can no longer be directed against our two qubit
quantum computer.

\subsection{Detecting entanglement in two qubit
states}\label{section-entanglement-2-qubit}

One of the main claims made in the forthcoming chapters is that we
achieve entangled states in liquid state NMR. In this respect, one
might ask the following questions. How can entanglement be
\textit{detected} in our two qubit system?
Finally, can we go one step beyond detection, and actually
\textit{quantify} the amount of entanglement? This section deals
with these questions. Useful primers on the subject of detecting and
quantifying entanglement are the references
\cite{SeparabilityPrimerLewenstein,EntanglementCharacterizeBruss}.

\subsubsection{Positivity of partial
transpose}\label{section-PPT-test}

For a two qubit state, entanglement can be detected, for example,
through the so-called positivity of partial transpose (PPT) test,
first proposed and exemplified in \cite{PeresWernerSep}. Consider
the separable mixed state \eqref{separable-mixed}. Taking the
transpose of the reduced density matrix \cite{CohenTannoudji1},
$\rho_A$, results in an overall state, 
\begin{equation}\label{PPT-1}
\rho^{T_A}=\,\sum_i\,p_i\,\bigl(\rho_A^T\otimes\rho_B\bigr)_i,\quad\text{with\
} p_i\ge0{\text{\ and\ }\sum_ip_i=1},
\end{equation}
where $T$ denotes the transpose operation. As the reduced matrix
$\rho_A$ is hermitian (being a valid density operator
\cite{NielsenChuang}), $\rho^{T}_A=\rho_A^*$ is also an hermitian,
unit trace matrix and hence another legitimate density operator.
This, in turn, implies that the state $\rho^{T_A}$ is also a valid
density operator and has non-negative eigenvalues. The PPT test,
therefore, detects non-separability by searching for negative
numbers in the eigenvalue spectrum of the partial
transpose\footnote{It is equally valid to find $\rho^{T_B}$ instead
of $\rho^{T_A}$.} of $\rho$.

The PPT is a simple operational test and I illustrate it with a
relevant example. Consider the Werner state,
\begin{eqnarray}\label{Werner-singlet}
\rho_{\varepsilon}=&\,(1-\varepsilon)\frac{\mathbf{1}_4}{4}+\varepsilon\ket{\psi^-}\bra{\psi^-}\\
=&\,\begin{pmatrix}
\frac{1-\varepsilon}{4} & 0 & 0 & 0 \\
0 & \frac{1+\varepsilon}{4} & -\frac{\varepsilon}{2} & 0\\
0 & -\frac{\varepsilon}{2} & \frac{1+\varepsilon}{4} & 0\\
0 & 0 & 0 &
\frac{1-\varepsilon}{4}\label{Werner-singlet-matrixform},
\end{pmatrix},
\end{eqnarray}
and taking the partial transpose with respect to the first qubit,
\begin{equation}
\rho^{T_A}=\,
\begin{pmatrix}
\frac{1-\varepsilon}{4} & 0 & 0 & -\frac{\varepsilon}{2} \\
0 & \frac{1+\varepsilon}{4} & 0 & 0\\
0 & 0 & \frac{1+\varepsilon}{4} & 0\\
-\frac{\varepsilon}{2} & 0 & 0 &
\frac{1-\varepsilon}{4}\label{Werner-singlet-PT},
\end{pmatrix},
\end{equation}
we obtain an eigenvalue spectrum,
$\{(1+\varepsilon)/4,(1+\varepsilon)/4,(1+\varepsilon)/4,(1-3\varepsilon)/4\}$.
For legitimate values of $\varepsilon$, the first three eigenvalues
are always positive, whereas the fourth eigenvalue will be
non-negative only if $\varepsilon\le 1/3$. This immediately suggests
that an \text{in}separable or entangled $\rho_\varepsilon$ will
always have $\varepsilon>1/3$. In fact, the $1/3$ threshold can also
be deduced by setting $n=2$ in Braunstein's upper bound,
\eqref{upper-bound-Braunstein}. Horodecki, Horodecki and Horodecki
went on to show \cite{HorodeckiSeparability} that the positivity of
the partial transpose is not only a necessary, but also a sufficient
requirement for separability of systems of dimensions $2\times 2$
and $2\times 3$, but the result cannot be extended to higher
dimensional systems. (For example, there exist
\cite{BoundEntanglement} entangled states which satisfy the PPT
test; such states are called \textit{bound} entangled states, and
are undistillable \cite{BoundEntanglementUndistillable}.) 
Multi-partite entanglement, is in fact, an open theoretical
challenge in present-day quantum information science, and many
general results are still unknown or are being investigated (for
example, see \cite{MultiPartiteEntDur}). However, for two qubits (a
$2\times 2$ system), the situation is rosier, and luckily, we shall
be primarily concerned with two qubit (four dimensional) quantum
systems in \textit{most} parts of this thesis.

\subsection{Quantifying entanglement in two qubit
systems}\label{section-entanglement-quantify}

As indicated earlier, \textit{qualifying} entanglement is an arduous
task, especially for higher dimensional mixed systems. How can we
expect the \textit{quantification} to be any easier? Several
conditions for a ``good'' entanglement measure have been compiled
\cite{NielsenChuang,EntanglementCharacterizeBruss,QuantifyingEntVedral},
and there do exist a range of measures, satisfying some or all of
these conditions. Once again, for two qubit systems, the situation
is relatively simple and well understood.

\subsubsection{Concurrence and entanglement of
formation}\label{section-concurrence-eof}

\begin{figure}
\begin{center}
\includegraphics[scale=0.8]{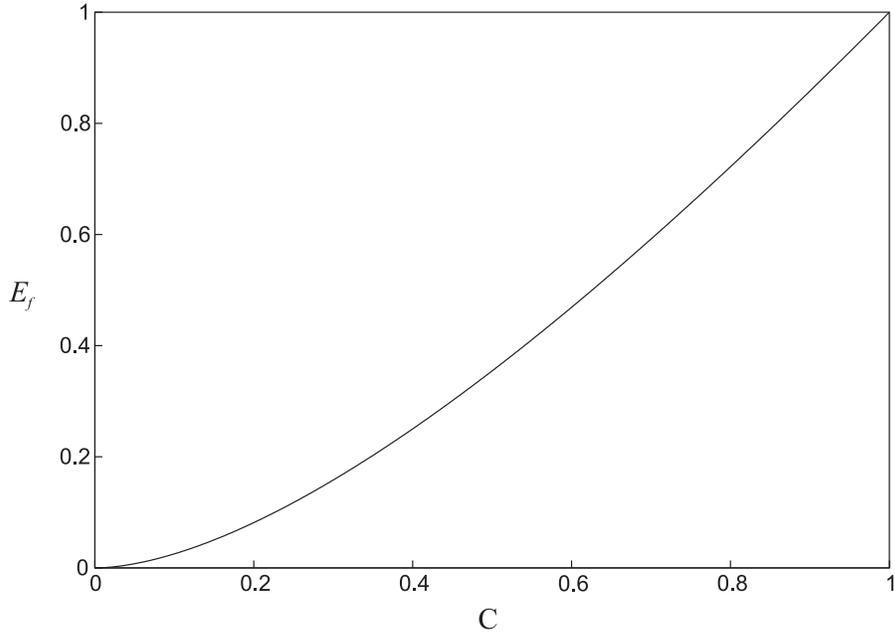}
\caption{The entanglement of formation $E_f$ as a function of
concurrence $C$.}\label{Figure-conc-eof}
\end{center}
\end{figure}

According to the PPT criterion, negative eigenvalues of $\rho^{T_A}$
are a signature of entanglement; the negativity of the most negative
eigenvalue is one straightforward measure. However, for another
entanglement measure, called the entanglement of formation (EOF)
\cite{BennetMixedEntPRA}, there also exists a closed form
mathematical expression \cite{concurrence} (in the case of two
qubits). We denote this quantity as $E_f$, and is our measure of
choice for detecting and quantifying entanglement: a non-zero $E_f$
indicating the existence of entangled states. A prescription for
finding $E_f$ \cite{concurrence} for a state $\rho$
is outlined below.
\begin{enumerate}
\item Determine the ``spin-flipped'' operator, $\tilde{\rho}=(\sigma_y\otimes\sigma_y
)\rho^*(\sigma_y\otimes\sigma_y )$, where $\rho^*$ is the complex
conjugate of $\rho$.
\item Find the non-Hermitian operator, $\rho '=\rho\tilde{\rho}$.
\item Compute the positive square roots of the eigenvalues of $\rho
'$ and sort them in decreasing order,
$\{\lambda_1,\lambda_2,\lambda_3,\lambda_4\}$.
\item Find the concurrence defined as $C=maximum(0,\lambda_1-\lambda_2-\lambda_3-\lambda_4)$.
\item The value of $C$ is then used to calculate the value of
\begin{equation}
x=\frac{1}{2}(1+\sqrt{1-C^2})\label{concurrence-def}.
\end{equation}
\item We then use $x$ to calculate the entanglement of
formation, through the formula,
\begin{equation}\label{eof-def}
E=-x\lg_2{x}-(1-x)\lg_2{(1-x)},
\end{equation}
reminding us of the classical binary entropy \cite{NielsenChuang}
function. Note that in computing \eqref{eof-def}, we use the
convention $0\,\lg_2{0}\equiv 0$.
\end{enumerate}
The concurrence  and entanglement of formation $E_f$, can both be
used as legitimate measures of entanglement. Both measures are zero
for a separable state and non-zero for entangled states. For
maximally entangled states, $E_f$ is $1$ and $C$ is $1$.
Furthermore, $E_f$ is a monotonically increasing convex function of
$C$ \cite{concurrence}, the functional dependence is shown in Figure
\ref{Figure-conc-eof}.

\subsection{Entanglement in singlet-triplet
mixtures}\label{section-entanglement-singlet-triplet}

In this section, I determine $E_f$ for a special class of two qubit
density matrices: mixtures involving four states
$\ket{\psi^-}\bra{\psi^-}\equiv S_0$,
$\ket{\psi^+}\bra{\psi^+}\equiv T_0$, $\ket{00}\bra{00}\equiv T_1$
and $\ket{11}\bra{11}\equiv T_{-1}$. The first two ($S_0$ and $T_0$)
are the Bell states \eqref{Bell-psi} and the other two ($T_1$ and
$T_{-1}$) are separable states. $S_0$ is conventionally called the
singlet and $T_0$, $T_{\pm 1}$, the triplets. I also define $T_m$
as,
\begin{equation}\label{Tm-define-1}
T_m=\,\frac{1}{2}(T_1+T_{-1}),
\end{equation}
and observe that $T_m$ is also an equal mixture of the Bell states
$\ket{\phi^+}\bra{\phi^+}$ and $\ket{\phi^-}\bra{\phi^-}$.

These $S/T$ mixtures have a special place in the present work, as
our initialization experiment, to be discussed in Chapter
\ref{pure}, prepares states containing combinations of only $S_0$,
$T_0$ and $T_m$. The ideal experiment prepares $S_0$, but the
triplets mix in, in varying proportions, resulting in an $S/T$
mixture. Whether the resulting mixture is entangled or not, depends
on what is mixed in and in what amounts. Based on $E_f$
measurements, we can outline some important results. These results
are helpful in determining the correct threshold for entanglement.

\begin{figure}
\begin{center}
\includegraphics[scale=0.7]{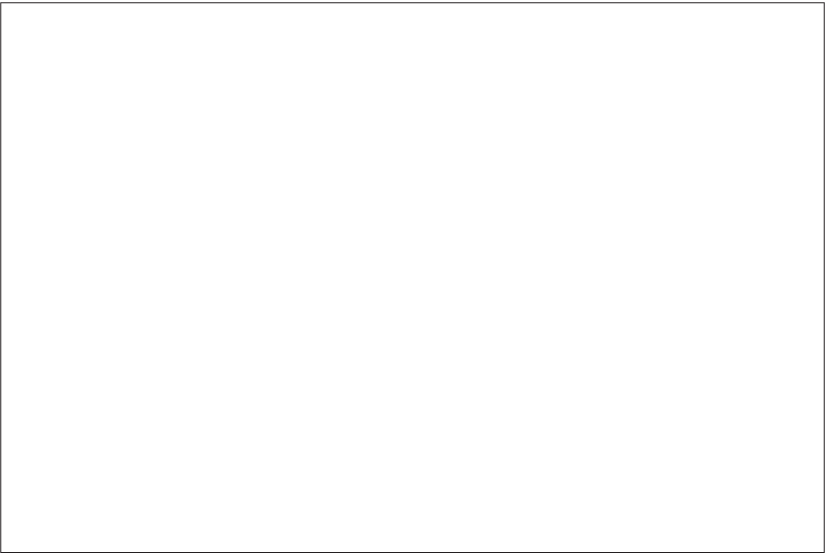}
\caption{Entanglement of formation $E_f$ for singlet triplet
mixtures:  (a) $S_0$ and $T_0$, (b) $S_0$ and $T_1$, (c) $S_0$ and
$T_m$ and (d) $S_0$, $T_0$ and $T_m$. The $E_f$ versus $\varepsilon$
plot for an $S_0/T_{-1}$ mixture is identical to
(b).}\label{Figure-eof}
\end{center}
\end{figure}

\begin{enumerate}
\item The $E_f$ for the pure $S_0$ is obviously $1$.
\item Now consider the ``binary'' $S_0/T_0$ mixture,
$\varepsilon S_0+(1-\varepsilon)T_0$, where $\varepsilon\le 1$. The
resulting $E_f$ as a function of the amount of singlet,
$\varepsilon$, is shown in Figure \ref{Figure-eof}(a), showing that
$E_f$ is always non-zero, except when the the amounts of $S_0$ and
$T_0$ are perfectly balanced, \ie\ when $\varepsilon=1/2$.
Interestingly, adding an entangled state ($T_0$) to another
entangled state ($S_0$) reduces the overall amount of entanglement.
This can also be seen in light of the convexity property
\cite{EntanglementCharacterizeBruss} of the entanglement of
formation,
\begin{equation}\label{convexity-eof}
E_f(\varepsilon S_0+(1-\varepsilon)T_0)\le\,\varepsilon
E_f(S_0)+(1-\varepsilon)E_f(T_0),\quad\text{for\ }\varepsilon\le 1.
\end{equation}
The take-home lesson is that the $S_0/T_0$ mixture will
\textit{always} be entangled if $\varepsilon>1/2$.
\item Figure \ref{Figure-eof}(b) shows the $E_f$ for an
$S_0/T_1$ mixture: like the case for $T_0$, adding $T_1$ also
reduces the entanglement but it is not as effective in ``pushing''
the state to the separable region, and the mixture is entangled for
$\varepsilon>0$.
\item If we mix in $T_m$, the state is entangled for
$\varepsilon>1/2$; the $E_f$ is plotted in Figure
\ref{Figure-eof}(c). This indicates that $T_m$, a balanced mixture
of separable states, is more effective in destroying the
entanglement of the singlet, as compared to adding $T_1$ or $T_{-1}$
on its own. In fact, $T_0$ and $T_m$ are the most effective states
in destroying the entanglement in the singlet state. The motivation
for this argument is also discussed in Appendix
\ref{app-entanglement}, Section \ref{section-attack-singlet}.
\item Finally, the ``ternary'' $S_0/T_0/T_m$ mixture,
\begin{equation}\label{ternary-S0T0Tm}
\rho=\,\varepsilon S_0+(1-\varepsilon)\bigl(\chi
T_0+(1-\chi)T_m\bigr),\quad{\text{where}\ }0\le\varepsilon ,\chi\le
1,
\end{equation}
will \textit{always} be entangled for $\varepsilon>1/2$,
irrespective of the value of $\chi$. This is illustrated in the
two-dimensional plot, Figure \ref{Figure-eof}(d), which also points
towards additional subtleties, as now, $\varepsilon>1/2$ is not a
necessary but only a sufficient condition. This is suggested by the
non-zero $E_f$ for large values of $\chi$, even when
$\varepsilon<1/2$, indicating that this time, the mixture is
``reflecting'' the entanglement present in the $T_0$ state.
\item For arbitrary mixtures, the state is always entangled when $\varepsilon>1/2$; $E_f$ depends
only on $\varepsilon$, and is independent of the \textit{relative}
proportions of $T_0$ and $T_m$. A simple proof to this hypothesis is
presented in Appendix \ref{app-entanglement}, Section
\ref{section-eof-independent-T0Tm}.
\item I now come back to the Werner state, \eqref{Werner-singlet},
for which I have already shown that $\rho_{\varepsilon}$ will be
entangled for $\varepsilon>1/3$. In fact, the Werner state can also
be viewed as a ternary mixture, by decomposing the maximally mixed
state into the singlet and triplets,
\begin{equation}
\frac{\mathbf{1}_4}{4}=\,\frac{S_0+T_0+2T_m}{4},
\end{equation}
enabling us to re-write $\rho_\varepsilon$ as,
\begin{eqnarray}
\rho_\varepsilon=&\,(\frac{1-\varepsilon}{4})(S_0+T_0+2T_m)+\varepsilon
S_0\nonumber\\
=&\,(\frac{1-\varepsilon}{4})(T_0+2T_m)+(\frac{3\varepsilon+1}{4})S_0,\label{Werner-singlet-fraction}
\end{eqnarray}
where the co-efficient of $S_0$, signifying the amount of singlet is
now $(3\varepsilon+1)/4$, instead of $\varepsilon$. The
$\varepsilon>1/3$ condition deduced above, then translates to the
condition $(3\varepsilon+1)/4>1/2$, which is identical to the
threshold derived above for $S_0/T_0/T_m$ mixtures. Thus, states
involving $S_0$, $T_0$ and $T_m$, (including the Werner states),
will always be entangled if the \textit{amount of the singlet}
exceeds the \textit{entanglement threshold} of $1/2$. (A formal term
for the amount of singlet in a state $\rho$ is the \textit{singlet
fraction} \cite{BennetMixedEntPRA} and is defined as,
\begin{equation}\label{singlet-fraction-def}
F=\,\expectation{\psi^-}{\rho}{\psi^-}.)
\end{equation}
\end{enumerate}

A more concise proof of these arguments can also be found in
\cite{HorodeckiInfoThSep}: the reduced density matrices
\cite{CohenTannoudji1} for the state $\rho$, \eqref{ternary-S0T0Tm},
are the maximally mixed states, $\mathbf{1}_2$. It has been shown
that a state $\rho$ with maximally disordered subsystems, will be
separable if and only if its largest eigenvalue is at the most,
$1/2$. The state $\rho$ has an eigenvalue $f$, ensuring
\textit{inseparability} for $f>1/2$.

\section{Methods for enhancing
polarization}\label{section-hyper-polarization}

It is clear that pure initial states, also lying above the
entanglement threshold cannot be achieved without resorting to some
means for enhancing the polarization. Several methods have been
proposed and implemented for reaching non-Boltzmann spin
populations, and most of them have been motivated by the need for
enhanced signal-to-noise ratios in conventional NMR. I now briefly
review some of these methods in the context of quantum information
processing.

\subsection{High fields and low
temperatures}\label{section-high-field-low-temp}

One of the most obvious methods for increasing the polarization,
achieving almost pure states, would be to increase
$\mathcal{B}=\hbar\omega/kT$. Calculations with the single qubit
fractional populations at room temperature, suggest that to achieve
a purity of $99$\% ($n_{\alpha}=0.99$ and $n_\beta=0.01$), a
magnetic field of roughly $200000$~T is needed which is clearly
impossible to achieve in the foreseeable future. Alternatively,
using a reasonable field of $9.5$~T requires a temperature as small
as $10$~mT, which is technologically possible, but no liquid exists
at this temperature! Therefore, it seems that the na\"{\i}ve
approach is not feasible and a more indirect method is needed.

\subsection{Polarization transfer from nuclear
spins}\label{section-polarization-transfer-nuclear-spins}

One method to increase the polarization of a qubit would be to
``borrow'' it from another more polarized qubit. Polarization
transfer techniques exploit the small population differences in
thermal states and the gains in sensitivity are proportional to
$\gamma_I/\gamma_S$, the gyromagnetic ratios of the sensitive and
insensitive nuclei. These techniques, for example, involve the
incoherent nuclear Overhauser effect
\cite{LevittSpinDynamics,FreemanChoreo,Toolkit} or the coherent
INEPT (\textbf{i}nsensitive \textbf{n}uclei \textbf{e}nhanced by
\textbf{p}olarization \textbf{t}ransfer) sequence and its refocussed
version, \cite{LevittSpinDynamics,FreemanChoreo,Toolkit}; the latter
two effectively acting as a kind of SWAP \cite{SWAPConcMagRes}
between the sensitive and insensitive nuclei. However, in these
experiments, not only are the gains small for most nuclei of
interest, but also, it is not clear how they can be of any use in
enhancing polarizations in ^1\H\ nuclei, which already have the
largest gyromagnetic ratios (besides ^3\H). Due to these very
limitations, these methods are likely to be of little value in
realizing hyperpolarization for NMR quantum computers.

\subsection{Polarization transfer from electron
spins}\label{section-polarization-transfer-electron-spins}

Instead, it is more promising to make use of the hyperfine
interaction: highly polarized electrons cannot only facilitate in
indirectly detecting nuclear spins (Section
\ref{section-measurement}), but can also contribute in the converse
manner by transferring polarizations \textit{to} the nuclei. These
techniques have been given the generic name of \textit{dynamical
nuclear polarization} (DNP) \cite{Abragam}, and are closely related
to the original Overhauser effect \cite{Overhauser1,Overhauser2}:
one of the spins in a coupled pair is perturbed from equilibrium and
cross-relaxation effects \cite{LevittSpinDynamics,FreemanChoreo}
result in enhanced polarization for the unperturbed spin. In DNP,
the pair of spins are a nucleus and an unpaired electron. Another
related experiment is ENDOR (\textbf{e}lectron \textbf{n}uclear
\textbf{do}uble \textbf{r}esonance) \cite{ENDORRev}, which
traditionally involves transferring electron polarizations to the
nuclei, performing an NMR experiment and detecting the NMR
transitions indirectly, by transferring nuclear polarizations back
to the electrons. For example, using the ENDOR approach, Mehring
\etal\ recently demonstrated \cite{MehringNuclearELectronEnt}
preparation of pseudoentangled states between nuclear and electron
spins. However, all of these methods are mostly employed in the
solid state and their real potential in the liquid phase is only
recently being explored; for example, it has been demonstrated that
during rapid dissolution of a solid sample polarized by the DNP
effect, the nuclear polarization is preserved, and can result in
enhancements of four orders of magnitude \cite{DNPLiquids}.

\subsubsection{Optical pumping and spin
exchange}\label{section-optical-pumping}

Another important method that produces non-Boltzmann spin
distributions is optical pumping followed by spin exchange
\cite{LaserPolarizedGoodson,LaserPolarizedBruner}: circularly
polarized laser light is used to pump electrons in alkali metal
vapour (such as \Rb) into a non-equilibrium distribution, increasing
the electronic spin polarization by as much as four orders of
magnitude. The electronic polarization is then transferred to the
nuclear spins in a noble gas such as ^{129}\Xe\ via the hyperfine
interaction. Optical pumping is, therefore, another example of
transferring polarization from electrons to nuclei. The method,
however, is limited in its applicability to quantum computing,
because only a few noble gas isotopes are known to be polarizable by
optical pumping, and being chemically inert, they do not normally
participate in bond formation, resulting in isolated systems
unsuitable for universal quantum information processing. In
principle though, the spin polarization of the \Xe\ nuclei
\textit{can} be transferred to protons using SPINOE
(\textbf{s}pin-\textbf{p}olarization \textbf{i}nduced
\textbf{n}uclear \textbf{p}olarization)
\cite{SPINOESciencePines,SPINOEConcMagRes,SPINOEPRL}---without the
need for irradiating the spins, as in a conventional Overhauser
experiment---but the transfer efficiency remains very low (nearly
$10$\%). One demonstration of using optical pumping to obtain a
ten-fold increase in polarization over the thermal state, and
subsequently implementing a quantum algorithm is given in
\cite{ChuangXenon}.

\subsection{Chemically induced dynamical nuclear
polarization}\label{section-CIDNP}

CIDNP (\textbf{c}hemically \textbf{i}nduced \textbf{d}ynamical
\textbf{n}uclear \textbf{p}olarization)
\cite{FreemanChoreo,HandbookNMRFreeman} is a misleading name for
another technique for generating enhanced magnetic ordering; the
method is unrelated to the dynamical nuclear polarization discussed
above. The effect is based on the experimental observation that
nuclear spins can ``steer'' the course of a chemical reaction; this
seems quite a remarkable fact because the nuclear spin Zeeman
energies are negligible as compared to bond energies, yet the former
can exert a decisive influence on bond formation. I present a simple
description of this phenomenon. Suppose we are given a pair of free
radicals $\textbf{A}^*$ and $\textbf{B}^*$, where $*$ represents an
unpaired electron and $\textbf{B}$ is a spin-$1/2$ nucleus, that can
exist in the state $\ket{\alpha}$ or $\ket{\beta}$. Also suppose
that the radicals are ``caged'' together in solution, and the pair
of lone electrons are in a triplet state ($T_0,T_{\pm 1}$).
Recombination \textit{inside} the cage (geminate recombination)
takes place only if the pair of electrons can be brought into the
singlet state ($S_0$), as required in a stable, bonding molecular
orbital \cite{PhysicalChemistryAtkins}. The reaction,
\begin{equation}
\textbf{A}^* + \textbf{B}^*\,\rightarrow \textbf{AB},
\end{equation}
therefore, can happen only if there is a concomitant triplet to
singlet conversion of the electronic states. How can such a
conversion take place? The nucleus $\textbf{B}$ is hyperfine coupled
with its electron and causes the electronic spin state to precess at
a certain frequency, allowing the parallel $T_0$ configuration to
evolve towards the anti-parallel $S_0$, and the conversion will be
more favoured if $\textbf{B}$ is in one particular spin state rather
than the other. Suppose the $\ket{\alpha}$ state encourages the
$T_0\rightarrow S_0$ conversion. This implies that most geminally
recombined radicals, that is the molecules $\textbf{AB}$, will have
the $\textbf{B}$ nuclear spins predominantly in the $\ket{\alpha}$
state. Likewise, the remaining molecules will eventually escape the
cage and will recombine instead with the solvent or some other
species, possibly forming another molecule $\textbf{C}\textbf{B}$,
this time with preferential magnetic ordering in the $\ket{\beta}$
state. At the end, we have two distinct manifolds comprising
spin-polarized molecules. For QIP, the prospects of CIDNP are as yet
unexplored. For example QIP requires multiple polarized spins in a
\textit{single} molecule and it is not clear how this can be
realized by CIDNP.

\subsection{\textit{Para}-hydrogen derived hyperpolarization}\label{section-hyperpolarize-Pauli}

The \textit{para-hydrogen induced polarization} (PHIP) effect has
been used not only to obtain large enhancements in spectroscopy
\cite{SimonPHRev} and imaging \cite{PHImaging} applications, but has
also demonstrated initial success in quantum information processing
\cite{AnwarPH,AnwarDeutsch,Hubler,AnwarPHReview,AnwarGrover} in the
liquid state. The following three chapters of this thesis will
describe the PHIP effect (Chapter \ref{phip}); employing it to
prepare pure, entangled states (Chapter \ref{pure}); and the first
demonstrations of quantum algorithms on the pure state NMR quantum
computer (Chapter \ref{QCimplement}).

\chapter{\textit{Para}-hydrogen induced polarization}\label{phip}

``\ldots NMR is the premier spectroscopic example, not of quantum
mechanics, but of quantum \textit{statistical} mechanics
\cite{Warren} \ldots .'' The fact that most high temperature NMR
quantum (statistical) computers, are in fact, ``ensemble
processors'' \cite{CoryProcNatl}, with spins statistically
distributed in closely spaced energy levels, requires the generation
of pseudopure states. However, we have also seen that the pseudopure
approach cannot lead to computers of useful sizes. It appears that
quantum statistical mechanics limits the usefulness of NMR as a
viable technology for future computing, and the opening remark to
this chapter, is a reflection of this view.

Ironically, quantum statistical mechanics also helps in overcoming
some of the obstacles it creates for itself in the first place: the
connection between spin statistics and sensitivity enhancement in
NMR  \cite{WeitekampPersonal}, was known even before NMR celebrated
its debut in the quantum computing arena. One possible solution to
enhanced NMR sensitivity exploits the coupling between nuclear spin
and molecular rotation and the coupling obeys rules laid out by
quantum statistical mechanics. The origin of the hyperpolarization
lies in the almost perfect \textit{spin order}, sometimes also
called the \textit{symmetrization order} \cite{Bowers} of the
nuclear spin state derived from \pHH . Our motivation in using this
spin order is in producing nearly pure states in liquid NMR, thereby
initializing the quantum computer in a well-defined state and
realizing entanglement in high temperature liquids.

This chapter is a foray into the fundamentals of the spin order
which arises as a result of the strict correlation between the
nuclear spin and molecular rotational degrees of freedom. I shall
explore the symmetry requirements on the wavefunction of a diatomic
molecule, such as \dihyd . These requirements apparently restrict
``homonuclear freedom'', but in effect, cool the pair of nuclear
spins into a pure quantum state. The symmetrization postulate
\cite{Symmetrization96,Symmetrization98,ShankarQM} results in the
existence of two distinct spin isomers of the molecule; this effect
is discussed in Section \ref{spin-isomers}. Sections
\ref{statmech-ph} and \ref{prepare-ph} explain the relative
proportions of the spin isomers and outline some methods of
isolating the \textit{para} isomer. For quantum computing, the
exceptionally high purity ``locked'' in the \pHH\ molecule has to be
transferred to a suitable molecule in solution; the underlying
chemistry is described in Section \ref{chemical-parahydrogen}. A
thorough description of the nuclear spin states in \pHH\ and its
derivatives in solution requires a formal treatment in the language
of density operators. This formalism is spelled out in Section
\ref{operator-parahydrogen}. The hydrogenation can take place
outside or inside the strong magnetic field leading to different
descriptions. These experiments are outlined in Sections
\ref{ex-situ} and \ref{in-situ}. The signal enhancements in these
experiments are described in Section \ref{signal-enhance}. Depending
on the details of the hydrogenation and detection, variants of the
\textit{in situ} experiment are possible, which are explained in
Sections
\ref{pasadena-delay-subsection}--\ref{section-isotropic-pasadena}.
Finally, I present a brief summary in Section
\ref{section-phip-summary}. In short, this chapter prepares the
necessary theoretical background for the initialization experiment
to be discussed in Chapter \ref{pure}.

One of the most comprehensive reviews on \pHH\ induced polarization
(which I shall define later) is \cite{NattererPHRev}, while one of
the earliest is \cite{BowersAdvMagRes}. Another review focusing on
the chemical applications is \cite{SimonPHRev}.

\section{Spin isomers of hydrogen}\label{spin-isomers}

In common usage, the word ``hydrogen'' can refer to either the
simplest atom, \H, or the diatomic molecule, \H_2. To keep the
following discussion clear, I shall refer to the diatomic molecule
as \dihyd . We observe that even a simple molecule such as \dihyd\
can exhibit a vast richness of properties. One such property is its
nuclear spin isomerism \cite{SilveraRMP} and forms a central topic
of this chapter.

The spin isomers are a result of the symmetrization postulate and
the Pauli principle
\cite{Sakurai,Herzberg,Symmetrization96,Symmetrization98,SpectroscopyHollas},
which impose certain symmetry requirements on a composite system
comprising any number of identical particles: under the exchange of
any two particles, the \textit{total} wavefunction must be either
symmetric (even) or antisymmetric (odd). Furthermore, for particles,
which are called \textit{fermions}, the wavefunction must be odd
under the exchange operation, also called the parity operation
$\mathcal{P}$ \cite{CohenTannoudji1}.

Suppose we have two fermions in distinct positions; I label their
positions as $A$ and $B$ and assume their wavefunctions are $\chi$
and $\phi$. The overall wavefunction could then be written as
$\psi=\,\chi_A\phi_B$, and if we interchange their labels the
resulting wavefunction becomes
$\psi'=\,\mathcal{P}(\psi)=\phi_A\chi_B$. Now, the Pauli principle
simply requires the wavefunction to be odd, \ie\ $\psi'=-\psi$.
There is another interesting consequence of the odd symmetry of the
fermionic wavefunction. If the fermions have the same position, say
$A$, then $\phi_A\chi_A=-\chi_A\phi_A$ implies a vanishing
wavefunction, $\psi'=-\psi=0$: two fermions with identical
wavefunctions \textit{cannot} occupy the same position, they are
\textit{excluded} from being in the same state---this latter fact
being an expression of the Pauli Exclusion Principle.

For \dihyd , containing two protons, which are of course fermions,
the overall wavefunction $\psi_{total}$ must be odd for the same
reason. This wavefunction has contributions from several degrees of
freedom, all of which we assume act independently. These
contributions come from the electronic states, translational,
vibrational and rotational molecular motions and nuclear spin
degrees of freedom, 
\begin{equation}\label{wavefunctionH2}
{\psi}_{total}=\,{\psi_{e}}\,{\psi_{trans}}\,{\psi_{vib}}\,{\psi_{rot}}\,{\psi_{ns}}.
\end{equation}
In keeping with the Pauli principle, I analyze the parities of the
five independent contributions, and permit only those combinations
for which the overall parity comes out to be odd. For the dihydrogen
molecule the ground state electronic function $\psi_{e}$ is even
\cite{SpectroscopyHollas}. Exchanging the protons in the diatomic
molecule has no effect on the translational motion, which is simply
a centre of mass motion, and so the translational part
$\psi_{trans}$ is also even. Likewise, for its vibrational motion,
the molecule can be modelled as a pair of identical masses connected
through a spring; interchanging the masses leaves the system
unchanged, and so $\psi_{vib}$ is also even. The first three
contributions in \eqref{wavefunctionH2} being even, the parity of
$\psi_{total}$ is therefore determined by the combined parity of the
last two terms, $\psi_{rot}$ and $\psi_{ns}$: even rotational terms
must combine with odd nuclear spin terms and odd rotational terms
must combine with even spin terms. The reasoning is simple: odd
functions multiplied by even functions result in an odd product.
This explains the exact correlation between the molecular rotation
and nuclear spin degrees of freedom. The natural question to ask
next is how the individual wavefunctions $\psi_{rot}$ and
$\psi_{ns}$ behave under the exchange operation.

The rotational wavefunction of the molecule is a representation
\cite{CohenTannoudji1} of
the rotational state vector in the position space 
\begin{equation}\label{rotational-wavefunc}
\psi_{rot}(\theta ,\phi)=\,\braket{\theta ,\phi}{J,m_J},
\end{equation}
where $\theta$ and $\phi$ are angles in space and $J$ is the angular
momentum quantum number of the molecule, $m_J$ being its projection
along the quantization axis. The function $\psi_{rot}(\theta ,\phi)$
transforms under rotations like the spherical harmonic function
$Y_{m_J}^J$ \cite{Sakurai}, and the parity operator
rotates this function according to 
\begin{equation}\label{rotational-wavefunc-parity}
\mathcal{P}(\psi_{rot}(\theta ,\phi))\propto\,Y^J_{m_j}(\pi-\theta
,\phi +\pi)=\,(-1)^JY_{m_J}^J(\theta
,\phi)\propto\,(-1)^J\psi_{rot}(\theta ,\phi),
\end{equation}
which clearly shows that the parity is controlled by the quantum
number $J$ through the factor $(-1)^J$: an even $J$
($=0,2,4,\ldots$) results in symmetric rotational functions and odd
$J$ ($=1,3,5,\ldots$) gives rise to antisymmetric functions.

I now turn to the parities of the nuclear spin wavefunctions. The
kets corresponding to $\psi_{ns}$ for two spins
can assume four forms, 
\begin{eqnarray}\label{nuclear-spin-psi}
\ket{00}&=\,&\ket{T_1},\label{T1}\\
\ket{11}&=\,&\ket{T_{-1}},\label{Tminus1}\\
\frac{1}{\sqrt{2}}(\ket{01}+\ket{10})&=&\ket{T_0}\quad\mbox{and}\label{T0}\\
\frac{1}{\sqrt{2}}(\ket{01}-\ket{10})&=&\ket{S_0}\label{S0}.
\end{eqnarray}
The first three of these states (\ref{T1}-\ref{T0}) are
characterized by the spin quantum number $S=1$, and are referred to
as the spin \textit{triplet} states and are denoted in general as
$\ket{T_{sub}}$. The subscripts $1\mbox{,}-1\mbox{\ and\ }0\,$
correspond to the projection $m_S$ of the triplets along the spin
quantization axis. The state $\ket{S_0}$, given in \eqref{S0}, is
characterized by $S=0$ and is called the spin \textit{singlet}.
(Finding these nuclear spin vectors for a composite two-spin system
is a straightforward application of the problem of addition of two
angular momenta \cite{Sakurai}).
The proton is a spin-half particle (a fermion) and so for \dihyd\
$S_1=S_2=1/2$, implying that the total angular momentum quantum
number $S$ can be $1$ or $0$. The projections for $S=1$, are clearly
$m_1=\{1,-1,0\}$ and for $S=0$ the only possible projection is
$m_0=\{0\}$. Corresponding to the four distinct $\{S,m_s\}$ pairs,
four nuclear spin states are possible, which are exactly the states
\eqref{T1}--\eqref{S0}. These kets also completely span the four
dimensional Hilbert space of the two nuclear spins in \dihyd\ and a
basis comprising these basis elements is generally called the
\textit{symmetrical} basis.

For the purpose of establishing the correlation with the molecular
rotational states, we are interested in the symmetry properties of
these nuclear spin states. The triplets are symmetrical with respect
to proton interchange whereas the singlet is antisymmetrical. For
example, relabelling the spins in the singlet $\ket{S_0}$ would
result in
\begin{equation}\label{S0-interchange-odd}
\ket{S_0}=\frac{1}{\sqrt{2}}(\ket{01}-\ket{10})\xrightarrow{\mathcal{P}}\frac{1}{\sqrt{2}}(\ket{10}-\ket{01})=-\ket{S_0}.
\end{equation}
In short, the nuclear spin triplets $\ket{T_1}$, $\ket{T_{-1}}$ and
$\ket{T_0}$ have even spin wavefunctions and combine with odd
rotational states. The singlet $\ket{S_0}$ has an odd spin
wavefunction and must combine with even rotational states. This
correlation is summarized in Table \ref{correlation-rot-ns}.
\begin{table}
\begin{tabular}{ccccccccc}
\hline  $J$ && parity of $\psi_{rot}$ && parity of $\psi_{ns}$ && allowed spin vectors && form of \H_2 \\
 \hline
$0$ && even && odd && $\ket{S_0}$ && \textit{para} \\
$1$ && odd && even && $\ket{T_1},\ket{T_{-1}},\ket{T_0}$ && \textit{ortho} \\
$2$ && even && odd && $\ket{S_0}$ && \textit{para} \\
$3$ && odd && even && $\ket{T_1},\ket{T_{-1}},\ket{T_0}$ && \textit{ortho} \\
$\vdots$ &&&&&&&& \\
  \hline
\end{tabular}
\caption{Correlation between molecular rotational and nuclear spin
wavefunctions. \textit{Para} and \textit{ortho} spin manifolds of
\H_2.} \label{correlation-rot-ns}
\end{table}

Now suppose we can sort the \dihyd\ molecules according to their
values of $J$, keeping \textit{all} odd $J$ molecules in one
cylinder and \textit{all} even $J$ molecules in another. As
molecular rotation and nuclear spin are intertwined, the
``rotational sieve'' also acts as a ``spin sieve'', separating the
singlet and triplet nuclear spin states. As a result, the odd $J$
cylinder will be an equal mixture of the \textit{three} kinds of
triplet molecules, called \textit{ortho} and the even $J$ cylinder
will solely comprise singlet molecules --- their spin states are of
just \textit{one} kind and the molecules are termed \textit{para}.
Therefore if we can somehow partition the even and odd $J$
rotational manifolds, then in principle one of these compartments
will contain a pure quantum spin state. This pure quantum state is
the initial state of our quantum computer and will be discussed in
subsequent parts of this thesis.

\section{\textit{Para}-hydrogen from the perspective of statistical mechanics}\label{statmech-ph}

How do we in practice, achieve the spin sorting described above?
Remembering that zero is an even number, we can cool the molecules
to their rotational ground state $J=0$ and expect all nuclear spins
to be in the singlet state. This approach effectively freezes the
rotation of the molecules, locking them into the rotational ground
state, and as a result, traps the nuclear spins in the singlet
state. It is like manipulating the spin manifold through a more
congenial arbiter, the molecular rotational manifold---as we now
explore.

The rotational energy levels are populated according to the
Boltzmann equilibrium distribution
\cite{Sakurai,Herzberg,SpectroscopyHollas}, the fractional
population
$n_J$ in each state being given as, 
\begin{equation}\label{rotational-Boltzmann}
n_J=\,\frac{1}{Z}\bigl(g_Jg_S\exp{(-E_J/kT)}\bigr),
\end{equation}
$Z$ being the relevant partition function, $E_J$ the energy of the
state with degeneracy $g_Jg_S$, $k$ the Boltzmann constant and $T$
the absolute temperature. In the rigid rotor model
\cite{SpectroscopyHollas,SpectroscopyBanwell} of \dihyd , the energy
corresponding to the angular momentum $P_J$ is,
\begin{equation}\label{energy-J}
E_J=\,\frac{P_J^2}{2I}=\,J(J+1)\frac{\hbar^2}{2I},
\end{equation}
where $I$ is the moment of inertia of \dihyd . Defining the
\textit{rotational temperature} \cite{SpectroscopyHollas},
\begin{equation}\label{rotational-temperature}
\theta_r=\,\frac{\hbar^2}{2Ik},
\end{equation}
\eqref{rotational-Boltzmann} can also be written as,
\begin{equation}\label{rotational-Boltzmann-2}
n_J=\,\frac{1}{Z}\bigl(g_Jg_S\exp{(-J(J+1)\theta_r/T)}\bigr).
\end{equation}
Among gases, \dihyd\ has the largest rotational temperature of about
$85~$K (calculated from data for rotational constants presented in
\cite{SilveraRMP}). Furthermore, the energy difference between any
two levels $J$ and $J+1$,
\begin{align}\label{energy-J-difference}
(E_{J+1}-E_J)/k=&\,\theta_r(J+1)(J+2)-\theta_rJ(J+1)\\
=&\,2\,\theta_r(J+1)
\end{align}
is proportional to the rotational temperature and $J+1$. For
example, in \dihyd\ the energy gap between the lowest ($J=0$) and
first excited ($J=1$) rotational states corresponds to a temperature
of $2\,\theta_r=170~$K. It is therefore possible to cool to low
temperatures, ensuring that most molecules are in the $J=0$ state.
All other diatomic molecules have smaller rotational constants and
the $0\leftrightarrow 1$ gap is smaller, requiring substantially
lower temperatures to preferentially populate the $J=0$ state. For
example, in the homonuclear molecule comprising two deuterium
nuclei, ^2\H_2, the lowest rotational states are separated by
$2\theta_r=86~$K; furthermore, as the deuterium nucleus is a boson
(nuclear spin $I=0$), the lowest rotational ($J=0$) state would be
populated by \ortho -deuterium molecules.


In the absence of external electric and magnetic fields, the $J$th
rotational state is $(2J+1)$-fold degenerate ($g_J=2J+1$), and the
allowed
projections are given as 
\begin{equation}
m_J=\,-J,-(J-1),\ldots ,J-1,J,\quad\quad (J\ge 0).
\end{equation}
We have also seen that for even $J$, only one spin state (the
singlet) is allowed and for odd $J$, three spin states (triplets)
are allowed, and so the spin degeneracy $g_S$ in
(\ref{rotational-Boltzmann}) is one or three depending on the parity
of $J$. These degeneracies allow us to write the partition
function for the \dihyd\ molecule, 
\begin{equation}\label{partition-rotational-nuclear}
Z=\,\sum_{J=0,2,4,\ldots}\,(2J+1)\exp{(-J(J+1)\theta_r/kT)}+\,3\sum_{J=1,3,5,\ldots}\,(2J+1)\exp{(-J(J+1)\theta_r/kT)}.
\end{equation}

\subsection{\textit{Para}-\textit{ortho}
ratios}\label{section-para-ortho-ratio}

With these expressions, we can determine the relative ratio of the
\textit{para} and \textit{ortho} molecules, 
\begin{equation}\label{ratio-para-ortho}
\frac{N_{para}}{N_{ortho}}=\,\frac{\sum_{J=even}\,(2J+1)\exp{(-J(J+1)\theta_r/kT)}}{3\sum_{J=odd}\,(2J+1)\exp{(-J(J+1)\theta_r/kT)}}.
\end{equation}
The fraction is a function of temperature and can be calculated
numerically for different values of $T$, however analytical
expressions can also be found under certain assumptions. For
example, in the high temperature limit $T\gg\theta_r$, the spacing
between the levels is small compared to $kT$ and the discrete levels
can be treated as a continuum of levels, replacing the discrete sums
in \eqref{partition-rotational-nuclear}--\eqref{ratio-para-ortho}
with
continuous integrals. For example, consider the sum 
\begin{equation}\label{partition-high-temp-sum}
\sum_{J=0}^{\infty}(2J+1)\exp{(-J(J+1)\theta_r/T)}.
\end{equation}
Carrying out the substitutions $J(J+1)=x^2$, $2J+1=2\,x\,dx$ and
$\theta_r/T=\alpha$, the sum can be approximated as
\begin{equation}\label{partition-high-temp-intgrl}
\int_{x=0}^{\infty}\,2\,x\exp{(-\alpha x^2)}\,dx
\end{equation}
which is in fact a standard integral \cite{Arfken} with the solution
\begin{equation}\label{standard-integral}
\frac{1}{\alpha}=\,\frac{T}{\theta_r}.
\end{equation}
Therefore, the sum over only the even or only the odd $J$ states (in the high temperature limit) is, 
\begin{equation}\label{partition-high-temp-sum-even-odd}
\frac{1}{2}\frac{1}{\alpha}=\,\frac{1}{2}\frac{T}{\theta_r}.
\end{equation}
These integrals allow us to calculate the
\textit{para}:\textit{ortho} molecular ratio in the high temperature
limit. From (\ref{ratio-para-ortho}) and the values of the
integrals, we get,
\begin{equation}\label{ratio-para-ortho-high-temp}
\frac{N_{para}}{N_{ortho}}\xrightarrow{T\gg\theta_r}\frac{T/(2\theta_r)}{3\,T/(2\theta_r)}=\,1/3,
\end{equation}
showing that at temperatures above $2\theta_r$, \dihyd\ is
essentially a $\approx 1:3$ mixture of \textit{para} and
\textit{ortho} molecules.

\begin{figure}
\begin{center}
\includegraphics[scale=]{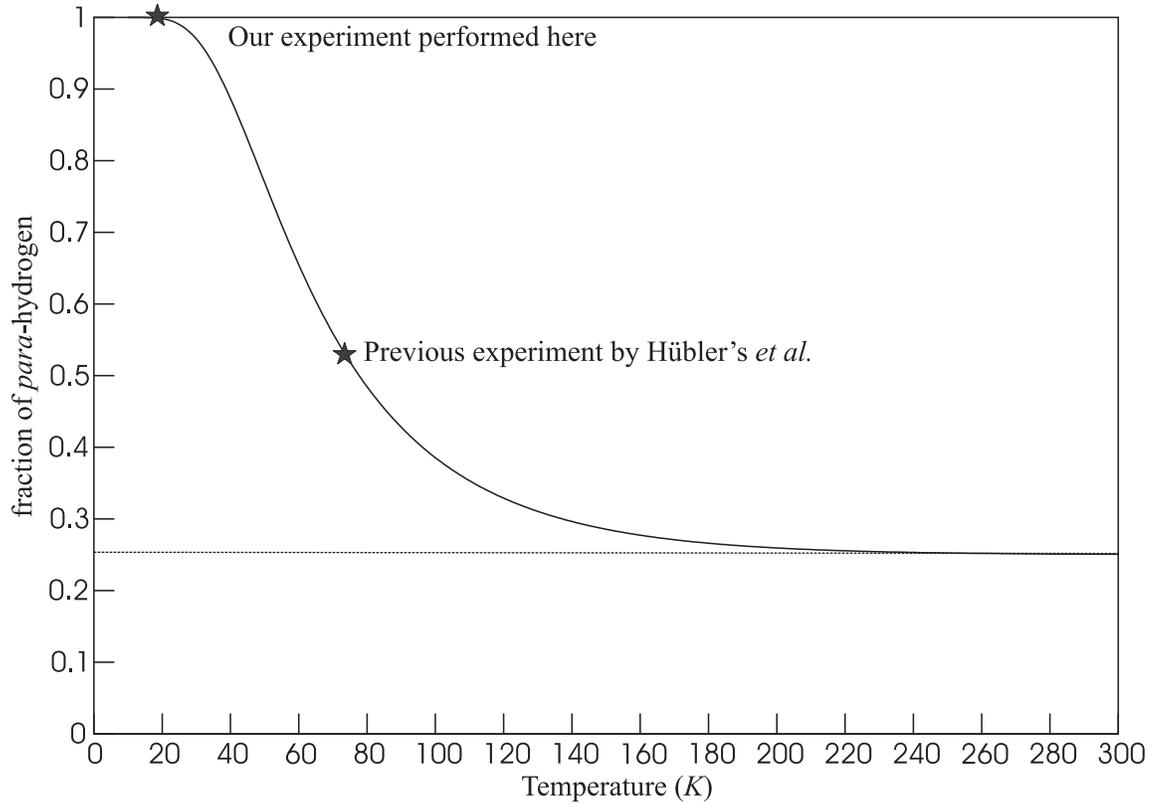}
\caption{Percentage of \pHH\ as a function of temperature $T$. The
plot also indicates the \pHH\ temperatures employed in our
experiment \cite{AnwarPH} and the previous work by H\"{u}bler \etal\
\cite{Hubler}. These experiments are described in Chapter
\ref{pure}. The dashed line is a guide to the eye, representing a
\pHH\ fraction of $1/4$, the limit at high temperature.
\label{Figure-para-ortho}}
\end{center}
\end{figure}

At lower temperatures, when $T\lesssim \theta_r$, the sums in
\eqref{partition-rotational-nuclear}--\eqref{ratio-para-ortho} must
be calculated explicitly. The series rapidly converges as only the
lowest $J$ states (such as $J=0,1,2$) are populated and it is both
permissible and customary to take
only the first few terms in, 
\begin{equation}\label{ratio-para-ortho-low-temp}
\frac{N_{para}}{N_{ortho}}=\,\frac{1}{3}\,\biggl(\frac{1+5\exp{(-6\theta_r/T)}+9\exp{(-20\theta_r/T)}+\ldots}
{3\exp{(-2\theta_r/T)}+7\exp{(-12\theta_r/T)}+11\exp{(-30\theta_r/T)}+\ldots}\biggr).
\end{equation}
This expression can be evaluated numerically to estimate the
percentage of \pHH\ in a
\para -\ortho\ mixture as a function of temperature. Results are
presented in Table \ref{ratio-para-ortho-table} and Figure
\ref{Figure-para-ortho}, showing that at the temperature of liquid
\N_2 ($77~$K) we have roughly an equal mixture of \para\ and \ortho
-hydrogen and at temperatures at or below $20~$K, we obtain
essentially pure \pHH .

\begin{table}
\begin{tabular}{ccc}
\hline
Temperature $T$ (K)&&  \% age of \pHH \\
\hline
$300$&&$25.06$\\
$200$&&$25.25$\\
$150$&&$28.56$\\
$100$&&$38.55$\\
$80$&&$46.45$\\
$77$ (liquid \N_2)&&$50.47$\\
$60$&&$65.46$\\
$40$&&$88.66$\\
$20$&&$99.81$\\
$18$&&$99.93$\\
\hline
\end{tabular}
\caption{Percentage of \pHH\ as a function of
temperature.\label{ratio-para-ortho-table}}
\end{table}

\section{Preparation of \pHH}\label{prepare-ph}

Table \ref{ratio-para-ortho-table} immediately suggests a method of
preparing pure \pHH : cooling the \dihyd\ molecules to around $20~$K
and waiting for sufficient time, expecting the \para -\ortho\
mixture to equilibrate into the pure \para\ form. However, this
picture is not accurate, to say the least. \textit{Ortho} to
\para\ conversion is normally forbidden by angular momentum
selection rules \cite{SilveraRMP,Herzberg,SpectroscopyHollas}, and
so we need a slightly more sophisticated approach. For \ortho\ to
\para\ conversion, $J$ must change by $\pm 1$ and \textit{at the
same time} the nuclear spins must reorient from the $S=1$ triplet
state into the $S=0$ singlet. This reorientation can take place for
example, in the presence of dipolar couplings with neighbouring
\ortho\ molecules \cite{OrthoParaConversionPravica98}, however the
intrinsic conversion is extremely slow
\cite{OrthoParaConversionPravica98} and takes place over a matter of
days. To expedite the equilibration between the \ortho\ and
\para\ manifolds, we can introduce an exogenous source of a magnetic
field gradient \cite{PHCatalyticConverter}, breaking the nuclear
symmetry of the \ortho\ molecules, and for this purpose a catalyst
with paramagnetic properties is normally used.

The \textit{ortho}$\,\leftrightarrow\,$\textit{para} conversion
takes place on or near the surface of the catalyst, rapidly
achieving equilibration and the resulting
\para :\ortho\ ratio reflects the temperature of the surface, in
accordance with (\ref{ratio-para-ortho-low-temp}). If the surface
temperature is about $20~$K, the \ortho\ to \para\ conversion
reaches almost $100$\% completion. Removing the catalyst and warming
the \pHH\ to room temperature does not significantly alter the
purity of \pHH\ as the \para\ molecules do not convert back into the
\ortho\ form, by virtue of the same rules that had prohibited their
generation in the first place. It appears as if the molecules
``remember'' the surface temperature, the ratio of the isomers
continues to reflect the catalyst surface temperature\footnote{This
suggests that the \textit{para}-\textit{ortho} ratio can act as a
reliable temperature probe. For example, studying this ratio for the
isotopomers of water \cite{OrthoParaWater} in different comets,
nuclear spin temperatures of $\approx 25~$K have been suggested
\cite{OrthoParaComets,OrthoParaTransitions}, significantly cooler
than the observed rotational temperatures. Due to the strongly
forbidden \textit{ortho}$\leftrightarrow$\textit{para} conversion,
we can glimpse the inter-stellar environment billions of years
ago---approaching the lifetimes of comets!}. Upon warming, the
molecules redistribute themselves in the even $J$ rotational
manifold according to Boltzmann rules; however their nuclear spin
state does not change from the antisymmetrical singlet.
Theoretically, we can store the \pHH\ for indefinite periods of
time; practically, for hours and whatever little conversion takes
place is either at the surface of the storage cylinder or due to
traces of impurities including residual \ortho\ molecules.

Several experimental setups are being used for the generation of
\pHH\ in laboratories around the world (for an overview of the
methods, see \cite{SimonPHRev}). Our collaborators in York have
devised a method that uses liquid \He\ as the refrigerant, cooling
\dihyd\ to temperatures as low as $7~$K, allowing production of pure
\pHH . The
\para -\ortho\ equilibration takes place on an activated charcoal
catalyst. The \H_2 outlet is equipped with valves that allow
evacuation via a vacuum pump, and sample filling via a Young's NMR
tube adapter. Furthermore, the apparatus is equipped with a
thermocouple, which allows the temperature of the \H_2 gas to be
controlled (our studies use temperatures in the range $18~$K to
$20~$K), as well as a hydrogen pressure regulator on the inlet line
and a pressure gauge on the outlet. The apparatus is left on
permanently, except when the liquid helium reservoir needs to be
re-filled, ensuring that high-purity \pHH\ is always available.

The pure \pHH\ results in extremely low spin temperatures
\cite{Abragam}, effectively in the mK range, while physically the
gas can still be at room temperature. The spin temperature is now
decoupled from the lattice temperature and from the perspective of
the spins, the lattice is at infinite temperature, an approximation
I shall use in modelling decoherence in our spin system (in Appendix
\ref{app-operator-sum-decoherence}). In other words, the spin
entropy is very close to zero while the thermodynamic entropy of the
gas remains very high. In effect, this is like placing a tiny spin
refrigerator inside a molecular oven, which is hotter by five orders
of magnitude.

\section{Chemical addition reactions with \pHH}\label{chemical-parahydrogen}

Despite the exceptional purity, the \pHH\ singlet state is
spectroscopically useless. The $S=0$ state is the solitary
inhabitant of the even $J$ rotational manifold and no other spin
states are available to which nuclear spectroscopic transitions can
occur \cite{Bowers}. So \pHH\ is NMR silent. Furthermore, \dihyd\ is
a symmetric molecule and the two protons are magnetically equivalent
\cite{LevittSpinDynamics}. There is no way to distinguish between
the nuclei and so, if we wish to exploit the very high purity of our
\pHH\ product, we must somehow break the symmetry and transfer the
singlet state to another more interesting molecule, which allows for
some greater variety. It is therefore clear that some kind of
chemical reaction is needed. In fact, chemical applications have
steered research in \pHH , which has now been used for a number of
years, in probing the structure of transition metal complexes and
the study of reaction pathways involving hydrogenation of
unsaturated organic molecules. (For example, see \cite{SimonPHRev}
and the references therein.)


\begin{figure}
\begin{center}
\includegraphics[scale=0.7]{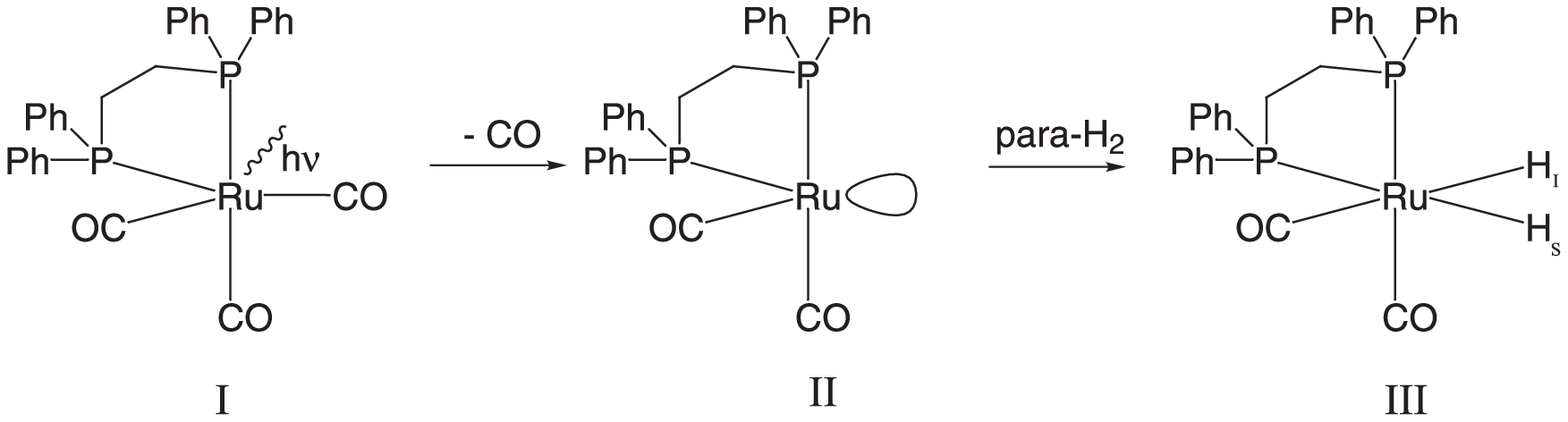}
\caption{Photolysis reaction scheme for hydrogenation of \dppestable
\ to produce \dihydride . The Ph represents an aromatic phenyl
group. The \C\ and \H\ atoms in the bidentate side chain are not
shown.}\label{reaction-dppe-photo}
\end{center}
\end{figure}

Figure \ref{reaction-dppe-photo} shows the reaction scheme for a
\textit{photolysis} reaction
\cite{LasersChemistry,GodardPHPhotolysis} involving the addition of
\pHH\ to a precursor molecule, \textbf{I}, \dppestable\ where (dppe)
indicates $1,2\text{--bis(diphenylphosphino)\,ethane}$
\cite{SchottPHRu}. The molecule is stable and does not normally add
\dihyd , but a UV photon knocks off one carbonyl group to generate
an unstable intermediate \textbf{II}, which immediately reacts with
\textit{dissolved} \pHH\ to give the desired product \textbf{III},
which I refer to as the \textit{dihydride}. The intermediate,
\textbf{II}, is a transitory species and is known to react with
\pHH\ almost instantaneously; the reaction, in fact, takes place on
the sub-microsecond timescale \cite{CroninPHInstant}. The dihydride,
\textbf{III}, contains two inequivalent hydrogen nuclei, \H_I and
\H_S, the former being trans to a carbonyl group and the other,
trans to a diphenylphosphino group. The two hydrogen nuclei form an
\textbf{AX} spin system in the presence of a strong magnetic field.

However, only breaking the symmetry in the \pHH\ molecule is not
enough. The addition of \pHH\ to the precursor \textbf{I}, must be
coherent, preserving the singlet spin state, both in \textit{form}
and \textit{magnitude}, quality and quantity. This means that the
pure $\ket{S_0}$ state must remain unblemished during the chemical
reaction, without dephasing into a mixture of states or decohering
into the maximally mixed state, which is always accompanied by a
loss of valuable polarization. The pair of hydrogen nuclei in the
dihydride, in other words, must \textit{fully inherit} the
\textit{pure} singlet state from the \pHH\ molecule. This is
possible only if the hydrogenation is \textit{pairwise}, meaning
that both protons bonding with a transition metal centre in a single
molecule, are derived from the same \pHH\ molecule. This concern for
preserving the singlet is a major concern in our experiments
involving \pHH\ and will be discussed in detail in the following
sections.

\begin{figure}
\begin{center}
\includegraphics[scale=0.7]{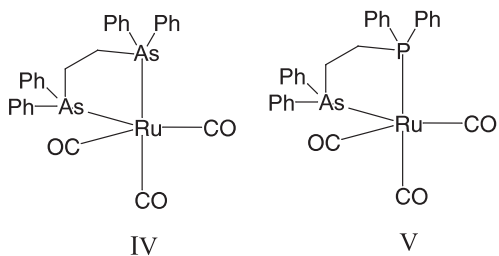}
\caption{Compound \textbf{IV} is
$1,2\text{--bis(diphenylphosphino)\,ethane}$ and \textbf{V} is a
``mixed'' phosphino-arsino compound
$1,2\text{--bis(diphenylphosphinoarsino)\,ethane}$}\label{dpae-mixed}
\end{center}
\end{figure}

In addition to the ``dppe'' compound \textbf{I}, we have also used
an analogue compound \textbf{IV} (see Figure \ref{dpae-mixed}),
$1,2\text{--bis(diphenylarsino)\,ethane}$, replacing the \P\ atoms
with \As. Collaborators in York are also investigating other
variants of \textbf{I} and \textbf{IV}, such as \textbf{V}, in which
only one \P\ atom is replaced with an \As, or deuterating the
compounds in the phenyl groups or the \C\H\ side chain, either fully
or partially. In the context of \pHH\ enhanced quantum information
processing, the search for new compounds with better decoherence
properties and more amenable spectral features continues, and
chemistry continues to guide our way.

The reaction depicted in Figure \ref{reaction-dppe-photo} is
photo-induced, which means that the reaction only takes place in the
presence of a UV trigger. A very short ($12~$ns) flash of UV
radiation ($308~$nm) from a pulsed excimer laser
\cite{LasersChemistry} is used to initiate the photochemical
reaction. The reaction is triggered by the laser flash and is
complete within one $\mu$s \cite{CroninPHInstant}. This allows
precise control over the reaction, which can now be initiated and
terminated at precise times. The laser effectively behaves as a
switch, preparing pure singlet states in an \textbf{AX} system, as
and when they are needed. Instead of using a single burst of
radiation from a pulsed laser, we can also apply radiation for an
extended period of time, using a continuous wave (CW) laser. In
fact, we have performed experiments using both the pulsed and CW
lasers, the former, for purposes of quantum computation and the
latter, for purposes of demonstration and investigating the time
dynamics of the spin system, especially over extended periods of
time. The CW experiments, for example, can be useful in studying the
decoherence properties of the system.

\section{Operator description of the nuclear spin states in the \pHH\ experiment}\label{operator-parahydrogen}

Investigating the \pHH\ enhanced systems from a quantum mechanical
perspective \cite{NattererPHRev,BowersThesis}, requires a
description of the nuclear spin states in the language of density
operators. The allowed spin states for the \dihyd\ molecule are
given in (\ref{T1}-\ref{S0}). Committing a slight abuse of notation,
I drop the kets and use the same
symbols to now represent the corresponding density matrices, 
\begin{equation}\label{density-matrix-notation}
T_1\equiv\,\ket{T_1}\bra{T_1},\quad
T_{-1}\equiv\,\ket{T_{-1}}\bra{T_{-1}},\quad
T_0\equiv\,\ket{T_0}\bra{T_0}\,\,\text{and}\quad
S_0\equiv\,\ket{S_0}\bra{S_0}.
\end{equation}
Writing these density operators (I call them states\footnote{The
exact meaning of the word `state' should be apparent
from context.}) in the computational basis, we obtain, 
\begin{align}
T_1=&\,\diagonal{1,0,0,0},\quad T_{-1}=\,\diagonal{0,0,0,1},\nonumber\\
T_0=&\,\frac{1}{2}\begin{pmatrix}
0 & 0 & 0 & 0\\
0 & 1 & 1 & 0\\
0 & 1 & 1 & 0\\
0 & 0 & 0 & 0\\
\end{pmatrix}\quad\text{and}\quad
S_0=\,\frac{1}{2}\begin{pmatrix}
0 & 0 & 0 & 0\\
0 & 1 & -1 & 0\\
0 & -1 & 1 & 0\\
0 & 0 & 0 & 0\\
\end{pmatrix}.
\end{align}
These states can also be written in product operator notation as,
\begin{align}
T_{\pm 1}=\,&\frac{\mathbf{1}_4}{4}+\frac{1}{2}(\pm I_z\pm S_z+2\,I_zS_z),\label{Tpm1-PO}\\
T_{0}=\,&\frac{\mathbf{1}_4}{4}+\frac{1}{2}(2\,I_xS_x+2\,I_yS_y-2\,I_zS_z)=\,\frac{\mathbf{1}_4}{4}+(ZQ_x-I_zS_z)\quad\text{and}\label{T0-PO}\\
S_{0}=\,&\frac{\mathbf{1}_4}{4}+\frac{1}{2}(-2\,I_xS_x-2\,I_yS_y-2\,I_zS_z)=\,\frac{\mathbf{1}_4}{4}+(-ZQ_x-I_zS_z)=\,\frac{\mathbf{1}_4}{4}-\mathbf{I}\cdot\mathbf{S}\label{S0-PO}.
\end{align}

Pure \pHH\ is the nuclear singlet state and would therefore be
represented by $S_0$ given in (\ref{S0-PO}),
\begin{equation}\label{rho-para}
\rho_{para}=\,S_0=\,\frac{\mathbf{1}_4}{4}+(-ZQ_x-I_zS_z),
\end{equation}
and pure \ortho -hydrogen, being an equal mixture of the triplet
states has an operator form,
\begin{equation}\label{rho-ortho}
\rho_{ortho}=\,\frac{1}{3}(T_1+T_{-1}+T_0)=\,\frac{\mathbf{1}_4}{4}+\frac{1}{3}(ZQ_x+I_zS_z).
\end{equation}
Similarly a mixed state with a singlet fraction $F$ will have the
form,
\begin{align}
\rho_F=&\,F\rho_{para}+(1-F)\rho_{ortho}\nonumber\\
=&\,\frac{\mathbf{1}_4}{4}+(\frac{1-4F}{3})(ZQ_x+I_zS_z),\label{rho-mixture-para-ortho}
\end{align}
which reproduces the states in (\ref{rho-para}-\ref{rho-ortho}) by
setting $F=1$ and $F=0$ respectively.

\section{\textit{Ex situ} hydrogenation}\label{ex-situ}

There are two kinds of experiments involving addition of \pHH\ to a
precursor, depending on \textit{when} and \textit{where} the
reaction takes place \cite{NattererPHRev,BowersAdvMagRes}. The first
method involves hydrogenating the precursor outside the magnet and
slowly transferring it into the strong field of the spectrometer
\cite{Altadena}; slow enough to ensure that the transfer is quantum
mechanically adiabatic \cite{Sakurai}, but fast enough to be rapid
in comparison with the $T_1$ recovery towards the Boltzmann
populations. This experiment was called ALTADENA (\textbf{a}diabatic
\textbf{l}ongitudinal \textbf{t}ransport \textbf{a}fter
\textbf{d}issociation \textbf{e}ngenders \textbf{n}et
\textbf{a}lignment) by its inventors \cite{Altadena}.

\begin{figure}
\begin{center}
\includegraphics[scale=0.8]{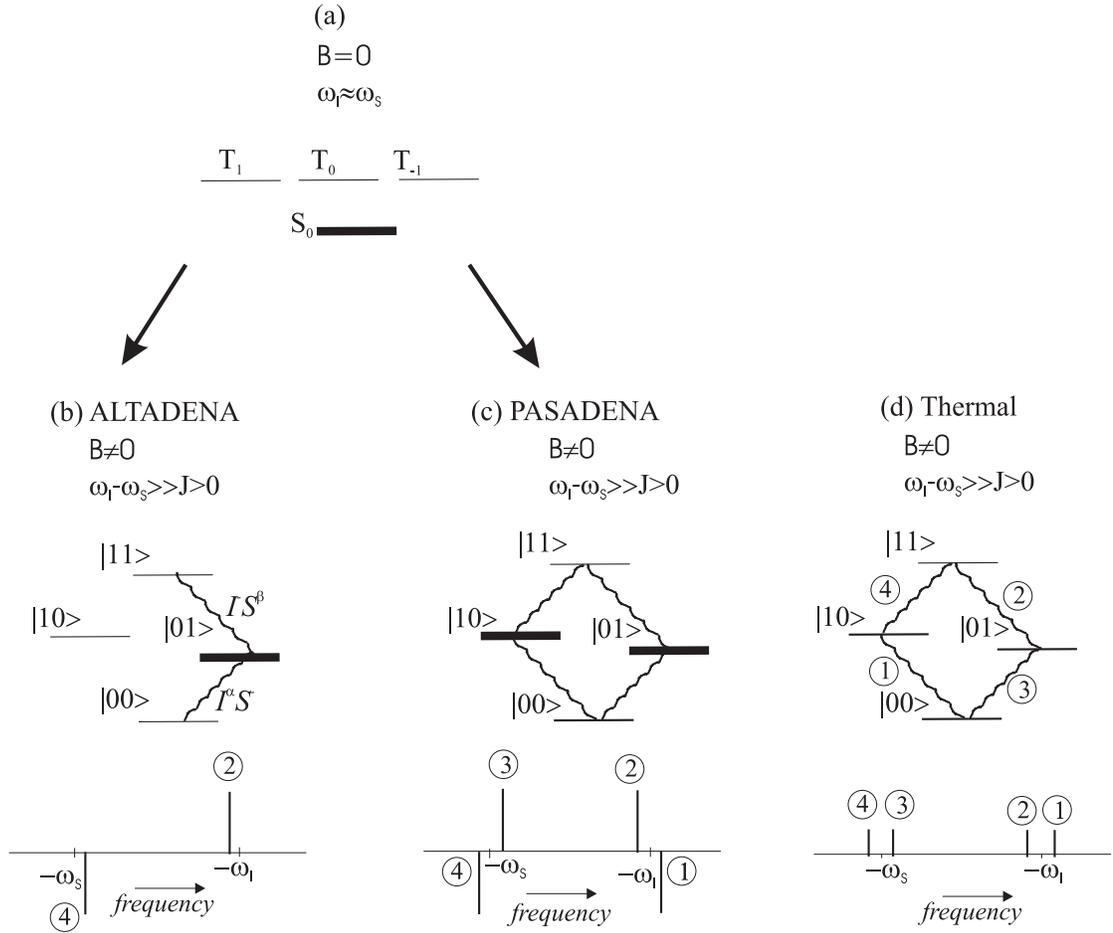}
\caption{PASADENA and ALTADENA experiments: (a) is the energy level
diagram for the dihydride product outside the magnetic field. Only
$S_0$ is populated and the triplets are triply degenerate. Diagram
(b) includes the ALTADENA population diagram and spectrum for small
$\theta$, \eqref{signal-altadena-small}. The diagram assumes that
$\omega_S>\omega_I$, resulting in only the $\ket{01}$ state being
populated. All spectra are drawn assuming $J>0$ and follow the sign
conventions described in \cite{SignsLevitt}. Part (c) shows the
population diagram and spectrum for the PASADENA experiment with
selective excitation, \eqref{signal-S0-90Iy}. Finally, (d) shows the
population diagram and spectrum for the thermal state,
\eqref{signal-thermal}, also serving as a reference for labelling
the coherences. In all spectra, frequency is positively increasing
from left to right and the vertical heights of the spectra are not
to scale.\label{pasadena-altadena}}
\end{center}
\end{figure}

In ALTADENA, the dihydride is formed outside the magnet, spins $I$
and $S$ acquire almost indistinguishable frequencies
$\omega_I\approx\omega_S\implies (\omega_I-\omega_s)\ll 2\pi J$, and
therefore constitute a strongly coupled \textbf{A}$_2$ spin system.
Only the $S_0$ state will be populated and the triply degenerate
triplets $\{T_{\pm 1},T_0\}$ will be separated from $S_0$ by the $J$
coupling \cite{BowersThesis,PHIPReactionPathwayJACS}, (Appendix
\ref{app-S0T0},) as shown in Figure \ref{pasadena-altadena}(a). The
dihydride is then adiabatically transferred into the intense field
with the result that the system becomes weakly coupled,
$(\omega_I-\omega_S)\gg 2\pi J$, forming an \textbf{AX} system. The
eigenstates in the weakly coupled system are the product states and
out of these four states, only $\ket{01}$ and $\ket{10}$
have a non-zero overlap with the singlet, 
\begin{equation}\label{altadena-overlap-1}
|\frac{1}{\sqrt{2}}\braket{01}{01-10}|^2=|\frac{1}{\sqrt{2}}\braket{10}{01-10}|^2=\frac{1}{2},
\end{equation}
whereas,
\begin{equation}\label{altadena-overlap-2}
|\frac{1}{\sqrt{2}}\braket{00}{01-10}|^2=|\frac{1}{\sqrt{2}}\braket{11}{01-10}|^2=0.
\end{equation}
Due to the adiabatic condition, the system ends up in one of these
states and the details are described in Appendix \ref{app-S0T0}. The
experiment is schematically depicted in Figure
\ref{pasadena-altadena}(b). As only one state (assume
$\ket{\alpha\beta}=\ket{01}$) is populated, two transitions are
allowed, one for each spin, as is shown in the Figure. What the
diagram does not faithfully depict is the fact that if dephasing and
decoherence are ignored, then the population difference between the
populated and unpopulated levels will be of the order of unity; the
spin order and state purity is conserved and the signal strengths
will be very high as compared to the thermal signal intensities,
shown somewhat generously, in part (d) of the same Figure.

The ALTADENA experiment is related in spirit, to the remote
detection experiment \cite{RemoteDetectionPines}, in which the
magnetization is encoded in a low-field environment and then the
sample is physically transported to a high-field magnet, conditions
optimized for detection. However, such techniques which involve
shuttling the sample back and forth are instrumentally demanding and
for purposes of quantum information processing, we can achieve the
same spin purity using an experimentally simpler approach.

\section{\textit{In situ} hydrogenation}\label{in-situ}

The ALTADENA approach relies on slowly varying the Hamiltonian so
that the spin system always remains in an eigenstate of the new
Hamiltonian. What if the Hamiltonian changes suddenly, for example,
by switching on the magnetic field $B_0$ instantaneously? This is
the basis of a second kind of experiment, called PASADENA
(\textit{\textbf{p}ara}-hydrogen \textbf{a}nd \textbf{s}ynthesis
\textbf{a}llow \textbf{d}ramatic \textbf{e}nhancement of
\textbf{n}uclear \textbf{a}lignment), and was discussed in the
original prediction \cite{Bowers} of signal enhancements that are
achievable using \pHH . The effect can be explained using the
``sudden'' approximation \cite{FanoDensityMatrixRMP} in quantum
mechanics. Consider a state $\rho(t)$ and suppose that at time
$t=0$, the Hamiltonian suddenly changes from $H_1$ to $H_2$. The
approximation tells us that the state $\rho(t=0^-)$ does not change
at the very moment the perturbation takes place, \ie\
\begin{equation}\label{sudden-approx}
\rho(t=0^+)=\,\rho(t=0^-).
\end{equation}
Furthermore, for $t>0^+$, the initial state $\rho(t=0^+)$
evolves under the new Hamiltonian $H_2$, 
\begin{equation}\label{sudden-approx-2}
\rho(t>0^+)=\,\exp{(-iH_2t)}\,\rho(t=0^+)\,\exp{(+iH_2t)},\quad\text{for\
}t>0^+.
\end{equation}
In zero-field conditions, the state of the two spin system is
clearly $\rho_1=\,\rho_{para}$ (as in \ref{rho-para}) and whatever
the Hamiltonian suddenly changes to, the new state of the system at
time $t=0^+$ remains unchanged,
\begin{equation}\label{PASADENA-instant}
\rho(t=0^+)=\,\rho(t=0^-)=\,\rho_{para}=\,S_0.
\end{equation}
Therefore $\rho_{para}=\,S_0$ is indeed an accurate initial state
and can be directly used for determining the dynamics of our two
qubit computer after the Hamiltonian switches from one form to
another.

One means to instantaneously change the Hamiltonian is to carry out
the experiment \textit{inside} a strong field: the dihydride product
is formed inside the magnet, instantaneously changing the spin
system from \textbf{A}$_2$ to \AX\ and therefore changing the
Hamiltonian. This is the scheme we employed in our experiments. An
\textit{in situ} hydrogenation, can be achieved, quite conveniently,
with our photolytic reaction, already illustrated in Figure
\ref{reaction-dppe-photo}, in which a laser is used to trigger the
formation of the dihydride, at times of \textit{our choosing}.
Furthermore, the reaction takes place sufficiently fast, justifying
the validity of the sudden approximation, which dictates that the
two spins before and \textit{immediately after} the \textit{in situ}
hydrogenation are in the state $S_0$. This results in both the
$\ket{01}$ and $\ket{10}$ levels being populated, as shown in the
energy diagram in Figure \ref{pasadena-altadena}(c), from which we
deduce the possibility of four transitions, resulting in the
characteristic PASADENA spectrum, a pair of anti-phase doublets. The
exact form of the spectrum, however, depends on the details of the
experiment and the excitation pulse, both of which I shall discuss
shortly. It must also be remembered that population diagrams
\cite{FreemanChoreo}, such as the one shown in Figure
\ref{pasadena-altadena}(c), may sometimes lead us to the wrong
impression: for example, it is not evidently clear whether the
population distribution sketched out in Figure
\ref{pasadena-altadena}(c) suggests an incoherent mixture of states,
$1/2(\ket{01}\bra{01}+\ket{10}\bra{10})$, or the coherent state
$1/2(\ket{01}-\ket{10})(\bra{01}-\bra{10})=S_0$. What we know from
the PASADENA density matrix calculations, corroborated by the
spectra, is that it is indeed the latter.

\section{PHIP signal forms and enhancements}\label{signal-enhance}

I have indicated that in ideal cases both ALTADENA and PASADENA
preserve the purity of the quantum state. The perfect spin order of
the pair of ^1\H\ nuclei in the dihydride product, results in very
strong signals, when compared with the signals derived from the
thermal state. These exceptional polarizations, whether obtainable
from ALTADENA or PASADENA are often captured in the term
\textit{\pHH\ induced polarization} or simply PHIP
\cite{NattererPHRev}. In fact, signal enhancements in NMR
experiments involving hydrogen were observed even before the
theoretical prediction of PHIP was made, and they were wrongly
attributed  \cite{BryndzaCIDNPPH} to CIDNP effects. Puzzlingly, the
enhancements increased when hydrogen was stored in liquid \N_2
temperatures for longer periods before the reaction, and only later,
researchers realized the origin of these increased intensities and
that storing the \dihyd\ in ultra-cold environments simply increased
the mole fraction of \pHH . (For a historical review of the PHIP
effect, see \cite{NattererPHRev}.)

The form and intensity of the PHIP signal depends upon the details
of the experiment and the flip angle of the detection pulse.
Consider a state $\rho$ subject to a detection pulse $P_d$; the
state transforms to $\rho'=\,P_d\,\rho\,P_d^{\dag}$ and consider the
signal vector, \eqref{signal-vector}, from $\rho'$, which I denote
$\textbf{Sg}(\rho,P_d)$. This vector indicates the \textit{relative}
directions (emission or absorption) and \textit{relative} strengths
of spectral lines; but additional information, including reference
spectra, are needed to associate individual vector terms with
individual spin resonances.

\subsubsection{Thermal signal}\label{thermal-signal}

For a two spin \textit{homonuclear} system, the density matrix
$\rho_{th}$ in the high temperature limit $kT>>\hbar\omega$ can be
approximated\footnote{An equally valid approximation is
$\mathbf{1}_4/4-\mathcal{B}/4(I_z+S_z)$ in accordance with
\eqref{equilibrium-rho-2-qubit}; but the
sign of $\mathcal{B}$ is immaterial for our purposes.} as 
\begin{equation}\label{rho-thermal}
\rho_{th}=\,\frac{\mathbf{1}_4}{4}+\frac{\mathcal{B}}{4}(I_z+S_z).
\end{equation}
At $295~$K and for a proton frequency of $400~$MHz, the Boltzmann
factor, $\mathcal{B}$, is $6.48\times 10^{-5}$. A $90_y$ hard pulse
converts $\rho_{th}$ to the state 
\begin{equation}\label{rho-therm-90ypulse}
\rho_{th}'=\,\frac{\mathbf{1}_4}{4}+\frac{\mathcal{B}}{4}(I_x+S_x),
\end{equation}
corresponding to the signal,
\begin{equation}\label{signal-thermal}
\mathbf{Sg}(\rho_{th},90_y)=\frac{\mathcal{B}}{8}\{1,1,1,1\}.
\end{equation}

\subsubsection{ALTADENA enhancement}\label{altadena-enhance}
The ideal ALTADENA experiment results in a state (\textit{e.g.}),
\begin{equation}\label{altadena-state}
\rho_{alt}=\,\ket{01}\bra{01}=\,\frac{\mathbf{1}_4}{4}+\frac{1}{2}(I_z-S_z-2\,I_zS_z),
\end{equation}
which can be excited by a $\theta_y$ detection pulse to yield the
observable terms, \begin{equation}\label{altadena-90y}
\frac{1}{2}(\sin{\theta})\,(I_x-S_x)-\frac{1}{2}(\sin{\theta}\cos{\theta})\,(2\,I_zS_x+2\,I_xS_z).
\end{equation}
If $\theta=\pi/2$, the vector becomes,
\begin{equation}\label{signal-altadena-90}
\mathbf{Sg}(\rho_{alt},90_y)=\,\frac{1}{4}\{1,1,-1,-1\},
\end{equation}
showing that the spectrum comprises two in-phase doublets, one being
in positive and the other in negative absorption. If the excitation
angle $\theta$ is small, $\sin{\theta}\approx\theta$ and
$\cos{\theta}\approx 1$ and the signal vector can be approximated
as,
\begin{equation}\label{signal-altadena-small}
\mathbf{Sg}(\rho_{alt},\text{small\
}\theta)\approx\,\{0,\theta/2,-\theta/2,0\},
\end{equation}
showing that only the inner two transitions will be excited, in
accordance with the ALTADENA spectrum predicted in Figure
\ref{pasadena-altadena}(b). Comparing (\ref{signal-thermal}) and
(\ref{signal-altadena-90}), we can compute the relative
enhancement of the signal intensities, 
\begin{equation}\label{enhancement-ALTADENA}
\eta_{alt}=\,\frac{1/4}{\mathcal{B}/8}=\frac{2}{\mathcal{B}},
\end{equation}
which is $\approx 30864$ for a proton frequency of $400~$MHz at
$295~$K, showing that the ALTADENA signals will be about four orders
of magnitudes stronger than the thermal signal. This remarkable
signal-to-noise enhancement motivated research in the detection of
minor concentrations of dihydride isomers, sometimes appearing as
intermediates, in catalytic hydrogenation reactions. The enhancement
is itself a proof of the coherent addition of the pair of ^1\H\
nuclei to the precursor: the hydrogens add pairwise, both nuclei
coming from the same \dihyd\ molecule.

\subsubsection{PASADENA enhancement}\label{pasadena-enhancement}

The state after the PASADENA style \para -hydrogenation is given by
the singlet state
(\ref{S0-PO},\ref{rho-para},\ref{PASADENA-instant}) and to compute
the enhancements, we must first consider a suitable detection pulse.
A simple hard pulse will not work here: the singlet is an isotropic
state, with a deviation component
$1/2(-2\,I_xS_x-2\,I_yS_y-2\,I_zS_z)=-\mathbf{I}\cdot\mathbf{S}$,
which is symmetric in the $I$ and $S$ spin operators. Hard pulses,
also have symmetric spin Hamiltonians of the form
$I_\alpha+S_\alpha$ and therefore commute with the isotropic state, 
\begin{equation}\label{commute-hard-singlet}
[\mathbf{I}\cdot\mathbf{S},\theta(I_\alpha+S_\alpha)]=\,0,\quad\text{where\
}\alpha =x,y,z.
\end{equation}
So, the singlet remains invariant and therefore undetectable under
hard pulses, suggesting the need to apply an asymmetric RF
Hamiltonian. Soft pulses such $\theta_1 I_y+\theta_2 S_y$ with
$\theta_1\ne\theta_2$, asymmetrically excite the spin system and
could therefore be used for detection. In our experiments, we use a
soft $90\,I_y$ pulse, selectively exciting \textit{only} the $I$
spin. We can deduce the observable terms using the product
operator notation, 
\begin{align}
-\mathbf{I}\cdot\mathbf{S}=&\,\frac{1}{2}(-2\,I_xS_x-2\,I_yS_y-2\,I_zS_z)\nonumber\\
\xrightarrow{90\,I_y}&\,\frac{1}{2}(2\,I_zS_x-2\,I_yS_y-2\,I_xS_z)\nonumber\\
\xrightarrow{\text{obs}}&\,\frac{1}{2}(2\,I_zS_x-2\,I_xS_z),\label{detect-S0-90y-PO}
\end{align}
suggesting that the spectrum comprises two opposite anti-phase
doublets. This is also captured in the corresponding
signal vector, 
\begin{equation}\label{signal-S0-90Iy}
\mathbf{Sg}(\rho_{para},90\,I_y)=\,\frac{1}{4}\{-1,1,1,-1\},
\end{equation}
which suggests \textit{seeing} a spectrum of the form $\{-,+,+,-\}$
or $\{+,-,-,+\}$ depending on the identities of $I$ and $S$ spin
resonances, where ``$+$'' and ``$-$'' refer to positive and negative
absorption lines respectively. The enhancement for the PASADENA
$90\,I_y$ experiment is identical to the ALTADENA $90\,I_y$
detection, \ie\
\begin{equation}\label{enhancement-PASADENA}
\eta_{pas}=\,\frac{1/4}{\mathcal{B}/8}=\frac{2}{\mathcal{B}}.
\end{equation}
Similarly, if we have a mixed state of the form
(\ref{rho-mixture-para-ortho}), the signal vector and enhancement
will be rescaled by a factor of $(4F-1)/3$, indicating that the
enhancements are maximum for pure \pHH .

In short, if we trigger the reaction inside the magnetic field using
a short laser pulse and detect the singlet immediately after the
product formation using a selective $90^\circ$ pulse, we should
observe two anti-phase doublets. Subsequently, allowing the spins to
relax to the thermal state and detecting with a hard $90^\circ$
excitation pulse, we shall see two in-phase doublets, and the
intensities of the thermal signal would be smaller by a factor of
$2/\mathcal{B}=2\hbar\omega/kT$. This factor is $\approx 30864$ for
a proton frequency of $400~$MHz at $295~$K.

\begin{figure}
\begin{center}
\includegraphics[scale=0.8]{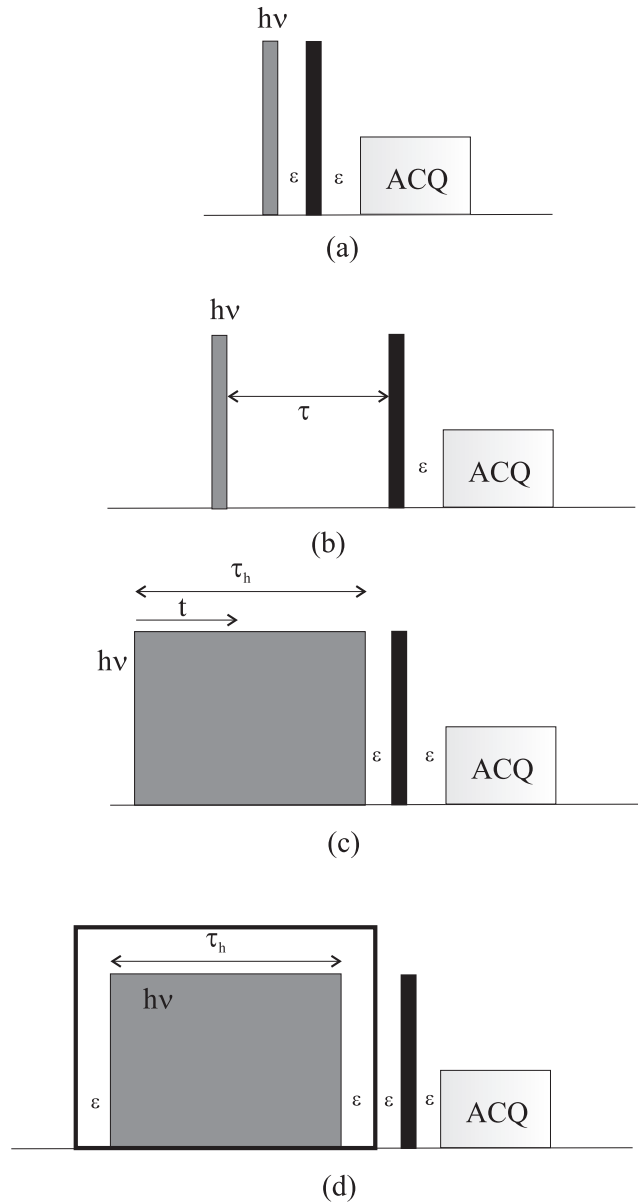}
\caption{Different modifications of the PASADENA experiment: (a)
instantaneous PASADENA involves detection right after a short laser
flash from a pulsed laser, (b) the delayed version of PASADENA when
the singlet is allowed to evolve for a duration $\tau$ before being
detected, (c) incoherent PASADENA, representing hydrogenation for an
extended period of time $\tau_h$, which can be achieved using our CW
laser, (d) isotropic PASADENA when extended hydrogenation takes
place in the presence of an isotropic mixing sequence, shown here by
an outlined rectangle. In each of these diagrams, the solid black
rectangle represents a detection pulse of some kind, the grey
rectangle represents laser irradiation, ACQ represents acquisition,
and periods marked by $\varepsilon$ are infinitesimal delays, shown
only for illustrative clarity.\label{pasadena}}
\end{center}
\end{figure}

As a part of a research collaboration, we have investigated four
kinds of experiments exploiting the PHIP effect.
\begin{enumerate}
\item In the \textit{instantaneous} PASADENA experiment, we detect the singlet
immediately after it is formed. The laser shot and the detection
pulse are controlled through a common spectrometer trigger,
synchronizing the two events. Instantaneous PASADENA is
schematically sketched in Figure \ref{pasadena}(a).
\item We also performed a \textit{delayed} PASADENA: detecting the state at a time $\tau >0$
after the dihydride product is formed. This is depicted in Figure
\ref{pasadena}(b).
\item Another modification is \textit{incoherent} PASADENA,
which involves hydrogenation over an extended period of time
$\tau_h$, shown in Figure \ref{pasadena}(c).
\item The extended hydrogenation can also take place in the presence
of an isotropic mixing sequence (to be described in Section
\ref{section-isotropic-pasadena}). I call this experiment
\textit{isotropic} PASADENA, and is shown in Figure
\ref{pasadena}(d).
\end{enumerate}

To understand these different versions of the PASADENA, we must
first look at the evolution of the singlet in the presence of strong
fields.

\section{Singlet dynamics in strong field}\label{singlet-dynamics}

The dihydride \textbf{AX} spin system will evolve in the presence of
the \AX\ Hamiltonian,
\begin{equation}\label{Hamiltonian-weak-coupling}
H=\,\Omega_II_z+\Omega_SS_z+\pi J\,2\,I_zS_z,
\end{equation}
where $\Omega_I$ and $\Omega_S$ are rotating frame angular
frequencies of the two spins. Defining average and differential
frequencies, 
\begin{align}
\Sigma=&\,\frac{\Omega_I+\Omega_S}{2}\label{sum-freq}\quad\text{and}\\
\Delta=&\,\frac{\Omega_I-\Omega_S}{2}\label{diff-freq},
\end{align}
the Hamiltonian can be rewritten as,
\begin{align}
H=&\,(\Sigma +\Delta)I_z+(\Sigma -\Delta)S_z+\pi J\,2\,I_zS_z\nonumber\\
=&\,\stackrel{H_c}{\overbrace{\Sigma (I_z+S_z)+\pi
J\,2\,I_zS_z}}+\stackrel{H_{nc}}{\overbrace{\Delta (I_z-S_z)}}\label{Hamiltonian-strong-coupling-breakapart}\\
=&\,H_c+H_{nc},\label{Hamiltonian-strong-coupling-breakapart-2}
\end{align}
where, in the last step, I have decomposed the Hamiltonian into two
parts, $H_c$ and $H_{nc}$, the first commuting and the second,
non-commuting with the isotropic singlet,
\begin{align}
[H_c\,,\,\mathbf{I}\cdot\mathbf{S}]=&\,0\quad\text{and}\label{Hc-commute}\\
[H_{nc}\,,\,\mathbf{I}\cdot\mathbf{S}]\ne&\,0.\label{Hnc-not-commute}
\end{align}
These Hamiltonians also commute between themselves,
\begin{equation}\label{commute-Hc-Hnc}
[H_c\,,\,H_{nc}]=\,0,
\end{equation}
allowing us to make the simplification,
\begin{equation}\label{Hamiltonian-decompose}
\exp{(-iH\tau)}=\,\exp{(-i(H_c+H_{nc})\tau)}=\,\exp{(-iH_{nc}\tau)}\exp{(-iH_{c}\tau)}.
\end{equation}
Important commutation relations
between several operator terms are summarized in Table \ref{Table-commutation}.
\begin{table}
\begin{center}
\begin{tabular}{cccc}
\hline \\
State $\rho$ & $[\rho,I_zS_z]$ & $[\rho,I_z-S_z]$ & $[\rho,I_z+S_z]$\\
 \hline \\
$ZQ_x$ & $0$ & $\ne 0$ & $0$\\
$DQ_y$ & $0$ & $0$ & $\ne 0$\\
$I_zS_z$ & $0$ & $0$ & $0$\\
 \hline
\end{tabular}
  \caption{Important commutation relations of different states $\rho$ with operator terms. \label{Table-commutation}}
  \end{center}
\end{table}

Considering only the deviation terms ($-ZQ_x-I_zS_z$), the singlet
$S_0$
evolves for time $\tau$ according to the prescription,
\begin{align}
-ZQ_x-I_zS_z\xrightarrow{H\tau}&\,\exp{(-iH_{nc}\tau)}\exp{(-iH_{c}\tau)}(-ZQ_x-I_zS_z)\exp{(iH_{c}\tau)}\exp{(iH_{nc}\tau)}\nonumber\\
=&\,\exp{(-iH_{nc}\tau)}(-ZQ_x-I_zS_z)\exp{(iH_{nc}\tau)}\nonumber\\
&\quad\quad\text{(as $H_c$ commutes with the singlet, from \eqref{Hc-commute})}\nonumber\\
=&\,\exp{(-i\Delta (I_z-S_z)\tau)}(-ZQ_x-I_zS_z)\exp{(i\Delta
(I_z-S_z)\tau)}.\label{evolution-S0}
\end{align}
The evolution can be simplified even further, as the $-I_zS_z$ term
commutes with $\Delta (I_z-S_z)$,
\begin{equation}\label{commute-deltaz-coupling}
[\Delta (I_z-S_z)\,,\,-I_zS_z]=\,0,
\end{equation}
and so the problem is reduced to considering the evolution of the
$ZQ_x$ term under the Hamiltonian term $\Delta (I_z-S_z)$.
The discussion leads us to two important observations.
\textit{First}, the singlet does not evolve under the weak coupling
$2\,I_zS_z$ Hamiltonian and \textit{second}, the evolution of the
singlet is completely determined by the evolution of the zero
quantum term $ZQ_x=I_xS_x+I_yS_y$, and that too, under only the
term $\Delta (I_z-S_z)$, while the $-I_zS_z$ term remains invariant,
\begin{equation}\label{evolution-S0-final}
-ZQ_x-I_zS_z\xrightarrow{H\tau}\,\{\exp{(-i\Delta\tau
(I_z-S_z))}(-ZQ_x)\exp{(i\Delta\tau (I_z-S_z))}\}-I_zS_z.
\end{equation}

\begin{figure}
\begin{center}
\includegraphics[scale=0.85]{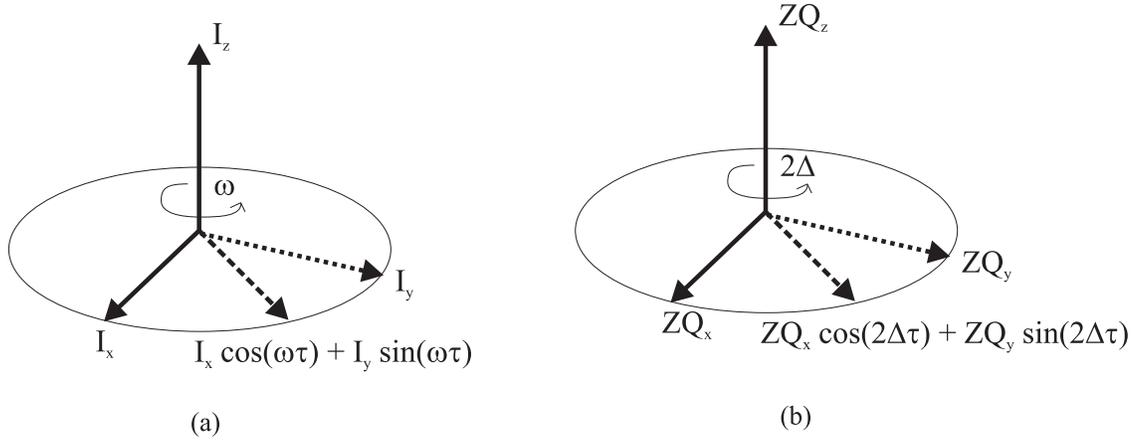}
\caption{Correspondence between the dynamics of (a) single spin
operators and (b) two spin operators in the zero quantum subspace.
The correspondence is $I_z\leftrightarrow ZQ_z$, $I_x\leftrightarrow
ZQ_x$, $I_y\leftrightarrow ZQ_y$ and $\omega\leftrightarrow
2\Delta$.\label{ZQevolve}}
\end{center}
\end{figure}

I now consider the evolution of \textit{only} $ZQ_x$ under
\textit{only} $\Delta (I_z-S_z)$. Suppose we place the transmitter
(Tx) frequency in the exact centre of the $I$ and $S$ spin
resonances; under these conditions, $\Omega_S=-\Omega_I$ and the
angular frequency $\Delta$ becomes
$(\Omega_I-(-\Omega_I))/2=\Omega_I$, which for simplicity, I denote
as $\Omega$. With this definition, the spin resonances will be
$2\,\Omega$ or $2\pi \delta$ radians s$^{-1}$ apart, where I have
defined $\delta$ as the difference between the spin frequencies in
Hz. (Frequency separation is $2\Omega=2\pi \delta~$rad s$^{-1}$.)
Now predicting the evolution of $-ZQ_x$ under the Hamiltonian
$\Delta
(I_z-S_z)$ is straightforward if we consider the commutation relation, 
\begin{equation}\label{commutation-ZQ}
[ZQ_x\,,\,ZQ_y]=\,i\frac{(I_z-S_z)}{2},
\end{equation}
and, defining\footnote{This definition is found in \cite{GlaserHH}
but it is not widely used in standard NMR literature.},
\begin{equation}\label{ZQz}
ZQ_z=\,\frac{(I_z-S_z)}{2},
\end{equation}
enables us to write
\begin{equation}\label{commutation-ZQ-2}
[ZQ_x\,,\,ZQ_y]=\,iZQ_z,
\end{equation}
which is in complete one-to-one correspondence with
$[I_x,I_y]=iI_z$. I rewrite the Hamiltonian $\Delta (I_z-S_z)$ as
$2\,\Delta (I_z-S_z)/2=2\,\Delta\,ZQ_z$, and predict the dynamics of $ZQ_x$, 
\begin{equation}\label{ZQx-evolve}
ZQ_x\xrightarrow{ZQ_z\tau}\,ZQ_x\cos{(2\,\Delta\tau)}+ZQ_y\sin{(2\,\Delta\tau)}.
\end{equation}
The evolution is illustrated in Figure \ref{ZQevolve} and shows that
the $ZQ_x$ term evolves in a subspace spanned solely by zero quantum
vectors, $\{ZQ_x,ZQ_y,ZQ_z\}$, continuously traversing the
``transverse'' zero quantum plane with a precession frequency
$2\,\Delta=2\pi \delta$, and sinusoidally interconverting between
$ZQ_x$ and $ZQ_y$.

\section{Delayed PASADENA}\label{pasadena-delay-subsection}

I have already derived the signal obtainable from instantaneous
PASADENA, \eqref{signal-S0-90Iy}, and I now predict the outcomes
from the other experiments. In delayed PASADENA, the pure singlet
evolves for a time $\tau$ under the Hamiltonian
\eqref{Hamiltonian-weak-coupling}, in accordance with 
\begin{align}
S_0\sim
-ZQ_x-I_zS_z\xrightarrow{H\tau}&\,-ZQ_x\cos{(2\,\Delta\tau)}-ZQ_y\sin{(2\,\Delta\tau)}-I_zS_z\label{para-evolve-full-BigDelta}\\
=&\,-ZQ_x\cos{(2\,\pi\delta\tau)}-ZQ_y\sin{(2\,\pi
\delta\tau)}-I_zS_z\sim\,\rho
(\tau)\label{para-evolve-full-smalldelta},
\end{align}
provided we place the transmitter in the exact centre of the
resonances. (As mentioned in Section
\ref{section-product-operator-define}, ``$\sim$'' denotes the
deviation part of the matrix.) The second of these equations,
\eqref{para-evolve-full-smalldelta}, re-parameterizes the frequency
in terms of the absolute frequency difference $\delta$. The equation
also shows that in time $\tau=1/(4\,\delta)$, $ZQ_x$ fully converts
into $ZQ_y$ and if we wait four times longer, \ie\ for
$\tau=1/\delta$, $ZQ_x$ would go full circle and end up where it
started from. This points towards the possibility of stroboscopic
observation: if decoherence is neglected,  $S_0$ would
\textit{appear} static at times which are multiples of the
precessional time period, $k/\delta$, for $k=0,1,2,\ldots$.
Furthermore, $ZQ_y=I_yS_x-I_xS_y$ is antisymmetric in the $I$ and
$S$ spin operators, and so we can detect the state with a hard
pulse. In Section \ref{section-tomography-pulse}, I shall describe
how we can use properly timed delays, fully converting the symmetric
$ZQ_x$ into antisymmetric $ZQ_y$ and hard pulses, to effectively
achieve a selective $90\,I_y$ pulse.

The state $\rho (\tau)$ (as in \eqref{para-evolve-full-smalldelta}),
can be detected using either a selective $90\,I_y$ pulse or a hard
$90^\circ$ pulse. The
selective pulse, results in the observable terms, 
\begin{equation}\label{pasadena-delay-tau-90Iy}
\rho
(\tau)\xrightarrow{90\,I_y,\,\text{obs}}\,I_zS_x\cos{(2\,\pi\delta\tau)}-I_zS_y\sin{(2\,\pi\delta\tau)}-I_xS_z,
\end{equation}
which compares with \eqref{detect-S0-90y-PO} by setting
$(2\,\pi\delta\tau)=0$. The resultant signal will take up the
form, 
\begin{equation}\label{signal-pasadena-delay-90Iy}
\mathbf{Sg}(\rho
(\tau),90\,I_y)=\,\frac{1}{4}\{-1,1,-\exp{(2\,\pi\delta\tau)},\exp{(2\,\pi\delta\tau)}\},
\end{equation}
indicating that the $I$ spin remains in the anti-phase absorption
mode, while the $S$ lines contain dispersive components, the phases
varying sinusoidally with $\tau$, although they always remain
$180^\circ$ apart within themselves. A hard $90_x$ detection pulse
applied to $\rho (\tau)$ results in the sate,
\begin{equation}\label{pasadena-delay-tau-90x}
\rho
(\tau)\xrightarrow{90_x,\,\text{obs}}\,(-I_zS_x+I_xS_z)\sin{(2\,\pi\delta\tau)},
\end{equation}
and the spectrum,
\begin{equation}\label{signal-pasadena-delay-90x}
\mathbf{Sg}(\rho
(\tau),90_x)=\,\frac{\sin{(2\,\pi\delta\tau)}}{4}\{1,-1,-1,1\}.
\end{equation}

\section{Incoherent PASADENA}\label{section-incoherent-pasadena}

\begin{figure}
\begin{center}
\includegraphics[scale=0.8]{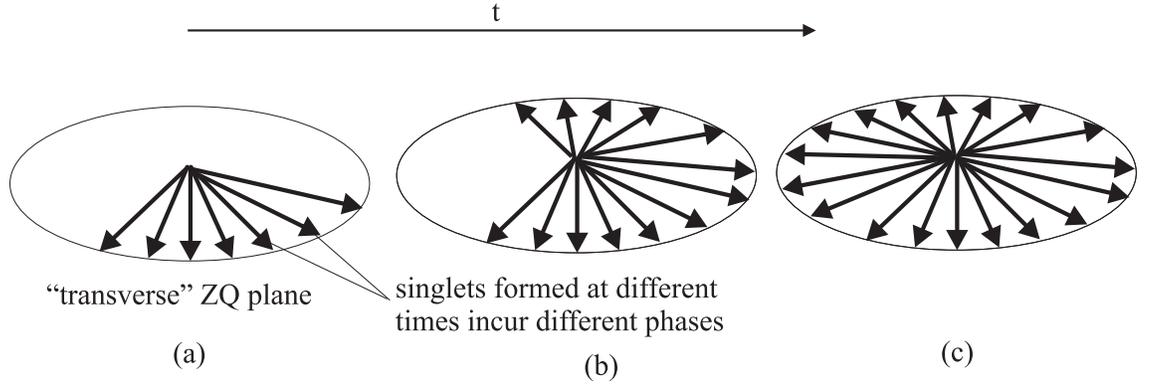}
\caption{The incoherent PASADENA experiment. As time $t$ increases
from (a) through (c), new singlets are continuously being created
and previously formed singlets begin to spread out in the $ZQ$
plane. At the end of the irradiation period, $t=\,\tau_h$, the
magnetization vectors are evenly distributed in the $ZQ$
plane.\label{ZQincoherent}}
\end{center}
\end{figure}

In incoherent PASADENA, the CW laser irradiates the precursor for an
extended duration $\tau_h$: singlets are formed at different times
and consequently, evolve by different amounts. If the duration
$\tau_h\gg 1/\delta$, the average state will be $\rho_{inc}=-I_zS_z$
as the terms in the $ZQ$ plane will have dephased completely, as
shown in Figure \ref{ZQincoherent}. Incoherent PASADENA is, in fact,
a time-distributed ensemble of coherent processes, the averaging in
time acts as decoherence in disguise \cite{QPTQFT}.

The state $\rho_{inc}$ is incoherent, yet the magnetic ordering is
very high. Traditional PHIP experiments usually generate this state,
after photochemical or thermal hydrogenation is allowed to proceed
for a protracted length of time. A selective
$90\,I_y$ pulse converts it into the state, 
\begin{equation}\label{incoherent-90Iy}
-I_zS_z\xrightarrow{90\,I_y}\,-I_xS_z,
\end{equation}
resulting in an anti-phase multiplet on the $I$ spin while $S$
remains completely ``silent'',
\begin{equation}\label{signal-incoherent-90Iy}
\mathbf{Sg}(\rho_{inc},90\,I_y)=\,\frac{1}{4}\{-1,1,0,0\},
\end{equation}
and therefore could be used to identify the $I$ or $S$ identities of
the multiplets. A hard $90_y$ pulse does not render the state
observable, $-I_zS_z\xrightarrow{90_y}I_xS_x$; however, a
$\theta_y$ pulse with $\theta\ne \pi/2$ converts $\rho_{inc}$ into the state, 
\begin{align}
\rho_{inc}\sim
-I_zS_z\xrightarrow{\theta_y}&\,-(I_z\cos{\theta}+I_x\sin{\theta})(S_z\cos{\theta}+S_x\sin{\theta})\nonumber\\
=&\,-(I_zS_z\cos^2{\theta}+I_zS_x\sin{\theta}\cos{\theta}+I_xS_z\sin{\theta}\cos{\theta}+I_xS_x\sin^2{\theta})\nonumber\\
\xrightarrow{\text{obs}}&\,\frac{1}{2}\sin{(2\theta)}\,(I_zS_x+I_xS_z),
\end{align}
with the resulting signal vector, 
\begin{equation}\label{signal-incoherent-90y}
\mathbf{Sg}(\rho_{inc},\theta_y)=\,\frac{1}{8}\sin{(2\theta)}\{1,-1,1,-1\}.
\end{equation}
The spectrum is of the form $\{-,+,-,+\}$ or $\{+,-,+,-\}$,
comprising two anti-phase multiplets, and the intensities are
maximized for the angle $\theta=\pi/4$. The $45_y$ hard pulse yields
the spectrum,
\begin{equation}\label{signal-incoherent-90y-maximum}
\mathbf{Sg}(\rho_{inc},45_y)=\,\frac{1}{8}\{1,-1,1,-1\}.
\end{equation}
and the enhancement over the thermal signal \eqref{signal-thermal}
is found out to be,
\begin{equation}\label{enhancement-incoherent}
\eta_{inc}=\,\frac{1/8}{\mathcal{B}/8}=\,1/\mathcal{B},
\end{equation}
which is $\approx 15432$ for a proton frequency of $400~$MHz at
$295~$K, and is half the enhancements in the ALTADENA
\eqref{enhancement-ALTADENA} and instantaneous PASADENA
\eqref{enhancement-PASADENA} setups.

\section{Isotropic PASADENA}\label{section-isotropic-pasadena}

Isotropic PASADENA is a variant of incoherent PASADENA:
hydrogenation is allowed to proceed for a duration $\tau_h$, but in
the presence of an \textit{isotropic mixing}
\cite{FreemanChoreo,Cavanagh} sequence. The experiment uses the CW
laser and is shown in Figure \ref{pasadena}(d). The singlet will
evolve under the Hamiltonian \eqref{Hamiltonian-weak-coupling}.
However, consider replacing $H$ with a new Hamiltonian,
\begin{eqnarray}
H_{J,\text{weak}}&=&\,\pi
J\,2\,I_zS_z\label{Hamiltonian-J-weak},\quad\text{or}\\
H_{J,\text{strong}}&=&\,\pi
J\,2\,\mathbf{I}\cdot\mathbf{S}\label{Hamiltonian-J-strong}.
\end{eqnarray}
The state $S_0$ commutes with these coupling Hamiltonians, and is
therefore, a stationary state, ideally remaining unchanged, even for
extended lengths of time. Isotropic mixing sequences, built around
inversion (180_{x,y}) pulses \cite{Claridge}, in general, achieve
the strong coupling Hamiltonian, \eqref{Hamiltonian-J-strong}, and
therefore, hydrogenation in the presence of an ideal isotropic
mixing preserves the singlet \cite{Hubler}.

The Hamiltonian, \eqref{Hamiltonian-J-strong}, comprises only a $J$
coupling and no Zeeman (or chemical shift) terms. In this case,
$\Omega_I=\Omega_S=0$, satisfying the so-called
\textit{Hartmann-Hahn} matching
\cite{Cavanagh,FreemanChoreo,GlaserHH} condition, under which the
polarizations are \textit{mixed} (shared) between the spins, much
like coupled pendulums continuously exchanging their kinetic
energies, while keeping the total energy constant \cite{Pendulums}.
This energy-matched polarization transfer is the basis of the
\textbf{to}tal \textbf{c}orrelation \textbf{sp}ectroscop\textbf{y},
TOCSY \cite{TOCSYOriginalErnst} experiment, which allows
polarization to periodically ``hop'' among spins, even if they are
not directly coupled.

\begin{figure}
\begin{center}
\includegraphics[scale=0.8]{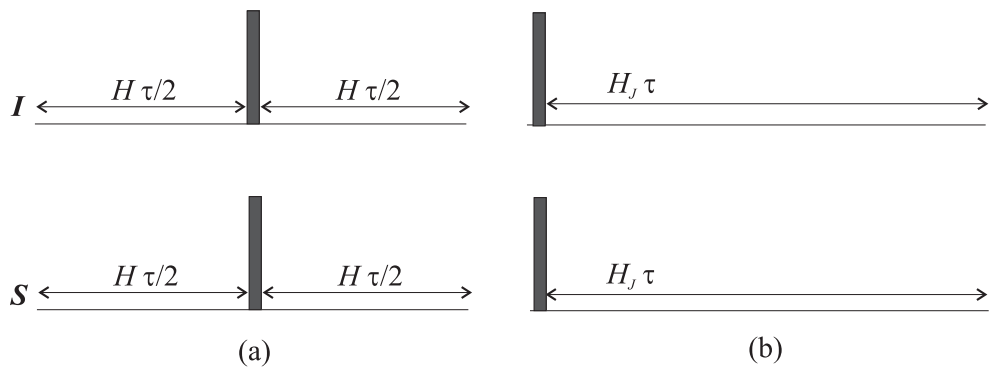}
\caption{Modifying the \AX\ Hamiltonian
\eqref{Hamiltonian-weak-coupling}, to make it ``look like''
$H_{J,\text{weak}}=\,2\pi J\,I_zS_z$. The pulse sequence (a) is the
spin echo and (b) is its equivalent description, provided
observation is stroboscopic. The black rectangle in each case,
denotes a $180_x$ hard pulse. The upper line is the spin $I$ and the
lower is $S$.\label{Figure-spin-echo}}
\end{center}
\end{figure}

The underlying concept in modifying a Hamiltonian is very simple and
can be understood using a familiar example: the homonuclear
spin-echo sequence \cite{LevittSpinDynamics,Toolkit}, shown in
Figure \ref{Figure-spin-echo}. The sequence, shown in part (a) of
the figure, sandwiches a $180_x$ pulse between two periods of free
precession under the weak coupling Hamiltonian,
\eqref{Hamiltonian-weak-coupling}. If the two spin system is
observed after the second $\tau/2$ period, then the evolution can
always be described by an effective Hamiltonian
$H_{J,\text{weak}}=2\pi J\,I_zS_z$, which involves only the scalar
coupling term. Successive inversion pulses can be applied at
intervals of $\tau$, which should be less than $|1/J|$, and if the
observation is synchronized with the repetition rate, the effective
Hamiltonian will be $H_{J,\text{weak}}$.

It is possible to ``sculpt'' more complicated Hamiltonians, using
recurrent combinations of pulses and interspersed delays of free
precession, or back-to-back pulses, and the mathematical theory of
coherent averaging (also called average Hamiltonian theory)
\cite{ErnstNMR,WaughCoherentAveraging}, remains a powerful tool for
analyzing such sequences. For example, different kinds of pulse
sequences can be designed to attain an effective Hamiltonian of the
form, \eqref{Hamiltonian-J-strong}. Most of these isotropic mixing
sequences use composite pulses
\cite{CompositeLevitt,CompositeCounsell,WimperisComposite}, instead
of the simple inversion pulses, as depicted in Figure
\ref{Figure-spin-echo}. The composite pulses can compensate
systematic errors and act over larger frequency ranges without
overly demanding power requirements. Different possibilities of
`composite pulse mixing'\footnote{I use this non-standard phrase in
comparison with the established phrase of `composite pulse
decoupling' \cite{Claridge,GlaserHH}.} sequences exist, and they are
thoroughly investigated in \cite{GlaserHH}.

Composite pulse mixing sequences are generally ``built up'' from
their constituent pulses by a more or less algorithmic procedure,
which is outlined in Appendix \ref{app-isotropic-mixing} for the
MLEV-16 sequence \cite{MLEV}. I numerically investigated the
performance of different na\"{\i}ve and sophisticated isotropic
mixing sequences, in terms of their ability to preserve an ideal
singlet state. These numerical results are also presented in
Appendix \ref{app-isotropic-mixing}. The next chapter describes our
experimental results from isotropic PASADENA demonstrating the
approximate preservation of the singlet.

\section{Summary of PHIP signals}\label{section-phip-summary}

\begin{figure}
\begin{center}
\includegraphics[scale=]{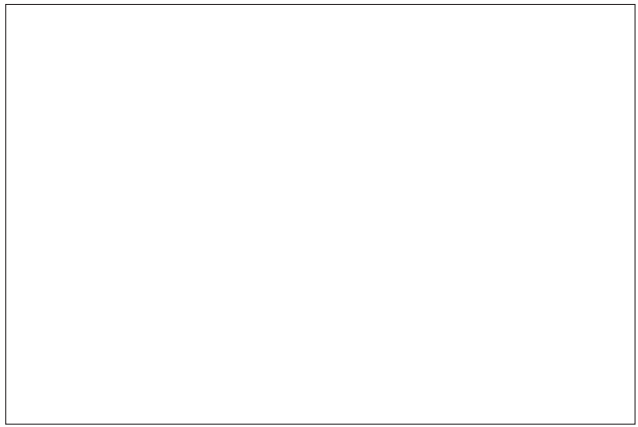}
\caption{Predicted spectra from (a) an ALTADENA experiment with a
$90_y$ detection pulse, (b) instantaneous PASADENA with $90\,I_y$,
(c) delayed PASADENA with $90_x$, (d) incoherent PASADENA with a
$45_y$ hard pulse and (e) incoherent PASADENA with a $90\,I_y$
pulse. The $I$ spin is \textit{assumed} to be towards the left of
the spectrum. All of these spectra are drawn to the same vertical
scale; with (a), (b) and (e) of similar size and twice as big as
(d). In plotting (c), I assumed $\sin{(2\pi\delta\,\tau)}=0.6$,
resulting in a signal smaller than (a) by a factor $1/0.6$.
\label{Figure-spectrums-predicted}}
\end{center}
\end{figure}

The main results including suitable detection strategies, spectral
forms and signal enhancements for different PHIP based experiments
are summarized in Table \ref{summary-alta-pasa}. Predicted spectra
for the different experiments are presented in Figure
\ref{Figure-spectrums-predicted}.

\begin{table}
\begin{tabular}{lcccc}
\hline  Experiment & excitation pulse & spectral form & predicted enhancement \\
 \hline
 ALTADENA & $90_y$ & $\{+,+,-,-\}$ & $2/\mathcal{B}$ \\
 Instantaneous PASADENA & $90\,I_y$ & $\{-,+,+,-\}$ & $2/\mathcal{B}$ \\
 Delayed PASADENA & $90_x$ & $\{+,-,-,+\}$ & $2/\mathcal{B}\sin{(2\,\pi\delta \tau)}$ \\
 Incoherent PASADENA & $45_y$ & $\{+,-,+,-\}$ &
 $1/\mathcal{B}$\\
 Incoherent PASADENA & $90\,I_y$ & $\{-,+,0,0\}$ &
 $2/\mathcal{B}$\\
  \hline
\end{tabular}
\caption{Summary of the main results for ALTADENA and the various
kinds of PASADENA style hydrogenations.} \label{summary-alta-pasa}
\end{table}

The next two chapters discuss our PHIP experiments.
\begin{enumerate}
\item Our primary motivation in harnessing the PHIP effect was to
prepare pure initial states that could be used for quantum
information processing. We experimentally demonstrate the
preparation of such states using a single flash of a pulsed laser,
utilizing instantaneous PASADENA detection. The resulting two qubit
system is almost pure and also lies above the entanglement
threshold. \item With the isotropic PASADENA, we can demonstrate
interesting dynamical characteristics of the singlet state,
including decoherence. \item Finally, we implement two quantum
algorithms (Deutsch and Grover) using pure initial states,
demonstrating for the first time the use of pure states in liquid
state NMR for quantum computation; the experiments are taken up in
Chapter \ref{QCimplement} .
\end{enumerate}

\chapter{Preparing pure initial states using \pHH}\label{pure}

Every computation must start off in a well defined initial state. As
seen in Chapter \ref{critique}, one of the major objections against
NMR as a scalable quantum computing technology, is the problem of
initialization. This chapter deals with our initialization
experiment, which is an important first and \textit{successful}
attempt at realizing an essentially pure quantum state, also capable
of performing universal quantum dynamics. The state also lies above
the entanglement threshold.

In our initialization experiment, pure \pHH\ is prepared. The
magnetic order is then coherently deposited into a molecule through
a precisely controlled, fast chemical reaction. The migration of
spin order from \pHH\ to the molecule of interest must take place
coherently and without appreciable loss, and at the end we must
determine the density matrix and thus its purity and entanglement of
formation. The preparation must also take place in a single
experiment, without the need for multiple runs (as in temporal
averaging schemes for pseudopure state preparation
\cite{KnillChuangEffectivePure}). Our experiment demonstrably
achieves all these requirements.

Section \ref{section-partial-twirl} describes a method of filtering
the state for quantum state tomography in a single experiment and
Section \ref{section-tomography-pulse} describes the actual pulse
sequence used for detection. The experiment itself, with all
accompanying technical details is surveyed in Section
\ref{section-experimental-details}. The acquired spectra are then
processed and interpreted; the data processing and the final results
are presented in Section \ref{section-tomography-results}. Finally,
Section \ref{section-CW} shows some results from the isotropic
PASADENA experiment, utilizing the CW laser. A very concise
description of this experiment can also be found in our published
work, \cite{AnwarPH}.

Unsurprisingly, the \pHH\ route to achieving pure initial states has
also been attempted previously \cite{Hubler}, although the purity
reached in that case was $\varepsilon\approx 0.1$, well below the
entanglement threshold. H\"{u}bler and co-workers used the molecule
$\text{Ir}(\text{H})\text{Cl}(\text{H})(\text{CO})(\text{PPh}_3)_2$
as their quantum computer, which they obtained from the addition of
a $1:1$
\para -\ortho\ mixture to their precursor molecule
$\text{Ir}\text{Cl}(\text{CO})(\text{PPh}_3)_2$ (Vaska's catalyst).
Our results show an almost nine-fold improvement in the purity, and
we attribute this increase to two factors. \textit{First}, we use
almost pure \pHH , prepared at $20~$K and using Equation
\eqref{rho-mixture-para-ortho} with singlet fractions $F=1$ and
$F=1/2$, gives us a three-fold enhancement in the magnetic order
straight away (see Figure \ref{Figure-para-ortho}). \textit{Second},
the original experiment permitted hydrogenation for an extended
period of time, resulting in signal loss due to the rotating frame
spin-lattice relaxation, $T_{1}\rho$ \cite{DictionaryNMR} (in the
presence of the isotropic mixing sequence \cite{MLEV}); and we
believe relaxation diminished their purity by another factor of
three.
Our experiment overcomes this second problem by letting the reaction
proceed to completion almost instantaneously. Pure \pHH\ gas and
faster reaction times, therefore, act together in bringing our
purity levels very close to one.

\section{One-shot tomography}\label{section-partial-twirl}

Full state tomography requires multiple copies of $\rho$ and so we
have come up with a so-called partial ``twirl''\footnote{The twirl
is the subject of Chapter \ref{twirl}.} operation, which prepares a
state having only two independent parameters, and is therefore
characterizable in a single experiment. Our procedure characterizes
the state obtained \textit{after} the assumedly perfect partial
twirl, and not the initial state prepared by the laser flash. In
this sense, the tomography is restricted to only a constrained
state, in which several independent parameters have been zeroed out.
(As I will discuss later, our experiments indicate that the
constrained state is almost indistinguishable from the initial
state.)

\begin{figure}
\begin{center}
\includegraphics[scale=]{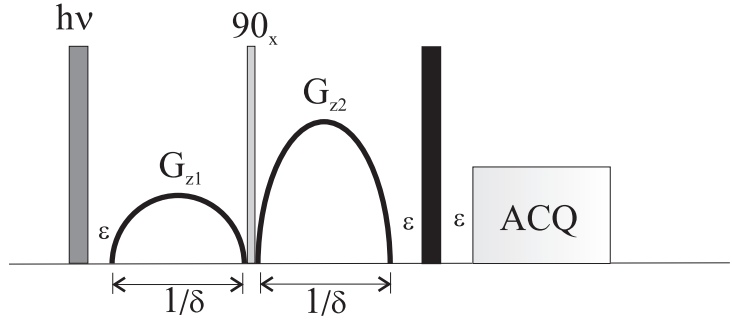}
\caption{Scheme of the initialization experiment. The dark grey
rectangle represents a single flash from the pulsed laser, the black
rectangle indicates a $90\,I_y$ detection pulse. The arcs show
strong gradient fields, sandwiching a light grey rectangle,
representing a $90_x$ pulse. The gradients and the intervening pulse
comprise the partial twirl, also called the \textit{filtration}
sequence.}\label{filtration}
\end{center}
\end{figure}

The ideal initialization experiment, which is instantaneous
PASADENA, produces the singlet state, $S_0$. If this were the case,
no tomography is needed, as the output would be known before hand.
However, more realistically, impurity terms mix in with $S_0$. This
could be due to several possible reasons: the small but finite
reaction times, $ZQ_x\leftrightarrow ZQ_y$ mixing under the \AX\
Hamiltonian, the appearance of $T_1$ and $T_{-1}$ terms due to
spin-lattice relaxation and reaction with residual \ortho -hydrogen.
The partial twirl takes an input state, $\rho_{in}$, with all these
error terms, and simplifies it into a known form, $\rho_{out}$,
\begin{equation}\label{partial-twirl-scheme}
\rho_{in}\xrightarrow{\text{partial twirl}}\,\rho_{out}.
\end{equation}
A key requirement of the partial twirl is that it leaves the singlet
unchanged, both in form and fraction. Our partial twirl satisfies
this requirement and outputs an $S_0/T_0/T_m$ mixture (see Section
\ref{section-entanglement-singlet-triplet}), a state of the form,
\begin{equation}\label{filter-output-symmetrical}
\rho_{out}=\,a\,S_0+b\,T_0+c\,(T_1+T_{-1}).
\end{equation}
The coefficients $a$, $b$ and $c$ are the singlet and triplet
fractions, \ie\ $a=\langle S_0\rangle=F$, $b=\langle T_0\rangle$ and
$c=\langle T_1\rangle=\langle T_{-1}\rangle$. For a valid density
matrix, $a+b+2\,c=1$, acting as a constraint on the $\{a,b,c\}$
trio; as a result, only two coefficients are linearly independent
and one-shot tomography is possible. An alternative representation
of $\rho_{out}$ is in the product operator-multiple quantum basis,
\begin{equation}\label{filter-output-PO}
\rho_{out}=\,\frac{\mathbf{1}_4}{4}+p\,ZQ_x+q\,I_zS_z,
\end{equation}
which parameterizes the state in terms of $p$ and $q$, the two
independent parameters, which are directly related to the singlet
and $T_0$ triplet fractions, 
\begin{align}
p=&\,-a+b\quad\text{and}\label{filter-output-relation-1}\\
q=&\,1-2a-2b\label{filter-output-relation-2},
\end{align}
and conversely,
\begin{align}
a=&\,\frac{1}{4}(1-2p-q)\label{filter-output-relation-3}\\
b=&\,\frac{1}{4}(1+2p-q)\label{filter-output-relation-4}.
\end{align}
For pure $S_0$, $a=1$, $b=c=0$ and $p=q=-1$. Tomography translates
into determining the values of these coefficients. The
initialization experiment therefore involves a single laser flash
for making the dihydride, filtering it to bring it into the desired
form, and finally measuring the state achieved; the overall scheme
is shown in Figure \ref{filtration}.

\subsection{Implementing the partial
twirl}\label{partial-twirl-implement}

The partial twirl comprises a pair of strong $z$ gradients
(crushes), $\mathbf{G}_{z1}$ and $\mathbf{G}_{z2}$, of unequal
strengths, surrounding a $90_x$ pulse. I assume all pulse elements
to be ideal. We can understand the working of the
$[\mathbf{G}_{z1}\thickspace 90_x\thickspace \mathbf{G}_{z2}]$
sequence by following the fate of the various terms in the density
matrix as the sequence progresses. Defining the intermediate states
$\rho_1$ and
$\rho_2$, the states transforms as, 
\begin{equation}\label{filter-progress}
\rho_{in}\xrightarrow{\mathbf{G}_{z1}}\,\rho_1\xrightarrow{90_x}\,\rho_2
\xrightarrow{\mathbf{G}_{z2}}\,\rho_{out}.
\end{equation}
I expand the deviation part of $\rho_{in}$ in the product
operator-multiple quantum coherence basis,
\begin{equation}\label{filtration-input}
\begin{split}
\rho_{in}\sim\,&e_1(I_x+I_y+S_x+S_y+I_xS_z+I_yS_z+I_zS_x+I_zS_y)\\
&+e_2(I_z+S_z)+e_3(DQ_x+DQ_y)+e_4(ZQ_y)+\{l(ZQ_x)+m(I_zS_z)\}.
\end{split}
\end{equation}
This is not a completely arbitrary state, as several terms are tied
up with the same coefficient; however, the state is sufficiently
general for our purposes. The terms with the $e$ coefficients are
the error terms and must therefore be prevented from reaching the
output stage, $\rho_{out}$, while $ZQ_x$ and $I_zS_z$ should survive
the filtration.

The first gradient crushes all the single and double quantum terms,
shown with coefficients $e_1$ and $e_3$, resulting in,
\begin{equation}\label{filtration-stage-1}
\rho_{in}\xrightarrow{\mathbf{{G}}_{z1}}\rho_{1}\sim\,e_2(I_z+S_z)+e_4(ZQ_y)+\{l(ZQ_x)+m(I_zS_z)\}.
\end{equation}
The Zeeman $(I_z+S_z)$ and anti-symmetric $ZQ_y$ terms are
transformed into single quantum coherences under the
$90_x$ pulse and are crushed by the second gradient, 
\begin{eqnarray}\label{filtration-stage-2-zeeman}
e_2(I_z+S_z)\xrightarrow{90_x}&\,e_2(-I_y-S_y)\xrightarrow{\mathbf{G}_{z2}}\text{decohere},\quad\text{and}\\
e_4(ZQ_y)=e_4(I_yS_x-I_xS_y)\xrightarrow{90_x}&\,e_4(I_zS_x-I_xS_z)\xrightarrow{\mathbf{G}_{z2}}\text{decohere}.\label{filtration-stage-2-ZQy}
\end{eqnarray}
Finally, for the $ZQ_x$ and $I_zS_z$ terms, the $[90_x\thickspace
\mathbf{G}_{z2}]$ sequence mixes the $l$ and $m$ coefficients,
\begin{align}
ZQ_x=\,I_xS_x+I_yS_y&\xrightarrow{90_x}I_xS_x+I_zS_z=\frac{1}{2}(ZQ_x+DQ_x)+I_zS_z\label{ZQZ1}\\
&\xrightarrow{\mathbf{G}_{z2}}\frac{1}{2}ZQ_x+I_zS_z\label{ZQZ2}\quad\text{and}\\
I_zS_z&\xrightarrow{90^{\circ}_x}I_yS_y=\frac{1}{2}(ZQ_x-DQ_x)\label{ZQZ3}\\
&\xrightarrow{\mathbf{G}_{z2}}\frac{1}{2}ZQ_x,\label{ZQZ4}\\
\implies
l(ZQ_x)+m(I_zS_z)&\xrightarrow{90_x}\xrightarrow{\mathbf{G}_{z2}}\frac{1}{2}(l+m)(ZQ_x)+l(I_zS_z)\,\sim\rho_{out}.\label{ZQZ5}
\end{align}
If we compare the above output, $\rho_{out}$, with
\eqref{filter-output-PO}, we deduce that $p=(l+m)/2$ and $q=l$,
immediately suggesting that if $l=m=-1$, then $p=q=-1$, showing that
the singlet component in $\rho_{in}$ is completely conserved, as we
desired. I now consider the effects of finite duration gradient
pulses.

\subsection{Finite duration gradient
pulses}\label{section-filtration-gradient-duration}

It is not possible to implement zero length gradients, so we can
instead think about selecting the \textit{proper} rather than just
the \textit{shortest} timings. For this purpose, we can use the
results in Section \ref{singlet-dynamics}: $ZQ_x$ evolves in the
transverse zero quantum plane according to \eqref{ZQx-evolve}, and
therefore, setting the duration of each gradient to
$t_g=(1/\delta)$, we ensure that $ZQ_x$ is practically ``frozen'' in
time. Furthermore, the $I_zS_z$ term does not precess at all, and
neither of these terms evolves under the scalar coupling. For a
ball-park figure of $\delta=500$~Hz, $t_g=1/\delta=2$~ms, and the
complete filtration sequence lasts for about $2(1/\delta)=4$~ms, so
no appreciable signal loss due to relaxation takes place in this
small amount of time. (For example, considering a $T_2$ of $500$~ms,
the signal decreases by only $0.8$\%.)

In our initialization experiment, we use two schemes: one without
(Figure \ref{pasadena}(a)) and the other with (Figure
\ref{filtration}) the partial twirl. State tomography shows very
similar results for the two experiments, indicating that the error
terms (with coefficients $e$ in \eqref{filtration-input}) are very
small, and that the laser flash generates a state which is
\textit{close} to a Werner singlet.

We can also adjust the length of the gradient pulses to determine
the $I$ and $S$ spin identities of the spectral multiplets. Halving
the duration of each gradient from $(1/\delta)$ to $(1/2\delta)$
prepares the longitudinal state $I_zS_z$, 
\begin{align}
-ZQ_x-I_zS_z\xrightarrow{\mathbf{G}_{z1},\,t_g=1/2\delta}&\,ZQ_x-I_zS_z=I_xS_x+I_yS_y-I_zS_z\nonumber\\
\xrightarrow{90_x}&\,I_xS_x+I_zS_z-I_yS_y=DQ_x+I_zS_z\nonumber\\
\xrightarrow{\mathbf{G}_{z2},\,t_g=1/2\delta}&\,I_zS_z,\label{filtration-half-gradient}
\end{align}
which is converted to $I_xS_z$ by a $90\,I_y$ detection pulse,
resulting in the spectrum, 
\begin{equation}\label{signal-filtration-half-gradient}
\mathbf{Sg}(I_zS_z,90\,I_y)=\,\frac{1}{4}\{1,-1,0,0\}.
\end{equation}
The $I_zS_z$ state formed above is the same state, (of course with a
negative sign,) as $\rho_{inc}$, \eqref{incoherent-90Iy}, and the
partial twirl with the gradient durations reduced to half, is
therefore, a \textit{way of making the incoherent state from the
coherent state}. A selective pulse renders observable only one spin,
which is unambiguously assigned the label $I$ and the signal
\eqref{signal-filtration-half-gradient} together with
\eqref{signal-S0-90Iy} can then be used to determine the positive
and negative amplitude directions in the spectral outputs.

\section{The detection pulse}\label{section-tomography-pulse}

We have devised a detection pulse sequence, determining the
parameters $p$ and $q$, all that is needed to completely reconstruct
the density matrix, $\rho_{out}$.

In fact, the detection pulse is simply the selective $90\,I_y$ pulse
already described in context of singlet detection in the PASADENA
experiment (Section \ref{pasadena-enhancement}). The singlet (with
$p=q=-1$) results in the spectrum \eqref{signal-S0-90Iy} while the
general state, \eqref{filter-output-PO}, with arbitrary $p$ and $q$,
results in,
\begin{equation}\label{pulse-tomography-90Iy}
p(I_xS_x+I_yS_y)+q(I_zS_z)\xrightarrow{90\,I_y}\xrightarrow{\text{obs}}\,p(-I_zS_x)+q(I_xS_z).
\end{equation}
The coefficients do not mix and are seen on separate spins, the $p$
coefficient prepended to the $S$ observable and the $q$ coefficient
to the $I$ observable. The resulting signal vector is, 
\begin{equation}\label{signal-pulse-tomography-90Iy}
\mathbf{Sg}(\rho_{out},90\,I_y)=\,\frac{1}{4}\{q,-q,-p,p\}.
\end{equation}
The signals resulting from applying different kinds of detection
pulses are summarized in Table \ref{Table-pulse-tomography}.

Two comments about the detection pulse are not out of place here.
\begin{enumerate}
\item  The zero order phasing of the NMR spectrum is arbitrary.
The $y$ phase of the detection pulse results in absorption
Lorentzians but using for example, a $90\,I_x$ pulse instead
would result in the dispersive spectrum,
\begin{equation}\label{signal-pulse-tomography-90Ix}
\mathbf{Sg}(\rho_{out},90\,I_x)=\,\frac{1}{4}\{iq,-iq,-ip,ip\}=\,i\mathbf{Sg}(\rho_{out},90\,I_y),
\end{equation}
which could be easily brought into the absorptive form.
\item Similarly we could have also applied a $90\,S_y$ instead of a
$90\,I_y$ pulse without loss of generality. However, in our
experiments and mathematical analysis, we consistently followed the
$90\,I_y$ description.
\end{enumerate}

\begin{table}
\begin{center}
\begin{tabular}{ccc}
\hline  Excitation pulse $P$ && signal vector $\mathbf{Sg}(\rho_{out},P)$ \\
\hline \\
$90\,I_y$ && $1/4\,\{q,-q,-p,p\}$\\
$\theta_y$ && $1/8(p-q)\sin{(2\,\theta)}\{-1,1,-1,1\}$\\
$90_y$ && $\{0,0,0,0\}$\\
$45_y$ && $1/8(p-q)\{-1,1,-1,1\}$\\
  \hline
\end{tabular}
\end{center}\caption{Signal vectors from application of some hard and
selective pulses on the output of the filtration sequence
$\rho_{out}$ as given in
\eqref{filter-output-PO}.\label{Table-pulse-tomography}}
\end{table}

\subsection{Implementing the selective detection pulse}\label{subsection-jumpreturn-implement-selective}

There is a repertoire of methods available in the NMR literature
\cite{FreemanChoreo,Linden99,ShapedFreeman} to apply selective
pulses to one spin in a multi-spin system, the commonest being
shaped pulses. In all our quantum information processing
experiments, we have achieved spin selectivity using sequences based
on the Jump-and-Return method
\cite{Jonesapproxcount,JumpReturnOriginal}. These kind of sequences,
originally developed for water suppression, use only hard pulses and
interleaving delays.

Here I present an explicit construction for designing the
Jump-and-Return implementation of our $90\,I_y$ selective pulse.
The $I$ selective pulse can be written as, 
\begin{equation}
90\,I_y\equiv\,\Biggl\{
\begin{array}{c|c}
45\,I_y & 45\,I_y \\
45\,S_y & 45\,S_{-y}
\end{array},\label{jump-return-1}
\end{equation}
the top and bottom rows comprising pulses on the $I$ and $S$ spins
respectively, and time ordering in the pulse sequences being from
left to right. Now, we know that a $\theta_y$ pulse can be
decomposed\footnote{This identity can be easily proven by explicit
calculation.} as a composite rotation
\cite{FreemanChoreo,CumminsNJPComposite,CompositeLevitt},
\begin{equation}\label{jump-return-2}
\theta_y\equiv\,90_x\thickspace \theta_z\thickspace 90_{-x},
\end{equation}
allowing us to rewrite \eqref{jump-return-1} as,
\begin{equation}
90\,I_y\equiv\,\Biggl\{
\begin{array}{c|c|c|c}
45\,I_y & 90\,I_x & 45\,I_z & 90\,I_{-x} \\
45\,S_y & 90\,S_x & 45\,S_{-z} & 90\,S_{-x}
\end{array},\label{jump-return-3}
\end{equation}
which can be expressed in more compact notation as,
\begin{equation}\label{jump-return-4}
45_y\thickspace 90_x\thickspace 45_{\pm z}\thickspace 90_{-x},
\end{equation}
where the sequence now comprises only hard pulses and
``contra-axial'' ($\pm z$) pulses. This sequence can in fact be
simplified \cite{VandersypenNMRQCCtrl} further. Rearranging the
terms in \eqref{jump-return-2}, we are able to deduce the relation,
\begin{equation}\label{jump-return-5}
\theta_y\thickspace 90_x\equiv\,90_x\thickspace \theta_z.
\end{equation}
Furthermore, we can displace the $z$ rotation to the left or to the
right, while changing the phase of a neighbouring pulse. For
example, using the identity,
\begin{equation}\label{jump-return-6}
\alpha_\phi\thickspace \beta_z\equiv\,\beta_z\thickspace
\alpha_{\phi+\beta}
\end{equation}
the sequence \eqref{jump-return-5} becomes,
\begin{equation}\label{jump-return-7}
\theta_y\thickspace 90_x\equiv\,90_x\thickspace
\theta_z\equiv\,\theta_z\thickspace
90_{0+\theta}\equiv\,\theta_z\thickspace 90_{\theta}.
\end{equation}
This allows us to re-write \eqref{jump-return-4} in the form,
\begin{equation}\label{jump-return-8}
45_z\thickspace 90_{45}\thickspace 45_{\pm z}\thickspace 90_{-x},
\end{equation}
a sequence comprising a non-selective $z$ rotation, a pulse, a
contra-axial $\pm z$ rotation and a final pulse; both pulses being
hard. This sequence is a \textit{state independent} prescription for
implementing a $90\,I_y$ selective pulse. For the output of the
filtration sequence $\rho_{out}$ \eqref{filter-output-PO},
comprising only $ZQ_x$ and $I_zS_z$ terms, a further simplification
is possible, as both terms in $\rho_{out}$ commute with
non-selective $z$ rotations. In fact, the very purpose of the
transformation \eqref{jump-return-7} was to bring this $z$ rotation
to the beginning of the sequence; it has no effect on the state
$\rho_{out}$ and could therefore be simply dropped altogether. The
pulse sequence for fully characterizing the filter output can
therefore be written as,
\begin{equation}\label{jump-return-final}
90_{45}\thickspace 45_{\pm z}\thickspace 90_{-x},
\end{equation}
two hard pulses sandwiching a properly timed delay, in the spirit of
the Jump-and-Return style selective pulses. The hard pulses are
straightforward to implement, while the $45_{\pm z}$ rotation is
achievable by observing its propagator form,
\begin{equation}\label{delay-propagator}
\exp{\bigl(-i\frac{\pi}{4}(I_z-S_z)\bigr)}.
\end{equation}
From this, we immediately recognize that a suitable Hamiltonian to
implement this function would be the non-commuting part
$H_{nc}=\Delta (I_z-S_z)$ of the background Hamiltonian $H$ as
described in \eqref{Hamiltonian-strong-coupling-breakapart-2}. This
Hamiltonian is achieved by \textit{necessarily} placing the
transmitter in the exact middle of the $I$ and $S$ spin resonances
and allowing the two spin system to evolve under the background
Hamiltonian $H$ \eqref{Hamiltonian-weak-coupling} for a time $t_d$,
\begin{equation}
\Delta
t_d=\,\frac{\pi}{4}\implies\,t_d=\,\frac{\pi}{4\Delta}=\,\frac{\pi}{4\pi\delta}=\,\frac{1}{4\delta}\label{delay-time}.
\end{equation}
For a representative value of $\delta=500$~Hz, this delay is
$500~\mu$s.

\begin{figure}
\begin{center}
\includegraphics[scale=]{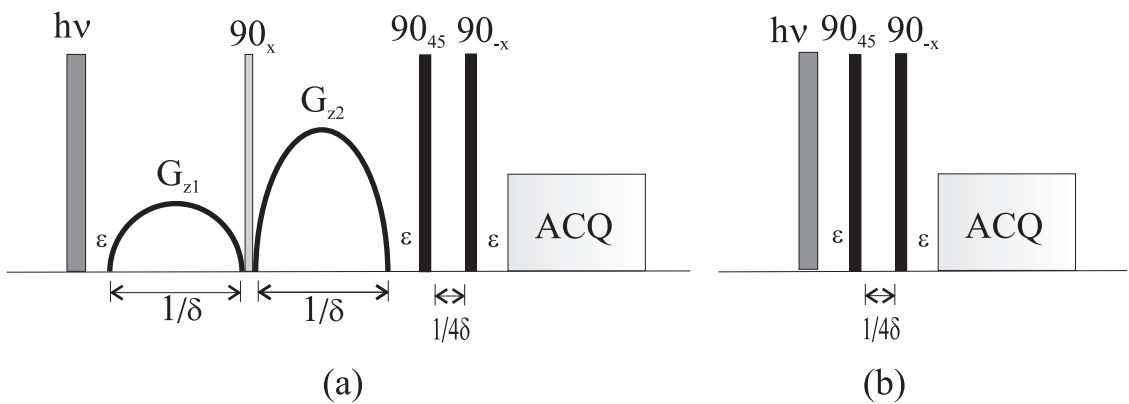}
\caption{The initialization experiment pulse sequences with (a) and
without (b) the partial twirl. All rectangles are appropriately
labelled. The transmitter frequency is placed in the middle of the
resonances.}\label{initialization}
\end{center}
\end{figure}

The complete pulse sequence for the initialization experiment can
therefore be written as,
\begin{equation}\label{initialization-pulse-sequence}
\text{flash}\thickspace \mathbf{G}_{z1,[t_g]}\thickspace
90_x\thickspace \mathbf{G}_{z2,[t_g]}\thickspace 90_{45}\thickspace
[t_d]\thickspace 90_{-x}\thickspace \text{acquire},
\end{equation}
schematically represented in Figure \ref{initialization}(a). If the
laser flash produces a hydride without any error terms in
$\rho_{in}$, we could rather leave out the partial twirl altogether;
the resulting scheme is shown in part (b) of the figure. We have
performed both kinds of experiments and observe no appreciable
difference between the results. Interestingly, if we \textit{knew}
that $\rho_{out}$ is indeed the singlet with $p=q=-1$, we could also
leave out the first hard pulse in the tomography sequence
\eqref{jump-return-final} as the singlet commutes with hard pulses.
However, for mathematical completeness, we retain this initial
pulse. A few additional comments about the detection pulse sequence
are the following.
\begin{enumerate}
\item Analysis shows that the inter-pulse delay, $t_d=1/(4\,\delta)$
maximizes the $ZQ_x$ to $ZQ_y$ conversion, acting as a $\pi/2$
rotation in the transverse zero quantum plane.
\item The sequence is most robust against pulse width errors when the phases of
the pulses are $135^{\circ}$ apart. The role of errors in this
sequence is considered in Appendix \ref{app-errors}.
\end{enumerate}

\section{Experimental system and methods}\label{section-experimental-details}

In this section, I describe the experimental details of our
instantaneous PASADENA experiment\footnote{All experiments discussed
in this and the next chapter, were performed by Dr.~S.~B. Duckett
and Dr.~D. Blazina at the University of York. This included
synthesis and characterization of all compounds, manufacture of the
\pHH\ plant, building and adjusting the optics for efficient
transfer of light to the sample, and finally implementing the pulse
sequences on the spectrometer. Myself and my supervisor, at Oxford,
were involved in pulse sequence design and analysis, data
processing, state tomography, decoherence modelling, simulations and
the theoretical design of the experiment. However, I visited York a
couple of times to assist in the experiments and also carried out
local experiments on simulated states.}.

\subsection{Chemical system}\label{section-chemical-system}

\begin{figure}
\begin{center}
\includegraphics[scale=]{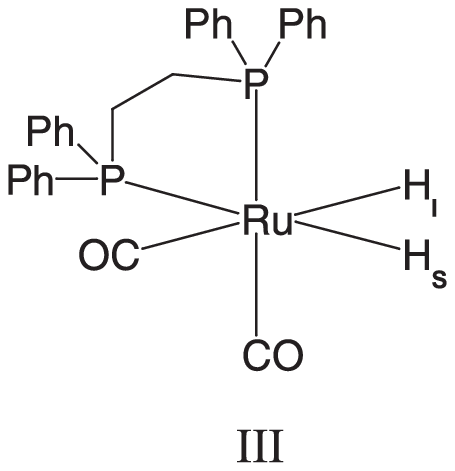}
\caption{The product of the photolysis reaction, \dihydride . The
two qubits in the quantum computer are the protons \H_I and
\H_S.}\label{Figure-dihydride}
\end{center}
\end{figure}

The molecules used in our experiment have already been identified in
Section \ref{chemical-parahydrogen} and the photolysis reaction
underpinning pure state preparation is illustrated in Figure
\ref{reaction-dppe-photo}. The stable precursor, \textbf{I},
\dppestable\ is photolyzed using a laser flash, converting it into
the transitory species, \textbf{II}, \dppeunstable , which
immediately reacts with dissolved \pHH\ producing \dihydride,
\textbf{III} (Figure \ref{Figure-dihydride}), the molecule of
interest. The reaction takes place \textit{inside} a magnet and is
an example of instantaneous PASADENA.

The dihydride \textbf{III} is a well-characterized system
\cite{SchottPHRu} and no secondary isomers are known to be produced
during or after the photolysis. The reverse reaction,
\textbf{III}$\,\rightarrow\,$\textbf{I} is extremely slow at the
operating temperature of $295$~K. (The backward reaction is
catalyzed by higher temperatures, above $340$~K, and by the presence
of \C\O .)

The precursor \textbf{I}, was dissolved in deuterated benzene
(\C_6\D_6) and placed in a $5$~mm (diameter) NMR tube; from which
dissolved gases were subsequently removed using freeze-pump-thaw
cycles, corresponding to, respectively, freezing with liquid \N_2,
degassing and finally melting the sample. The tube was covered in
\Al\ foil to protect against premature photolysis by ambient light.
\textit{Para}-hydrogen prepared at $20$~K was introduced into the
tube, at a pressure of about $3$~bar. Warming and subsequent shaking
ensured that the \textit{p}-\H_2 gas dissolved in the solvent. All
NMR studies were carried out with samples of approximately $1$~mM
concentration on a Bruker DMX-400 spectrometer with ^1\H\ at
$400.1$~and ^{31}\P\ at $161.9$~MHz, respectively. NMR properties of
the system and several other experimental parameters are summarized
in Table \ref{Table-system-method-parameters}.

\begin{table}
\begin{center}
\begin{tabular}{ll}
\hline \multicolumn{2}{c}{System NMR parameters} \\
\multicolumn{2}{c}{\dihydride}\\
\hline \\
resonance frequencies & $-6.32$ and $-7.55$~ppm \\
$\delta$ & $492$~Hz (on the $400$~MHz spectrometer)\\
^1\H\ $T_1$ (single quantum) & $1.7$~s \\
^1\H\ $T_2$ (single quantum) & $0.58$~s \\
$^2J_{\H\H}$ & $4.6$~Hz\\
&\\
  \hline
\multicolumn{2}{c}{Experimental parameters} \\
\hline \\
Proton frequency & $400$~MHz\\
System temperature ($T$) & $295$~K \\
$t_{\pi/2}$ & $8.25~\mu$s \\
Number of laser flashes ($F$) for control $^\star$ & $1000$ \\
Inter-flash delay ($t_f$) $^\star$ & $7$~s \\
Number of scans ($S$) $^\star$ & 3072 \\
Inter-scan delay ($t_s$) $^\star$ & $20$~s \\
Active volume fraction ($V_f$) $^\star$ & $12.5/34=0.368$ \\
Spectral width & $\approx 30$~ppm \\
Receiver gain & $128$ \\
 \hline
\end{tabular}
\end{center}\caption{Important system and experimental parameters
involved in the initialization experiment. Parameters marked with a
``$\star$'' are described in the
text.\label{Table-system-method-parameters}}
\end{table}

\subsection{Optical assembly}\label{Section-optical-arrangement}


\begin{figure}
\begin{center}
\includegraphics[scale=0.8]{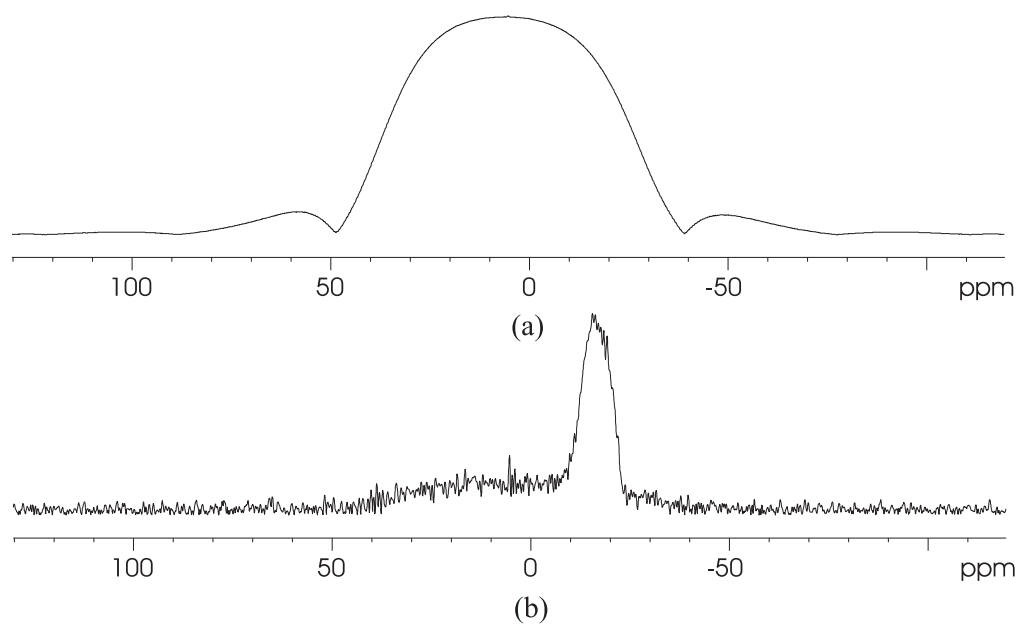}
\caption{One dimensional image of the (a) NMR tube filled with water
and (b) of the dihydride after the laser flash; showing that it is
indeed formed within the coil region. The ``hump'' in the active
region is mainly due to non-deuterated protons in the benzene
solvent. \label{Figure-profile}}
\end{center}
\end{figure}

The spectrometer was fitted with a special probe equipped for
\textit{in situ} photolysis \cite{GodardPHPhotolysis}, also designed
by colleagues at York.
For the single flash experiment, an MSX-250 pulsed \Xe\Cl\ excimer
laser radiating $308$~nm UV light was employed. The flash lasted for
$12$~ns and the flash-to-flash variability was known to be less than
$1$~ns. The specification sheet of the laser quotes a beamwidth of
$12\times 6~$mm and a beam divergence of $2\times 2~$mrad; and the
power per pulse was measured to be $32~$mJ. The single pulse of UV
light was then directed into the probe of the spectrometer using a
carefully aligned assembly of quartz prisms, an arrangement
resembling a periscope. The UV light was directed to fall in the
active region of the NMR tube, lying directly below the RF coils;
this is made possible by precise engineering of the probe head. The
laser is triggered by a control signal, initiated from within the
pulse programme. There is also a small ($\approx 10~\mu$s)
trigger-flash delay and the tomography pulse or filtration sequence
can ``wait'' to take into account this small time interval. The
dihydride is formed within the coil region because of the optical
positioning; this is also confirmed by a one dimensional image
\cite{Claridge} of the dihydride taken after the laser flash,
showing the relative positions of the hydride and the active region
and is shown in Figure \ref{Figure-profile}. The experiment was
performed in a darkened room, reducing the possibility of background
photodissociation of the compound, ensuring that all dihydride
formation is initiated by the UV laser.

For extended durations of irradiation, a Kimmon IK3202R-D $325~$nm
He-Cd $27~$mW CW UV laser was used; the optical assembly was left
unchanged. To enable a pre-determined amount of laser light to reach
the sample, the CW laser was employed in conjunction with a Vincent
Associates Uniblitz VMM-D1 shutter, which was opened and closed from
within the logic of the pulse programme. The response time of the
shutter was $1.5~$ms; and since our CW experiments involved minimum
irradiation times of $60~$ms, errors due to shutter timing were
negligible.

\subsection{Dihydride spectra}\label{section-spectrums}

\begin{figure}
\begin{center}
\includegraphics[scale=]{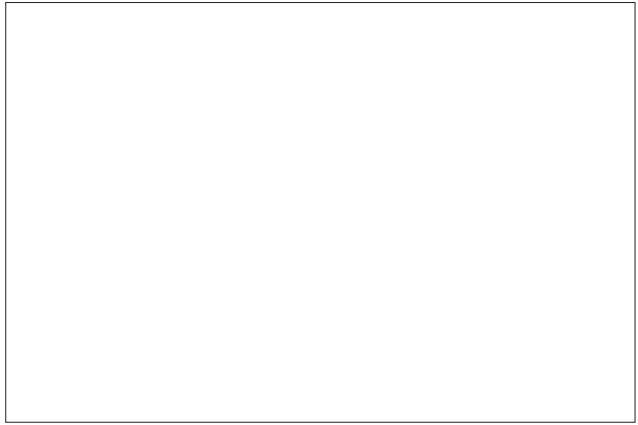}
\caption{NMR spectra from the single flash experiment. The spectrum
in (a) is the output from a $90_y$ hard pulse \textit{before} the
laser flash; (b) is the spectrum from the pulse sequence
\eqref{jump-return-final}, \ie\ without the filtration; (c) is
acquired after the sequence \eqref{initialization-pulse-sequence},
inclusive of the filtration; while in (d), the length of each
gradient is halved, exciting only the $I$ spin; and finally, (e) is
the spectrum obtained from the sequence \eqref{jump-return-final},
but without the initial $90_{45}$ hard
pulse.\label{Figure-coherent}}
\end{center}
\end{figure}

The NMR spectra for different pulse sequences, in conjunction with a
single flash of the pulsed laser, are shown in Figure
\ref{Figure-coherent}. The GARP broadband decoupling sequence
\cite{GARPShaka} was applied throughout signal acquisition to remove
couplings to ^{31}\P\ nuclei. For the initialization experiment, the
NMR tube must not contain any residual \dihydride\ \textit{before}
the flash is triggered. This is verified by applying a simple $90_y$
hard pulse prior to irradiation, no signal is seen, suggesting that
no dihydride exists before the flash or its concentration is too
minute to be detectable. The resulting spectrum, which is
essentially noise, is shown in part (a) of the figure. Figure
\ref{Figure-coherent}(b) is the result of performing instantaneous
PASADENA experiment depicted in Figure \ref{initialization}(b), a
selective Jump-and-Return sequence applied soon after the flash. The
spectrum shows the expected pair of anti-phase doublets, in
accordance with \eqref{signal-S0-90Iy}. Part (c) of the figure,
shows the spectrum resulting from the sequence
\eqref{initialization-pulse-sequence}, but this time also includes
the filtration sequence before the Jump-and-Return. The result is
very similar, and in fact apparently indistinguishable from (b),
suggesting that the \pHH\ addition is clean, and that error terms
mixing in with the singlet are very small. Modifying the filtration
sequence by reducing the duration of each gradient to
$1/(2\,\delta)$, results in only the $I$ spin being observed. Part
(d) of Figure \ref{Figure-coherent} shows exactly this,
authenticating the right multiplet as belonging to $I$. With the
identity of the spins now revealed, we can translate the spectral
pattern in Figure \ref{Figure-coherent}(b), (read leftwards as down,
up, up, down), into the signal vector $\{-,+,+,-\}$\footnote{I
assume `$+$' is upwards and `$-$' is downwards.}, where the leftmost
entries \textit{in the vector} correspond to spin $I$ and the
rightmost to $S$. For example, with this convention the spectrum (d)
translates to $\{+,-,0,0\}$. In short, the pair of lines
\textit{appearing} to the right in our displayed spectra belong to
spin $I$ and correspond to the two leftmost positions in the signal
vector. Finally, part (e) of the figure shows the spectrum obtained
from the pulse sequence $[t_d\thickspace 90_{-x}]$, dropping out the
first hard pulse from \eqref{jump-return-final} and the result is
apparently indistinguishable from (b). Hard pulses are irrelevant
for the singlet, indicating that the state prepared after the laser
flash is indeed close to being a perfect Werner singlet. Our
tomography results in Section \ref{section-tomography-results}, will
in fact, state the exact component-wise breakup of the quantum
state.

\begin{figure}
\begin{center}
\includegraphics[scale=]{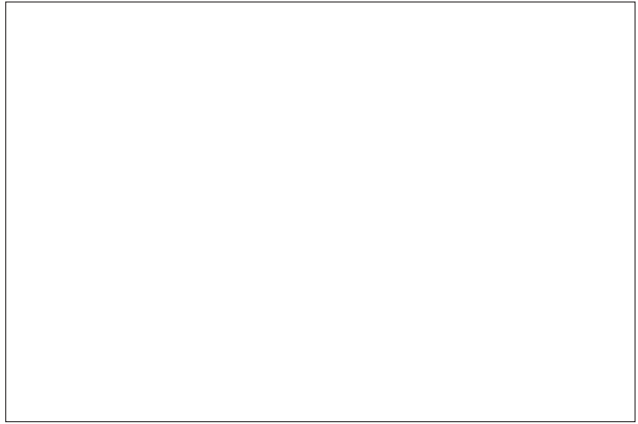}
\caption{Spectra from delayed PASADENA, in which the selective
detection pulse sequence is initiated with a variable delay $\tau$
after the laser trigger. The plotted spectra were obtained by
systematically varying $\tau$ under computer control: increasing in
steps of nearly $1/(4\delta)\approx 510~\mu$sec, from $3~\mu$s in
(a) to $2043~\mu$s in (e). Multiplet $I$ remains unchanged (ignoring
relaxation), while the components of the $S$ multiplet change phase
in steps of $\pi/2$; going from  (a) absorption $\{-,+,+,-\}$, to
(b) dispersion $\{-,+,i,-i\}$, (c) negative absorption $\{-,+,-,+\}$
to (d) negative dispersion $\{-,+,-i,i\}$ and back to (e) absorption
$\{-,+,+,-\}$ . This cyclic variation of the phase occurs at a
frequency $\delta$.\label{Figure-delayed}}
\end{center}
\end{figure}

Figure \ref{Figure-delayed} shows spectra from the delayed PASADENA
experiments (Figure \ref{pasadena}(b)), in which the selective
detection pulse is postponed for a variable delay $\tau$ after the
laser flash. The results are in complete agreement with
\eqref{signal-pasadena-delay-90Iy}, with the $I$ multiplet remaining
more or less unchanged (neglecting relaxation), and the phases of
the component lines in the $S$ multiplet varying sinusoidally, with
the expected frequency of $\delta$.

\subsection{Control experiment}\label{section-control-experiment}

The experimental spectra (Figure \ref{Figure-coherent}) are in
almost perfect agreement with the theoretical predictions. However,
we are interested in initializing the two qubit system in a pure
state, and to find the exact purity, it is essential to know not
only the form, but also the \textit{intensity} of the signal. As
absolute intensities are hard to come by in NMR, we need a reference
spectrum with a \textit{known} intensity. The thermal signal,
specified in \eqref{signal-thermal}, is a readily available
reference with an intensity deducible from first principles. Our
control experiment provides this intensity reference and thereby
calibrates the enhanced hyper-polarized signal strengths.

One straightforward means to implementing the control is allowing
the highly polarized dihydride to relax to the thermal equilibrium
state \eqref{rho-thermal} and applying a hard $90_y$ pulse, followed
by acquisition. Indeed this is a completely valid strategy, but the
problem is that the amount of dihydride, \dihydride , formed from a
\textit{single} laser flash is much too small and, after relaxation,
the thermal signal is too weak to be detected. For example, assume
we make a pure state in our initialization experiment, then if the
thermal signal from a single laser flash is to be detected over and
above the noise floor, the signal-to-noise ratio for the pure-state
spectrum should be of the order of $30000$: this is something
clearly not possible. Our spectra in Figure \ref{Figure-coherent}
show a signal-to-noise ratio of only about $40$.

\begin{figure}
\begin{center}
\includegraphics[scale=0.9]{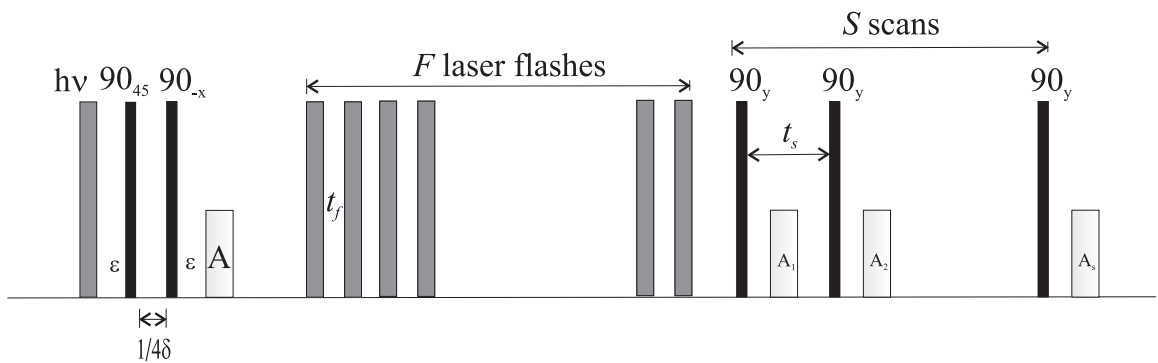}
\caption{Schematic outline of the calibration experiment. The first
laser flash, shown by a grey rectangle, produces \dihydride\ which
is detected and the signal acquired ($A$). Subsequently we apply $F$
laser flashes producing $F$ times as much dihydride. Adjacent
flashes are separated in time by $t_f$. The thermal signal was
measured by signal averaging $S$ times ($A_1$ through $A_s$), and
the averaged spectrum obtained after signal averaging, is in fact,
the \textit{sum} of the $S$ transients. Two successive scans are
separated by a delay of $t_s$.\label{Figure-control}}
\end{center}
\end{figure}

\begin{figure}
\begin{center}
\includegraphics[scale=]{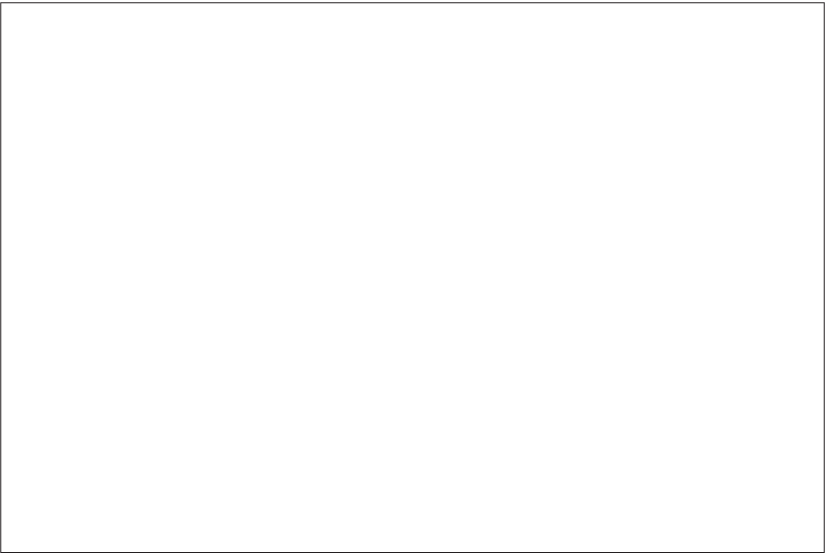}
\caption{The \pHH\ enhanced and calibration spectra of \dihydride.
The enhanced spectrum (a) is a single scan after a single laser
flash, and is identical to Figure \ref{Figure-coherent}(b), while
the calibration spectrum (b) is the sum of $3072$ scans after $1000$
laser flashes. The calibration spectrum has been divided by 3072000
and then multiplied by $30864$ (the theoretical maximum enhancement)
before plotting, so that the two spectra should show the same
intensity. In fact the \pHH\ enhanced spectrum (a) is even more
intense than na\"{\i}vely predicted.\label{Figure-calibration}}
\end{center}
\end{figure}

One way around this limitation is to increase the concentration of
the dihydride by applying more flashes. The initialization
experiment with $F$ flashes\footnote{The notation $F$ should not be
confused with the identical notation I have used for the singlet
fraction.} is schematically shown in Figure \ref{Figure-control}. In
practice, we applied $1000$ laser flashes in total, thereby
increasing the concentration of the dihydride a thousand-fold. Even
then, the signal-to-noise of the thermal spectrum was not
sufficient, and it was necessary to perform signal averaging
\cite{Claridge}. We performed $3072$ scans before a thermal spectrum
with a reasonable signal-to-noise ratio was obtained. The control
and enhanced spectra are shown side-by-side in Figure
\ref{Figure-calibration}. The resulting thermal signal is a sum of
$S$ scans and for meaningful comparison with the \pHH\ enhanced
spectrum from a single flash, must be divided by $S\times F$. We can
then compare the \pHH\ with the properly rescaled thermal spectrum
and determine the enhancement and hence the purity. This na\"{\i}ve
approach, however, yields an enhancement of about $77000$, well
above the theoretical maximum. The apparent discrepancy is addressed
and resolved by an active volume argument and is the subject of the
next subsection.

In the preceding discussion, we have assumed that $F$ flashes
produce $F$ times as much \dihydride\ than the amount produced by a
single shot, or in other words, the precursor \textbf{I} is not
\textit{depleted}. Our collaborators, in fact, performed a ^{31}\P\
quantitative experiment \cite{Claridge} to examine the extent of
depletion, which turns out to be very small. The underlying concept
of the quantitative experiment and results are outlined in Section
\ref{section-depletion}.

An inter-flash delay $t_f$ of $7~$s is sufficiently long to allow
\textit{different} precursor molecules, which are subject to
convective and diffusive motions, to cross the path of the laser
beam, and be irradiated by successive laser flashes. This ensures
that there is, more or less, a linear relationship between the
amount of \dihydride\ formed and the number of flashes
\cite{AnwarPHReview}. As a matter of principle, the selected $t_f$
should be large as compared to the characteristic timescales for
diffusion and convection. Furthermore, successive scans are
separated by about $20~$s, which is greater than $5T_1\approx 8.5~$s
and so saturation effects \cite{Claridge} can also be safely
ignored. With these values of $t_f$ and $t_s$, the total time
required for the experiment is about $19~$hours.

\subsection{Active volume
correction}\label{section-active-volume-correction}

Figure \ref{Figure-coherent} shows the spectrum from a single laser
flash alongside the calibration spectrum, which has been multiplied
by $30864/(S\times F)$ before plotting. For an almost pure state the
two spectra should be roughly the same size, but the \pHH\ enhanced
spectrum is significantly bigger than the thermal spectrum,
resulting in an apparent enhancement about $2.5$ times larger than
the maximum possible. This discrepancy is explained by considering
the fact that the NMR probe is not sensitive to the \textit{entire}
sample, but only to an ``active'' region, lying inside the RF coil
(refer to Figure \ref{Figure-profile}). Only the sample that lies
within the active volume can be detected, the rest remains
undetected, and therefore lies in what I call the ``passive''
region.

A single laser flash produces \dihydride\ lying wholly within the
active region and therefore, \textit{all} of it is detected and
contributes to the enhanced signal. This is not true for the
calibration signal. Applying the $1000$ laser flashes takes about
$2~$ hours during which the dihydride distributes itself throughout
the entire volume of the sample by diffusion and convection. As a
result, the calibration spectrum does not record \textit{all} of the
dihydride present in the NMR tube, but only registers the fraction
lying inside the active region; molecules in the passive region
yield no spectrum. As a result, the calibration signal is attenuated
by a factor $V_f$, the \textit{active volume fraction},
corresponding to the ratio of the active to the total volumes. The
true enhancement is therefore smaller than it appears, the
correction factor being $V_f$. As expected the directly measured
enhancement (that is, before correction) shows a linear dependence
on the total sample volume (data is shown in our submitted work
\cite{AnwarPHReview}), confirming our active volume hypothesis.

The active volume fraction can be estimated using simple geometrical
considerations. Considering a ratio of lengths as a good substitute
for a ratio of volumes, $V_f$ can be estimated from the length of
the RF coil and the depth of the NMR sample. In the experiment
described, these lengths were measured\footnote{The sample is long
because this improves the homogeneity of the magnetic field by
moving the susceptibility boundary away from the detection region.}
to be $12.5~$mm and $34~$mm, resulting in an approximate fraction
$V_f$ of $12.5/34$. This deduction of the $V_f$ fraction, however,
is prone to error, as it is based on an ``all or none'' assumption
of the probe sensitivity, that is, \textit{all} of the dihydride
inside the active region and \textit{none} in the passive region is
detectable. But as Figure \ref{Figure-profile}(a) shows, the
sensitivity drops off smoothly near the edges of the RF coil, rather
than in a sharp step-like manner. As a result, the sensitivity is
not uniform even within the active region, and our estimated $V_f$
remains, at best, only an estimate\footnote{My rough guesstimate for
the error in the active volume fraction would be $10$--$30$\%.} and
addressing this should be a priority in future experiments.

\subsection{Depletion of the precursor}\label{section-depletion}

The control experiment assumes that all flashes behave identically:
having identical photon fluxes, quantum yields and more importantly,
that the precursor (\textbf{I}) concentration is constant at all
times during the experiment. This section considers the effects of
depletion of the precursor molecules, ignoring any variability in
the light pulses.

\begin{figure}
\begin{center}
\includegraphics[scale=0.8]{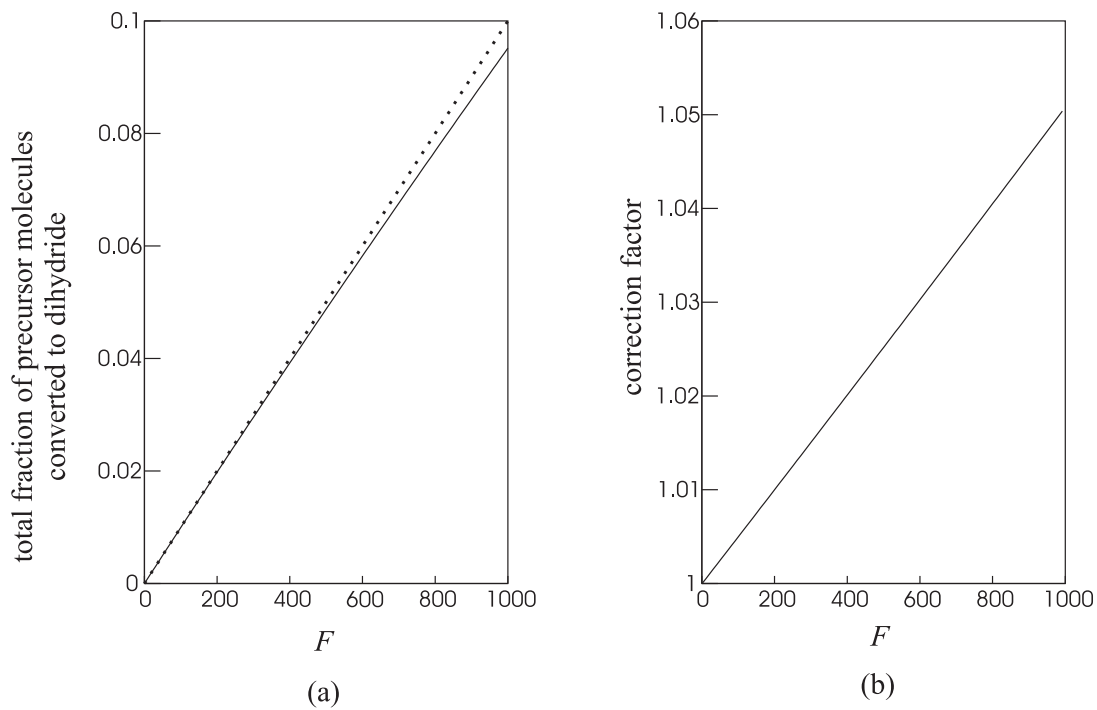}
\caption{Depletion of the precursor: in (a), the solid line shows
the overall fraction of molecules converted, $(1-(1-x)^F)$, as a
function of $F$. The dotted line is a plot of $Fx$, the total
fraction of molecules converted for a hypothetical pseudo-zero order
conversion and coincides with the depletion curve for small values
of $Fx$. Part (b) shows the depletion correction factor,
$Fx/(1-(1-x)^F)$, as a function of $F$. The fraction converted per
flash $x$ is assumed to be $0.01$\%.\label{Figure-deplete}}
\end{center}
\end{figure}

Under this assumption each flash delivers a \textit{fixed} number of
photons to the sample. Suppose this number is $R$. Each of the $R$
photons can potentially affect the conversion of a precursor
molecule to the dihydride, however the number of molecules converted
depends on the concentration of the precursor, $[C]$, existing just
\textit{before} the flash. A simple first order model for the
conversion can be written as,
\begin{equation}\label{depletion-rate}
\frac{d[C]}{dF}=\,-AR[C],
\end{equation}
where $A$ is a rate constant independent of the concentration. What
this constant signifies is that although the \textit{absolute
number} of photons converted will fall with each successive flash,
the \textit{fraction} of the molecules converted \textit{per flash}
remains constant. For example, the thousandth flash would convert a
smaller number of precursor molecules to the dihydride than the
first pulse, but in each case the fractional conversion would remain
the same.  Now suppose that each flash converts a constant $0.01$\%
of the precursor. If depletion was not an issue, then $1000$ flashes
would convert $1000\times 0.01$\%$=10$\% of the precursor, but
taking depletion into account, we predict an overall conversion of,
\begin{equation}
1-(1-0.0001)^{1000}
\end{equation}
which is $9.52$\%, smaller than the $10$\% as expected. Thus the no
depletion assumption \textit{underestimates} the thermal signal and
\textit{overestimates} the purity of the \pHH\ state by a factor of
$10/9.52=1.05$, which is our depletion correction factor. In the
hypothetical limit of an infinite precursor concentration, the rate
of conversion $d[C]/dt$ becomes more or less independent of $[C]$
and the conversion acts as a pseudo zeroth order reaction in which
the absolute number of molecules converted per flash
\textit{appears} to be constant. This is because the initial
concentration of the precursor is so high and the conversion per
flash so low, that $[C]$ does not substantially decrease on the
application of successive flashes. Mathematically, this implies that
all three parameters in the R.H.S. of \eqref{depletion-rate} act
like constants, resulting in a linear decrease in the precursor
concentration (or linear increase in the dihydride concentration).
These ideas are illustrated in Figure \ref{Figure-deplete}.

In fact, we carried out a simple experimental calculation to
determine the extent of precursor depletion. The two phosphorus
nuclei acquire different chemical shifts in \textbf{I} and
\textbf{III}, \cite{SchottPHRu}, resulting in distinct ^{31}\P\
signals for these compounds. This fact helps in performing a
quantitative experiment\footnote{The experiment involves direct
^{31}\P\ detection; uses $10000$ scans (the large number of scans
are needed because of the low concentrations of the precursor), an
inter-scan delay of $20~$s and proton-decoupling turned on only
during acquisition, preventing build-up of artificial intensities
due to the Nuclear Overhauser Effect during inter-scan delays.},
determing the relative concentrations of the two species after $F$
laser flashes, and therefore, the extent of conversion of \textbf{I}
to \textbf{III}. One such experiment (see footnote) showed that
after $1000$ flashes, about $4.46$\% of \textbf{I} was converted
into \textbf{III}. Suppose that after one flash, a constant fraction
$x$ of the precursor is converted, then after $1000$ flashes, the
fraction converted is $\bigl(1-(1-x)^{1000}\bigr)$, and equating
this with the experimentally determined fraction of $0.0446$, yields
a conversion per flash $x$ of $4.56\times 10^{-3}$\%. Therefore, to
correct for the effect of depletion, the \pHH\ enhancement must be
divided by,
\begin{equation}\label{depletion-factor}
\frac{1000\,x}{\bigl(1-(1-x)^{1000}\bigr)}\approx\,1.023.
\end{equation}

\section{Data processing}\label{section-tomography-results}

An estimate of
the purity of the two qubit (Werner singlet) state is, 
\begin{equation}\label{enhancement-first-estimate}
\frac{P}{T\times 1.023}\times S\times F\times V_f\times
\frac{1}{30864},
\end{equation}
where $P$ is \textit{some kind} of intensity of the enhanced \pHH\
signal and $T$ is \textit{some measure} of intensity of the thermal
signal acquired from the multi-scan, multi-flash control experiment.
This section considers the data processing steps we employed in
interpreting the spectra and determining the appropriate values of
$P$ and $T$ for the purity calculation and state tomography.

\subsection{Peak integration}\label{section-integration}

The enhanced and control spectra were processed
\cite{DataProcessingHochStern,LindonFerridgeDataProcess} using
home-written software\footnote{The data processing was done on a Sun
Blade 1000 machine running Solaris OS. Spectra were acquired on a
Bruker spectrometer at the University of York, and the data was then
converted and adapted for use on the Blade machine. I wrote the
programmes and scripts for data conversion, manipulation, $J$
matching, doubling and integration in C and in AWK \cite{AWK}. Other
processing tasks such as zero-filling, Fourier and inverse Fourier
transformations and plotting/display were performed using software
written by my supervisor. Finally, offset and baseline correction
were done with Mathematica and the Grace \cite{Grace} data
visualization package.} and analyzed by integration: measuring areas
between the absorptive Lorentzian peaks and a flat, zero mean
baseline. The integrals are a more reliable measure of signal
intensity than line heights due to two reasons. \textit{First}, line
heights critically depend on the \textit{frequency resolution}
\cite{DataProcessingHochStern,LindonFerridgeDataProcess} of the
spectrum, which is quite poor for our raw spectra; poorly digitized
spectra often resulting in underestimates of the height.
\textit{Second}, the total integral under a band-unlimited frequency
domain spectrum is equal to the first point in the time domain FID
(besides a normalization factor) \cite{FourierBracewell}, and is
therefore independent of RF inhomogeneities and $T_2$ relaxation
effects.

\begin{figure}
\begin{center}
\includegraphics[scale=0.7]{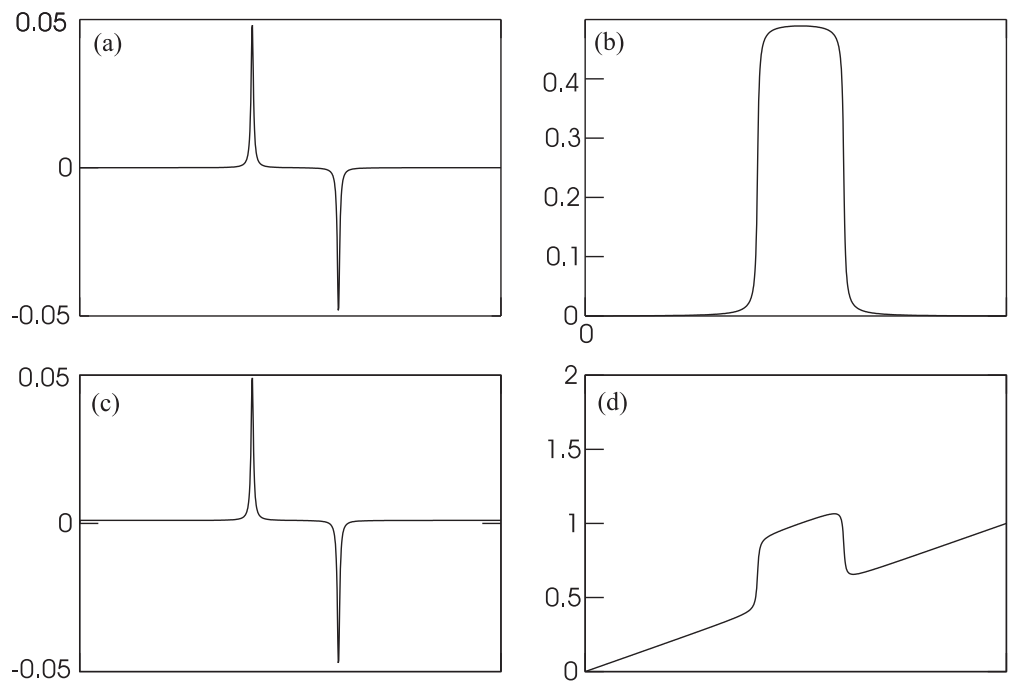}
\caption{Effect of a baseline zero-offset on the peak integrals. The
spectrum in (a) comprises an ideal anti-phase peak (maximum and
minimum are between $\approx\pm 0.05$) and (b) is the corresponding
integral (obtained by summing up all previous points, starting from
the left edge of the spectrum). The integral is uniformly zero in
the baseline region, reaches a maximum $\approx 0.5$ in the region
between the peaks and drops back to zero after the second peak. In
(c), the spectrum (a) is raised upwards by a small amount ($\approx
0.001$), and (d) shows the corresponding integral, indicating that
now, even the baseline region has a positive integral which
additively interferes with the peak integral. In this case, the peak
integrals will be an over-estimate and baseline correction becomes
necessary.\label{Figure-baseline-integral}}
\end{center}
\end{figure}

However, prior to integration, it is important to deal with some
spectral artifacts. A baseline offset or a sloping baseline
introduces errors; the effect is illustrated in Figure
\ref{Figure-baseline-integral}. I employed second order curve
fitting (baseline correction) prior to integration. Other important
factors to consider are the need for properly phased, absorptive
mode spectra and the number and extent of data points
\cite{DataProcessingHochStern}, over which the peaks are integrated.
The integrals also fluctuate due to noise and standard errors on the
integral values were measured by calculating the variance over
regions of baseline noise of identical spectral widths.

\subsection{$J$ processing}\label{section-J-processing}

In the \pHH\ experiment, the integrals of the anti-phase multiplet
partially cancel due to overlap. Any peak integrals we directly
measure, even after removing all the artifacts discussed above, will
therefore be an underestimate. There is however, one way of reducing
the overlap by ``dragging'' apart the peaks; the method is called
$J$\textit{-doubling} \cite{McIntyreJDbl,JMatchDblJones}. In
summary, the data processing steps involved in accurate peak
integration comprised the following steps. (I have described the
reduction in peak integrals due to overlap and $J$-doubling in
considerable detail in Appendix \ref{app-J-processing} and here I
give only the major results.)

\begin{figure}
\begin{center}
\includegraphics[scale=0.9]{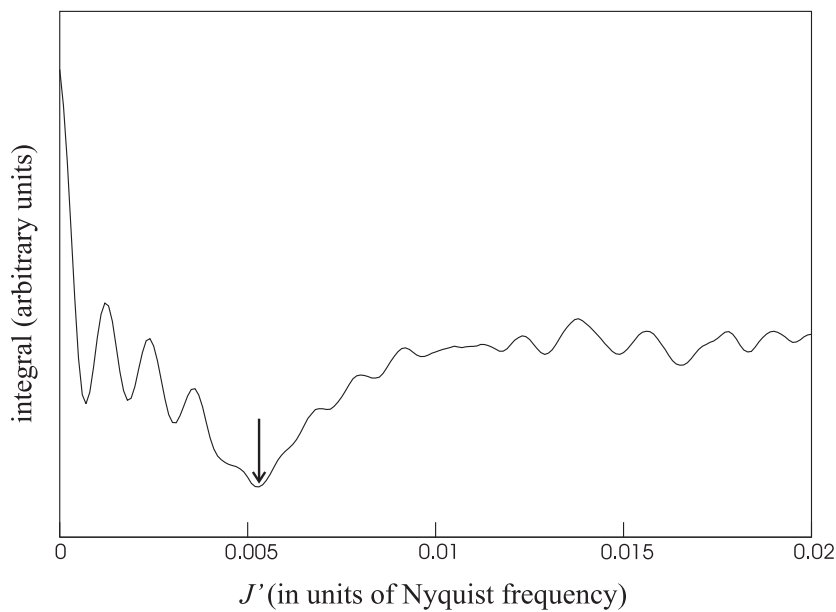}
\caption{Results of the computer programme searching for the global
minimum in the integrals of absolute $J'$ modulated spectra
(Appendix \ref{app-J-processing}). The trial values, $J'$ are given
in units of the Nyquist frequency. The global minimum of the
integral plot is indicated by an arrow and corresponds to the
matched $J'=J=0.0054$ value.\label{Figure-jmatchleft}}
\end{center}
\end{figure}

\begin{figure}
\begin{center}
\includegraphics[scale=0.9]{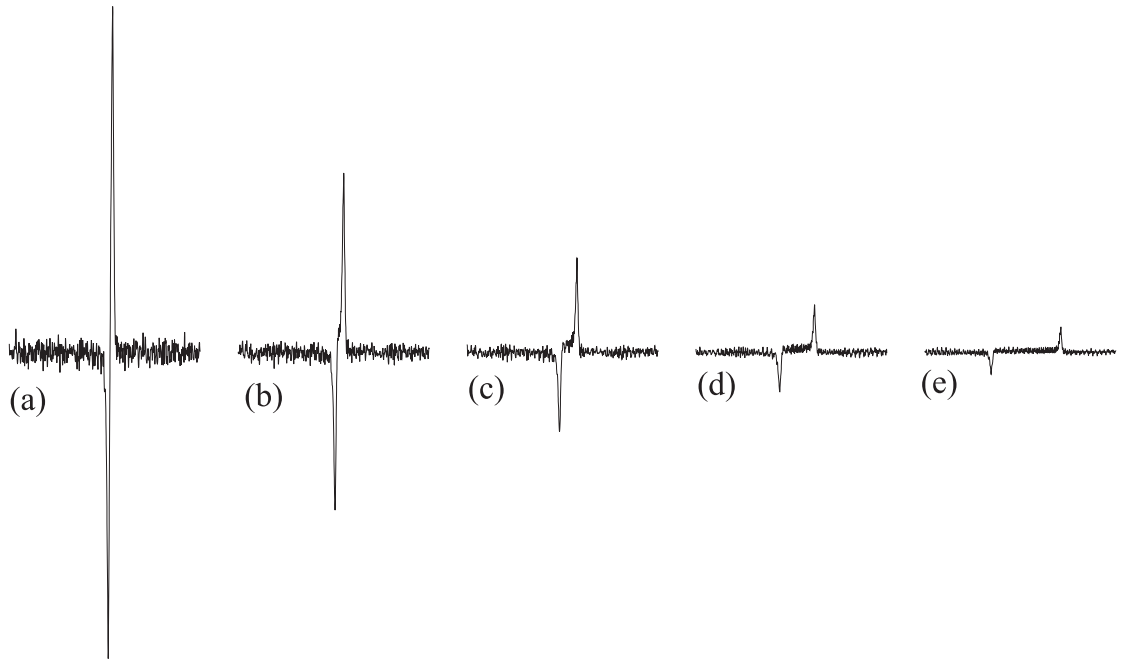}
\caption{Effect of $J$-doubling. Only the left ($S$) peak is shown.
Part (a) shows the original spectrum ($m=0$), and (b)--(e) show the
successively doubled spectra ($m=1,2,3,4$). The spectra are drawn to
the same vertical scale and the signal heights are seen to drop by
roughly half at each stage.\label{Figure-jleft}}
\end{center}
\end{figure}

\begin{figure}
\begin{center}
\includegraphics[scale=0.85]{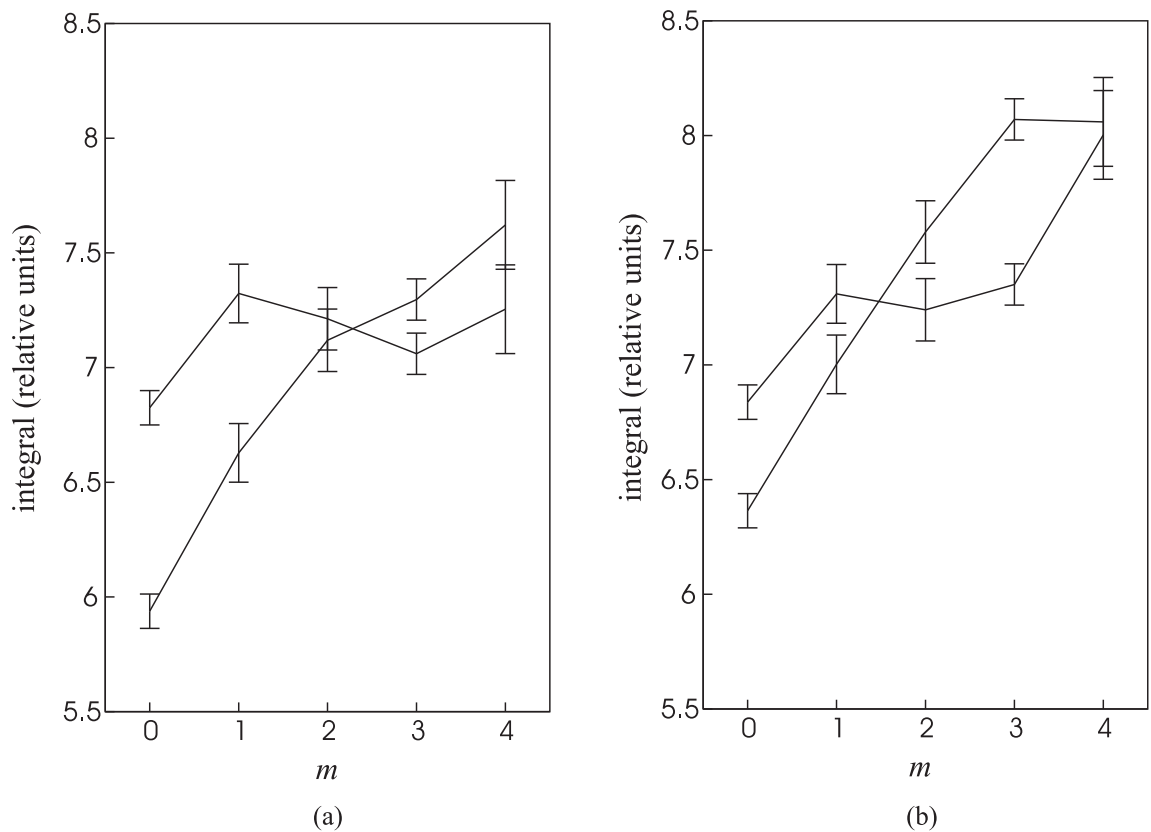}
\caption{Peak integrals as a function of the number of $J$-doubling
steps, $m$. The plot (a) shows the integrals for the $S$ spin and
(b) shows the integrals for the $I$ spin. The integrals, in general,
increase with increasing $m$ and for $m=4$, the peak integrals
within a doublet become statistically
indistinguishable.\label{Figure-integral}}
\end{center}
\end{figure}

\begin{enumerate}
\item We excised a small section from the overall spectrum that contained the dihydride
peaks.
\item The excised spectrum was inverse-Fourier transformed to yield a
time domain signal and zero-filled
\cite{DataProcessingHochStern,LindonFerridgeDataProcess} to increase
the apparent frequency resolution.
\item We implemented a computer algorithm scanning trial $J'$ values, looking for a global
minimum in the integrals of absolute $J'$-modulated spectra. This
step results in an estimate for the true splitting, $J$. A plot of
integrals of the $J'$-modulated spectra is shown in Figure
\ref{Figure-jmatchleft}.
\item The matched $J$ value is then used to double the original
spectrum. We double four times, substantially reducing the effects
of overlap but also degrading the signal-to-noise ratio. The
successively doubled spectra are shown in Figure \ref{Figure-jleft}.
\item Peak integrals are determined from the doubled spectra and
rescaled, taking into account the number of doublings employed.
These rescaled integrals are given in Table
\ref{Table-integrals-doubled}. As expected, after doubling a
sufficient number of times, the overlap disappears and the integrals
no longer destructively interfere. As a result the measured peak
integrals increase and the effect is shown in Figure
\ref{Figure-integral}. Another important consequence of the doubling
is to reduce the artificial imbalance between the peaks within a
doublet. We take the average between the statistically
indistinguishable peaks from the $m=4$ spectra, resulting in
accurate $I$ and $S$ peak integrals, which in this experiment are
$8.0310\pm 0.1936$ and $7.4328\pm 0.1936$ (calculated from the
entries in the last row in Table \ref{Table-integrals-doubled}).
\end{enumerate}

\begin{table}
\begin{center}
\begin{tabular}{cccccc}
\hline
$m$ & $S_l$ & $S_r$ & $I_l$ & $I_r$ & SD \\
\hline \\
$0$ & $5.9381$ & $6.8250$ & $6.3643$ & $6.8375$ & $0.075$ \\
$1$ & $6.6281$ & $7.3230$ & $7.0029$ & $7.3095$ & $0.128$ \\
$2$ & $7.1190$ & $7.2129$ & $7.5793$ & $7.2402$ & $0.136$ \\
$3$ & $7.2967$ & $7.0603$ & $8.0698$ & $7.3507$ & $0.090$ \\
$4$ & $7.6220$ & $7.2544$ & $8.0594$ & $8.0027$ & $0.194$ \\
  \hline
\end{tabular}
\end{center}\caption{Integrals calculated from the spectra, where $m$ is the number of
$J$-doublings employed. The subscripts ``l'' and ``r'' refer to the
left and right lines in the $I$ and $S$ spin peaks.
\label{Table-integrals-doubled}}
\end{table}

\subsection{State tomography
results}\label{section-subsection-tomography}

\begin{figure}
\begin{center}
\includegraphics[scale=0.7]{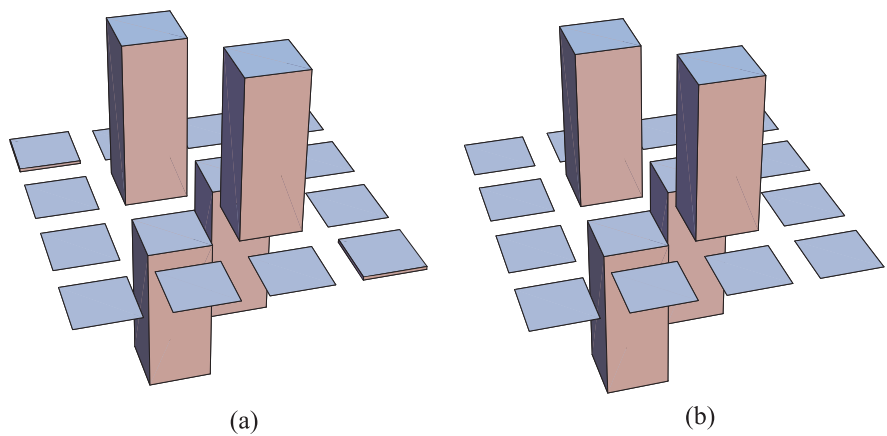}
\caption{Results from quantum state tomography, showing the density
matrices from (a) the state generated by the pulsed laser, compared
with (b) a perfect singlet. Despite small non-zero amounts of
$T_{\pm 1}$ in (a), the overall density matrices are very similar.
The elements in the bar chart represent magnitudes of the elements
in the density matrix written in the conventional
$\{\ket{\alpha\beta},\ket{\alpha\beta},\ket{\beta\alpha},\ket{\beta\beta}\}$
order. For example, the top left element in the graph represents the
magnitude of the element
$\expectation{\alpha\alpha}{\rho}{\alpha\alpha}$ and so on. All
elements in the matrix are real.\label{Figure-tomography}}
\end{center}
\end{figure}

The $I$ and $S$ integrals of the processed spectra are then
normalized by multiplying by the factor $(S\times F\times
V_f)/(T\times 1.023\times 30864)$, where $T$ is the ``raw'' average
integral under each of the four peaks, determined after baseline
correction but without any $J$-processing. In our case, we measured
a $T$ of $303.12\pm 4.44$. The other experimental parameters were:
$S=3072$, $F=1000$ and $V_f=12.5/34=0.368$. This calculation gives
us the $p$ and $q$ coefficients in the state representation
\eqref{filter-output-PO}, therefore achieving state tomography.
However, we prefer to express the state in the form,
\eqref{filter-output-symmetrical} and use the equations
\eqref{filter-output-relation-3}--\eqref{filter-output-relation-4}
to determine the singlet and triplet fractions ($a,b,c$) from $p$
and $q$. For example, in the particular experiment I am discussing,
the $I$ spin normalized integral corresponds to $q=-8.0310\pm
0.1936$ and the $S$ spin integral corresponds to $p=-7.4328\pm
0.1936$. For a pure $S_0$, we expect $p=q=-1$, however in our case,
the multiplets are slightly imbalanced, resulting in unequal
fractions of the triplets. Using
\eqref{filter-output-relation-3}--\eqref{filter-output-relation-4},
the exact state characterization is, 
\begin{eqnarray}
a=&&\,0.9371\label{fraction-S0},\\
b=&&\,0.0448\label{fraction-T0},\quad\text{and}\\
c=&&\,0.00905\label{fraction-T1},
\end{eqnarray}
not identical but very similar to a Werner singlet with balanced
triplets. The resulting density matrix is portrayed in Figure
\ref{Figure-tomography} and has a concurrence and entanglement of
formation of $0.874$ and $0.822$ respectively.

If we \textit{assume} that our quantum state is a Werner singlet,
$(1-\varepsilon)\mathbf{1}_4/4+\varepsilon S_0$, with a singlet
fraction equal to the $a$ calculated above, then the polarization of
the state is,
\begin{equation}
\varepsilon=\,0.916\pm 0.2,
\end{equation}
which I call the \textit{effective purity} and is consistent with a
spin temperature of $6.4~$mK or an effective magnetic field of
$0.45~$MT at room temperature.

\section{CW experiments}\label{section-CW}

\begin{figure}
\begin{center}
\includegraphics[scale=]{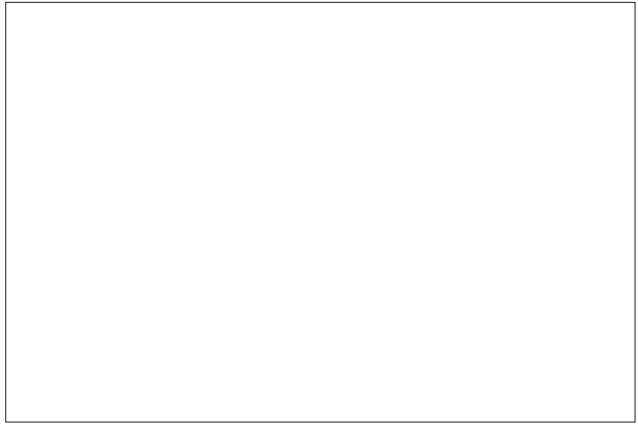}
\caption{Spectra from the isotropic PASADENA experiment, employing
MLEV-16 isotropic mixing. The spectrum in (a) is acquired after
$60~$ms of laser irradiation and (b) shows the spectrum obtained
after $300~$ms of irradiation. Na\"{\i}vely one would expect (b) to
be five times bigger as compared to (a), however the signal
intensities are smaller than expected. For comparison, (c) is a plot
of the spectrum (a) multiplied by five.\label{Figure-cw}}
\end{center}
\end{figure}

\begin{figure}
\begin{center}
\includegraphics[scale=0.7]{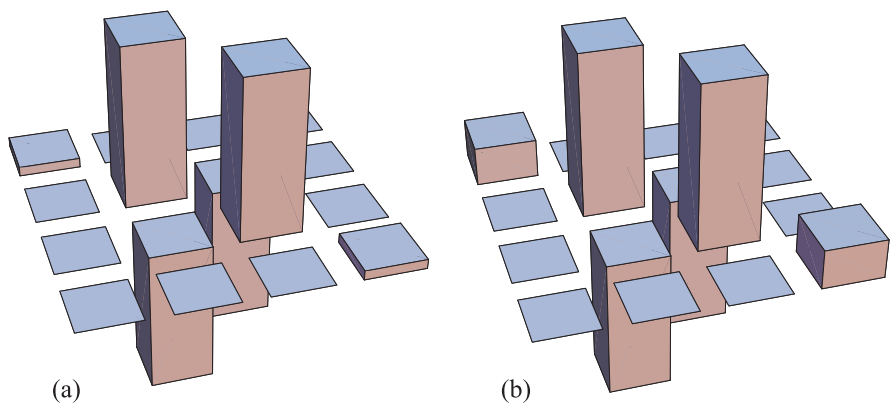}
\caption{Tomography results from the CW data, showing the density
matrices after (a) $60~$ms and (b) $300~$ms irradiation. Build-up of
significant $T_{\pm 1}$ error terms is evident from (b).
Interpretation and labelling of this figure are identical to those
in Figure \ref{Figure-tomography}.\label{Figure-tomography-cw}}
\end{center}
\end{figure}

This section discusses our isotropic PASADENA experiment utilizing
the CW laser. The spectra from extended hydrogenation periods of
$60~$ms and $300~$ms, in the presence of the MLEV-16 isotropic
mixing sequence are shown in Figure \ref{Figure-cw}. These spectra
show that the doubly anti-phase character of the spectrum is
maintained, indicating that the mixing sequence is very good at
preserving the form of the singlet. However, there is one problem:
na\"{\i}vely one would expect the $300~$ms spectrum to be five times
bigger than the $60~$ms spectrum, but it is not quite so large. We
attribute this discrepancy to the relaxation of the highly polarized
state, which can cause pronounced signal loss for extended durations
of hydrogenation. A rough calculation gives a value of about $0.6~$s
for the relaxation constant. It is important to note that this value
does not correspond to the single quantum $T_2$ relaxation normally
quoted in NMR, but rather it corresponds to a complicated decay of
an isotropic state towards the maximally mixed state in the presence
of a mixing sequence; this is sometimes described as ``spin-lattice
relaxation in the rotating frame'', $T_{1}\rho$
\cite{DictionaryNMR}. Quantum state tomography results are outlined
in Table \ref{Table-tomography-cw} and the density matrices for the
CW experiment are depicted in Figure \ref{Figure-tomography-cw}.

\begin{table}
\begin{center}
\begin{tabular}{ccc}
\hline
& $60~$ms & $300~$ms \\
\hline \\
$a$ & $0.9159$ & $0.8191$ \\
$b$ & $0.0565$ & $0.0687$ \\
$c$ & $0.0138$ & $0.05615$ \\
$\varepsilon$ & $0.8879$ & $0.7588$ \\
  \hline
\end{tabular}
\end{center}\caption{Tomography results from isotropic PASADENA
experiments, involving CW irradiation for $60$ and $300~$ms;
$\varepsilon$ is the effective purity. \label{Table-tomography-cw}}
\end{table}

\chapter{Implementation of quantum computation with the pure state quantum
computer}\label{QCimplement}

Chapter \ref{pure}\ described the experimental preparation of a two
qubit pure quantum state---the qubits are the two ^1\H\ nuclei in
the organometallic compound \dihydride , and the nuclei inherit the
nuclear spin singlet state, derived from pure \pHH. At this
juncture, the natural question to ask is, having \textit{prepared} a
pure state, can we go one step further and actually \textit{use}
this as an initial state for a quantum information processing task,
such as for implementing a quantum algorithm? The answer is yes and
this chapter describes our experiments achieving this.

The extremely high purity of the system, quite naturally, obviates
the need for assembling pseudopure states, a process requiring
non-unitary operations and generally lengthy or complicated pulse
sequences. Our experiments \cite{AnwarDeutsch,AnwarGrover}
constitute the first implementation of quantum algorithms in liquid
state NMR using pure initial states. As a result, they do not
prepare pseudopure states, instead they initialize the system
directly in an essentially pure quantum state (also lying above the
entanglement threshold). In related work, H\"{u}bler \etal\
implemented \cite{Hubler} the Deutsch algorithm on a \pHH\ derived
two qubit system, although with much smaller purities,
$\varepsilon\approx 0.1$. Our implementations extend their results,
to almost perfectly pure quantum states, with $\varepsilon$ very
close to $1$. Furthermore, in addition to achieving high purity and
the possibility for distilling entanglement
\cite{BennetMixedEntPRA}, our experiments are unique in the sense
that they prepare pure states \textit{on demand}; our laser flash
acts as a convenient switch, enabling us to generate the initial
state, as and when desired.

In this chapter, I describe the implementation of two quantum
algorithms using pure initial states. They are solving Deutsch's
problem of distinguishing between constant and balanced functions
and performing Grover's search on an unstructured database of four
quantum states. For the two algorithms, we employed two different
molecular systems: for Deutsch's algorithm, we used \dihydride
---a chemical system with which we are quite familiar by now---
and for Grover's search, we used its closely related arsenic
analogue, \ashydride , where dpae is 1,2-bis(diphenylarsino)ethane.
The important NMR parameters of the latter compound, whose structure
is also shown in Figure \ref{dpae-mixed}, are summarized in Table
\ref{Table-system-dpae}.

\begin{table}
\begin{center}
\begin{tabular}{ll}
\hline \multicolumn{2}{c}{System NMR parameters} \\
\multicolumn{2}{c}{\ashydride}\\
\hline
resonance frequencies & $-7.61$ and $-7.22$~ppm \\
$\delta$ & $160$~Hz (on the $400$~MHz spectrometer)\\
^1\H\ $T_1$ (single quantum) & $1.9~$s \\
^1\H\ $T_2$ (single quantum) & $0.67$~s \\
$^2J_{\H\H}$ & $4.8$~Hz\\
&\\
  \hline
\end{tabular}
\end{center}
\caption{Important NMR parameters of \ashydride\ used in the
implementation of Grover's search.\label{Table-system-dpae}}
\end{table}

Comparing the NMR parameters for \ashydride\ with \dihydride\ (see
Table \ref{Table-system-method-parameters}), two results are
obvious: \textit{first}, the \ashydride\ has a slightly longer $T_2$
relaxation time, resulting in better decoherence properties and
\textit{second}, the frequency separation is about three times
smaller in the ``dpae'' than in the ``dppe'' and this puts more
stringent requirements on spin selective pulse sequences in the
``dpae'' system. These properties are important in evaluating the
suitability of a molecular system for NMR quantum computing, and are
captured by the ratios $(\delta/J)$ and $(J/(1/T_2))$: a system is
suited for quantum information processing, if the condition
$\delta\gg J\gg 1/T_2$ is satisfied. These characteristic ratios for
the two molecules are compared in Table
\ref{Table-dppe-dpae-compare}, indicating that the ``dpae'' has
slightly better decoherence and worse frequency selectivity
properties.

\begin{table}
\begin{center}
\begin{tabular}{ccc}
\hline
Parameter & ``dppe'' & ``dpae'' \\
\hline
$\delta/J$ & $106.96$ & $33.33$ \\
$J/(1/T_2)$ & $2.67$ & $3.22$ \\
\hline \\
\end{tabular}
\caption{Frequency separation, splitting and decoherence parameters
for the ``dppe'' and ``dape''
compounds.\label{Table-dppe-dpae-compare}}
\end{center}
\end{table}

However, there is one clear cut advantage of using the \ashydride\
system: the initial state can be prepared with a purity
indistinguishable from one; our measurements indicate a purity of
$\epsilon\approx 1.06\pm 0.2$. Thus, we can describe our initial
states as being pure, without further qualification.

This chapter is organized as follows. Section \ref{section-Deutsch}
describes the Deutsch algorithm and Section \ref{section-Grover}
takes up Grover's search with the pure state quantum computer.
Experimental imperfections are visible in our spectra, and these can
be roughly explained using a simple model based on representing
quantum operations in terms of a family of Kraus operators. Appendix
\ref{app-operator-sum-decoherence} exemplifies the use of this
model, in context of the Deutsch spectra. The following two
sections, more or less, reproduce our published and submitted works,
\cite{AnwarDeutsch} and \cite{AnwarGrover}, and so include brief
discussions of many topics I have already discussed in previous
chapters.

\section{Deutsch's algorithm with pure quantum states}\label{section-Deutsch}


A published version of this Section can be found on the e-print
server: \newline http://www.arxiv.org/quant-ph/0406044.

\section{Grover's quantum search with pure quantum states} \label{section-Grover}

A published version of this Section can be found on the e-print
server: \newline http://www.arxiv.org/quant-ph/0407091.

\chapter{Practical implementation of twirl operations}\label{twirl}

Entanglement is an important resource in the quantum setting and has
been introduced in Chapter \ref{critique}. Communication protocols,
such as quantum teleportation
\cite{TeleportBennettOriginal,TeleportBeowmeesterExp}, work
optimally, when they use maximally entangled pure states. For two
qubits, these states are well known and are simply the four Bell
states, given as $\ket{\phi^{\pm}}=(\ket{00}\pm\ket{11})/\sqrt{2}$
and $\ket{\psi^{\pm}}=(\ket{01}\pm\ket{10})/\sqrt{2}$. In the
standard teleportation experiment, Alice and Bob share a pair of
maximally entangled pure states and the pre-existing entanglement,
in fact, constitutes a noiseless quantum channel. In real life
situations however, the channels are never perfect. The pure states
are subject to decoherence and degrade into mixed states, entailing
a decrease in both the purity and the entanglement. Entanglement
concentration or purification protocols
\cite{BennetPurificationPRL,Deutsch96} then pick up these
degenerated states and recover both the lost purity and
entanglement, accompanying a decrease in the ensemble size
\cite{PetersEntConcentration}. The first step in most of these
protocols is the ``twirl'' operation which converts an arbitrary
mixed state $\rho '$, not maximally mixed, into the Werner state
$\rho_\varepsilon$, \eqref{Werner-def}.
A useful fidelity measure \cite{BennetPurificationPRL} for such
states is their most significant overlap with the maximally
entangled state, 
\begin{equation}\label{fidelity-werner}
F(\rho)=\,\text{max}\expectation{\psi}{\rho}{\psi},
\end{equation}
where the maximum is taken over all four Bell states. It is well
known that entanglement can be distilled from the mixed $\rho$
whenever the state fidelity $F>1/2$ \cite{BennetPurificationPRL}.
Without loss of generality, we can also assume that the dominant
term in $\rho$ is the singlet state $\ket{\psi^-}$; the fidelity $F$
then becomes the overlap of the state with the singlet, which I have
referred to as the singlet fraction, \eqref{singlet-fraction-def}.
An equivalent parametrization of $\rho_\varepsilon$ can thus be made
in terms of $F$,
\begin{equation}\label{werner-F}
\begin{split}
\rho_{W}=&\,F\ket{\psi^-}\bra{\psi^-}\\
&+\frac{1-F}{3}(\ket{\psi^+}\bra{\psi^+}+\ket{\phi^+}\bra{\phi^+}+\ket{\phi^-}\bra{\phi^-}),
\end{split}
\end{equation}
the correspondence between the polarization $\varepsilon$ of the
Werner singlet and its singlet fraction being given as 
\begin{equation}\label{sf-polarization}
F(\rho_\varepsilon)=\,\frac{1+3\varepsilon}{4}.
\end{equation}
It can be checked that for the maximally mixed state $\varepsilon=0$
and $F=1/4$; whereas for the pure singlet $\varepsilon=1$ and $F=1$.
\par
Consider a mixed state $\rho '$ which satisfies two properties:
\textit{first}, it is not maximally mixed and \textit{second}, it
has a non-zero singlet fraction $F$. The twirl operation would take
such a state to $\rho_\varepsilon$, while conserving the singlet
fraction,
\begin{align}
&\rho ' \longrightarrow \rho_\varepsilon,\quad\text{such that}\\
&F(\rho_\varepsilon)=\,F(\rho ').
\end{align}
The restraints on $\rho '$ simply arise from the fact that maximally
mixed states remain unchanged under all quantum operations, and that
$\rho '$ must have a non-zero singlet fraction because the twirl
sequence does not \textit{create} new singlet, rather it only
\textit{preserves} the singlet, while averaging out all other terms,
converting them effectively to the maximally mixed state. In other
words, the twirl sequence ``Wernerizes'' the input state, by letting
the singlet pass unscathed, while scrambling other terms into the
maximally mixed
state. 

We have already come across the partial twirl in Section
\ref{section-partial-twirl} in context of preparing a state for
one-shot tomography. This chapter discusses the full twirl: the
different ways of practically implementing the operation, especially
on an ensemble quantum computer. Section \ref{section-twirl-intro}
introduces the implementation. Twirling is a kind of state averaging
and Section \ref{section-twirl-avg-one-qubit} discusses this concept
for a single qubit while Section \ref{section-twirl-two-qubits}
extends these ideas to the two qubit case. Finally, a step-by-step
NMR implementation is discussed in Section
\ref{section-twirl-experiment} and then I conclude in Section
\ref{section-twirl-conclusions}. This material is closely based on
our submitted manuscript, \cite{AnwarTwirl}.

An updated version of this chapter can be found on the e-print
server: \newline http://www.arxiv.org/quant-ph/0409142.

\chapter{Purity sharing}\label{sharing}

Preceding chapters dealt with different aspects of our \pHH\ derived
two qubit NMR quantum computer. Scaling up the prototype computer to
larger-sized systems is important if we wish to address more
challenging and practically useful mathematical problems. In this
spirit, one can come up with different ideas for extending the size
of our quantum computer. Some of these ideas seem promising but may
prove quite difficult to implement in practice.
\begin{enumerate}
\item For example, we can envisage a molecule (like an unsaturated organic polymer) which
has multiple sites where \H_2 molecules can add forming a
$2\mathbf{M}$ qubit quantum computer, $\mathbf{M}$ being the number
of dihydrogen molecules.
\item Another scheme can be visualized, involving the \textit{transfer} of purity from the
two pure hydrogen qubits onto other weakly polarized qubits. After
the polarization transfer, the hydrogens are allowed to detach from
the molecule and their place is taken up by fresh, pure qubits and
the transfer step is repeated. In this manner, purity can be
``pumped'' into the molecule. This scheme to some extent
\textit{resembles} the method of algorithmic cooling
\cite{AlgCoolProcNatlAcad} already mentioned in Section
\ref{section-schulman-vazirani-algorithmic}.
\item Both of
the above ideas involve bringing purity into the molecule from
outside. We can also think of another scenario: suppose we are
\textit{given} a molecule with $2\mathbf{M}$ pure or nearly pure
qubits, derived from $\mathbf{M}$ \pHH\ molecules. Instead of
``importing'' more purity into the molecule, we can consider the
question of \textit{sharing} out the purity on $k>2\mathbf{M}$
qubits. The question that I investigate in this chapter is whether
this sharing can be carried out, \textit{while remaining above the
entanglement threshold}. This question is related to purity
compression strategies outlined by Schulman and Vazirani
\cite{SchulmanVazirani}, but while the latter schemes
\textit{concentrate} purity onto a smaller number of spins, what I
consider here is ``diluting'' or \textit{sharing} out this purity
onto a larger quantum subspace.
\end{enumerate}

\par

An updated version of this chapter can be found on the e-print
server: \newline http://www.arxiv.org/quant-ph/0509036.

\clearpage
\bibliographystyle{jaip}

\begin{thebibliography}{100}

\bibitem{Abragam}
A.~Abragam,
\newblock {\em The Principles of Nuclear Magnetism},
\newblock Clarendon Press, Oxford, 1961.

\bibitem{ErnstNMR}
R.~R. Ernst, G.~Bodenhausen, and A.~Wokaun,
\newblock {\em Principles of Nuclear Magnetic Resonance in One and Two
  Dimensions},
\newblock Clarendon Press, Oxford, 1987.

\bibitem{LevittSpinDynamics}
M.~H. Levitt,
\newblock {\em Spin Dynamics: Basics of Nuclear Magnetic Resonance},
\newblock John Wiley and Sons, 2001.

\bibitem{FreemanChoreo}
R.~Freeman,
\newblock {\em Spin Choreography: Basic Steps in High Resolution NMR},
\newblock Oxford University Press, Oxford, 1998.

\bibitem{HandbookNMRFreeman}
R.~Freeman,
\newblock {\em A Handbook of Nuclear Magnetic Resonance},
\newblock Longman Scientific and Technical, 1988.

\bibitem{Cavanagh}
J.~Cavanagh, W.~J. Fairbrother, A.~G. Palmer, and N.~J. Skelton,
\newblock {\em Protein NMR Spectroscopy},
\newblock Academic Press, 1996.

\bibitem{Toolkit}
P.~Hore, J.~A. Jones, and S.~Wimperis,
\newblock {\em {NMR}: The Toolkit},
\newblock Oxford Chemistry Primers, Oxford, 2000.

\bibitem{NielsenChuang}
M.~A. Nielsen and I.~L. Chuang,
\newblock {\em Quantum Computation and Quantum Information},
\newblock Cambridge University Press, 2000.

\bibitem{BDiV1}
C.~H. Bennett and D.~DiVincenzo,
\newblock Nature {\bf 404}, 247 (2000).

\bibitem{RabiOscillations}
I.~I. Rabi,
\newblock Phys. Rev. {\bf 51}, 652 (1937).

\bibitem{DiVincenzoCrit}
D.~P. DiVincenzo,
\newblock Fortschr. der Physik {\bf 48}, 771 (2000).

\bibitem{Roadmap}
A {Q}uantum {I}nformation {S}cience and {T}echnology {R}oadmap,
{A}{R}{D}{A},
  http://qist.lanl.gov.

\bibitem{QCPhysicalRSoc}
Phil. Trans. R. Soc. Lond. A {\bf 361} (2003).

\bibitem{JonesPracticeNature}
J.~A. Jones,
\newblock Nature {\bf 421}, 28 (2003).

\bibitem{JonesRev}
J.~A. Jones,
\newblock Prog. NMR Spectrosc. {\bf 38}, 325 (2002).

\bibitem{CoryNMRQCReview}
D.~G. Cory et~al.,
\newblock Fortschr. der Phys. {\bf 48}, 875 (2000).

\bibitem{CoryProcNatl}
D.~G. Cory, A.~F. Fahmy, and T.~F. Havel,
\newblock Proc. Nat. Acad. Sci. USA {\bf 94}, 1634 (1997).

\bibitem{VandersypenNMRQCCtrl}
L.~M.~K. Vandersypen and I.~L. Chuang,
\newblock Rev. Mod. Phys. (in Press), {e-print: quant-ph/0404064}  (2004).

\bibitem{VandersypenThesis}
L.~M.~K. Vandersypen,
\newblock {\em Experimental Quantum Computation with Nuclear Spins in Liquid
  Solution},
\newblock PhD thesis, Stanford University, 2001.

\bibitem{VandersypenShor}
L.~M.~K. Vandersypen, M.~Steffen, G.~Breyta, C.~S. Yannoni, M.~H.
Sherwood, and
  I.~L. Chuang,
\newblock Nature {\bf 414}, 883 (2001).

\bibitem{JonesDeutsch}
J.~A. Jones and M.~Mosca,
\newblock J. Chem. Phys. {\bf 109}, 1648 (1998).

\bibitem{ChuangDeutsch}
I.~L. Chuang, L.~M.~K. Vandersypen, X.~Zhou, D.~W. Leung, and
S.~Lloyd,
\newblock Nature {\bf 393}, 143 (1998).

\bibitem{ChuangGrover}
I.~L. Chuang, N.~Gershenfeld, and M.~Kubinec,
\newblock Phys. Rev. Lett. {\bf 80}, 3408 (1998).

\bibitem{Jonesapproxcount}
J.~A. Jones and M.~Mosca,
\newblock Phys. Rev. Lett. {\bf 83}, 1050 (1999).

\bibitem{JonesGrover}
J.~A. Jones, M.~Mosca, and R.~H. Hansen,
\newblock Nature {\bf 393}, 344 (1998).

\bibitem{CollinsDeutschEnt}
D.~Collins, K.~W. Kim, and W.~C. Holton,
\newblock Phys. Rev. A {\bf 58}, R1633 (1998).

\bibitem{KumarGroverTomography}
R.~Das, T.~S. Mahesh, and A.~Kumar,
\newblock Chem. Phys. Lett. {\bf 369}, 8 (2003).

\bibitem{KumarDeutschEntangled}
K.~Dorai, A.~Kumar, and A.~Kumar,
\newblock Phys. Rev. A {\bf 63}, 034101 (2001).

\bibitem{KumarDeutsch2D}
T.~S. Mahesh, K.~Dorai, A.~Kumar, and A.~Kumar,
\newblock J. Magn. Reson. {\bf 148}, 95 (2001).

\bibitem{DeutschIonTraps}
S.~Gulde, M.~Riebe, G.~P.~T. Lancaster, C.~Becher, J.~Eschner,
H.~Haffner,
  F.~Schmidt-Kaler, I.~L. Chuang, and R.~Blatt,
\newblock Nature {\bf 421}, 48 (2003).

\bibitem{CohenTannoudji1}
C.~Cohen-Tannoudji, B.~Diu, and F.~Lalo\"{e},
\newblock {\em Quantum Mechanics}, volume~1,
\newblock John Wiley and Sons, 1977.

\bibitem{SorensenProdOp}
O.~W. S{\o}rensen, G.~W. Eich, M.~H. Levitt, G.~Bodenhausen, and
R.~R. Ernst,
\newblock Prog. NMR Spectrosc. {\bf 16}, 163 (1983).

\bibitem{Podkorytov}
I.~S. Podkorytov,
\newblock Conc. Magn. Reson. {\bf 9}, 117 (1997).

\bibitem{ProductOperatorABX}
L.~E. Kay and R.~E.~D. McClung,
\newblock J. Magn. Reson. {\bf 77}, 258 (1988).

\bibitem{Claridge}
T.~Claridge,
\newblock {\em High-resolution NMR Techniques in Organic Chemistry},
\newblock Pergamon, 1999.

\bibitem{SchmidtRohr}
K.~Schmidt-Rohr and H.~W. Spiess,
\newblock {\em {M}ultidimensional solid-state {NMR} and polymers},
\newblock Academic Press, London, 1994.

\bibitem{SolidNMR}
D.~D. Laws, H.~L. Bitter, and A.~Jerschow,
\newblock Angew. Chem. Int. Ed. {\bf 41}, 3096 (2002).

\bibitem{LiquidCrystalNMRQCVandersypen}
C.~S. Yannoni, M.~H. Sherwood, D.~C. Miller, I.~L. Chuang, L.~M.~K.
  Vandersypen, and M.~G. Kubinec,
\newblock Appl. Phys. Lett. {\bf 75}, 3563 (1999).

\bibitem{SolidSingleCrystalNMRQC}
G.~M. Leskowitz, N.~Ghaderi, R.~A. Olsen, and L.~J. Mueller,
\newblock J. Chem. Phys. {\bf 119}, 1643 (2003).

\bibitem{KumarSolidNMRQC}
A.~Kumar, K.~V. Ramanathan, T.~S. Mahesh, N.~Sinha, and K.~V.~R.
Murali,
\newblock Pranama J. Phys. {\bf 59}, 243 (2002).

\bibitem{SignsLevitt}
M.~H. Levitt,
\newblock J. Magn. Reson. {\bf 126}, 164 (1997).

\bibitem{CompositeLevitt}
M.~Levitt,
\newblock Prog. NMR Spectrosc. {\bf 18}, 61 (1986).

\bibitem{CompositeCounsell}
C.~Counsell, M.~H. Levitt, and R.~R. Ernst,
\newblock J. Magn. Reson. {\bf 63}, 133 (1985).

\bibitem{TyckoBroadBandPopInversion}
R.~Tycko,
\newblock Phys. Rev. Lett. {\bf 51}, 775 (1983).

\bibitem{WimperisComposite}
S.~Wimperis,
\newblock J. Magn. Reson. Ser. A {\bf 109}, 221 (1994).

\bibitem{CompositeZFreeman}
R.~Freeman, T.~A. Frenkiel, and M.~Levitt,
\newblock J. Magn. Reson. {\bf 44}, 409 (1981).

\bibitem{LogicGatesDiVincenzo}
D.~P. DiVincenzo,
\newblock Phil. Trans. R. Soc. Lond. A {\bf 454}, 261 (1998).

\bibitem{JonesNMRLogicGates}
J.~A. Jones, R.~H. Hansen, and M.~Mosca,
\newblock J. Magn. Reson. {\bf 135}, 353 (1998).

\bibitem{LogicGatesCory}
M.~Price, S.~S. Somaroo, C.~H. Tseng, J.~C. Gore, A.~F. Fahmy, T.~F.
Havel, and
  D.~G. Cory,
\newblock J. Magn. Reson. {\bf 140}, 371 (1999).

\bibitem{LogicGatesCoryPRA}
M.~Price, S.~S. Somaroo, A.~E. Dunlop, T.~F. Havel, and D.~G. Cory,
\newblock Phys. Rev. A {\bf 60}, 2777 (1999).

\bibitem{UniversalSWAP}
N.~Linden, H.~Barjat, E.~Kupce, and R.~Freeman,
\newblock Chem. Phys. Lett. {\bf 307}, 198 (1999).

\bibitem{MultiQubitGatesNMRCory}
M.~D. Price, T.~F. Havel, and D.~G. Cory,
\newblock New. J. Phys. {\bf 2}, 10.1 (2000).

\bibitem{KumarTransitionSelective}
K.~Dorai, A.~Kumar, and A.~Kumar,
\newblock Phys. Rev. A {\bf 61}, 042306 (2000).

\bibitem{CoryPhysicaD}
D.~G. Cory, M.~D. Price, and T.~F. Havel,
\newblock Physica D {\bf 120}, 82 (1998).

\bibitem{TransientBlochSiegertShift}
L.~Elmsley and G.~Bodenhausen,
\newblock Chem. Phys. Lett. {\bf 168}, 297 (1990).

\bibitem{VandersypenSimultaneousSoft}
M.~Steffen, L.~M.~K. Vandersypen, and I.~L. Chuang,
\newblock J. Magn. Reson. {\bf 146}, 369 (2000).

\bibitem{FreemanFreezingCoupling}
N.~Linden, B.~Herve, R.~J. Carbajo, and R.~Freeman,
\newblock Chem. Phys. Lett. {\bf 305}, 28 (1999).

\bibitem{AnwarPH}
M.~S. Anwar, D.~Blazina, H.~A. Carteret, S.~B. Duckett, T.~K.
Halstead, J.~A.
  Jones, C.~M. Kozak, and R.~J.~K. Taylor,
\newblock Phys. Rev. Lett. {\bf 93}, 040501 (2004).

\bibitem{AnwarDeutsch}
M.~S. Anwar, J.~A. Jones, D.~Blazina, S.~B. Duckett, and H.~A.
Carteret,
\newblock Phys. Rev. A {\bf 70}, 032324 (2004).

\bibitem{JumpReturnOriginal}
P.~Plateau and M.~Gu\'{e}ron,
\newblock J. Am. Chem. Soc. {\bf 104}, 7310 (1982).

\bibitem{CoryHamiltonianModulate}
E.~M. Fortunato, M.~A. Pravia, N.~Boulant, G.~Teklemariam, T.~F.
Havel, and
  D.~G. Cory,
\newblock J. Chem. Phys. {\bf 116}, 7599 (2002).

\bibitem{CumminsNJPComposite}
H.~K. Cummins and J.~A. Jones,
\newblock New J. Phys. {\bf 2}, 6.1 (2000).

\bibitem{JonesOffsetTailored}
H.~K. Cummins and J.~A. Jones,
\newblock J. Magn. Reson. {\bf 148}, 338 (2001).

\bibitem{JonesIsing}
J.~A. Jones,
\newblock Phys. Rev. A {\bf 67}, 012317 (2003).

\bibitem{JonesRobustPhilTrans}
J.~A. Jones,
\newblock Phil. Trans. R. Soc. Lond. A {\bf 361}, 1429 (2003).

\bibitem{BarencoQuantumGates}
A.~Barenco, C.~H. Bennett, R.~Cleve, D.~P. DiVincenzo, M.~Margolius,
P.~Shor,
  T.~Sleator, J.~Smolin, and H.~Weinfurter,
\newblock Phys. Rev. A {\bf 52}, 3457 (1995).

\bibitem{UniversalQC}
D.~Deutsch, A.~Barenco, and A.~Ekert,
\newblock Phil. Trans. Roy. Soc. Lond. A {\bf 449}, 669 (1995).

\bibitem{ZurekDecoherence}
W.~H. Zurek,
\newblock Phys. Rev. D. {\bf 26}, 1862 (1982).

\bibitem{RelaxationMcConnell}
J.~McConnell,
\newblock {\em Theory of Nuclear Magnetic Relaxation in Liquids},
\newblock Cambridge University Press, Cambridge, 1987.

\bibitem{JonesNMRcrit}
J.~A. Jones,
\newblock Fortschr. der Physik {\bf 48}, 909 (2000).

\bibitem{DecoherenceModelCory}
G.~Teklemariam, E.~M. Fortunato, C.~C. Lopez, J.~Emerson, J.~P. Paz,
T.~F.
  Havel, and D.~G. Cory,
\newblock Phys. Rev. A. {\bf 67}, 062316 (2003).

\bibitem{DecoherenceModelLindbaldCory}
N.~Boulant, T.~F. Havel, M.~A. Pravia, and D.~G. Cory,
\newblock Phys. Rev. A. {\bf 67}, 042322 (2003).

\bibitem{ViolaDynamicalCtrl}
L.~Viola,
\newblock e-print: quant-ph/0404038  (2004).

\bibitem{BangBangViolaKnillPRL}
L.~Viola, E.~Knill, and S.~Lloyd,
\newblock Phys. Rev. Lett. {\bf 82}, 2417 (1999).

\bibitem{DFSLidar}
D.~A. Lidar, I.~L. Chuang, and K.~B. Whaley,
\newblock Phys. Rev. Lett. {\bf 81}, 2594 (1998).

\bibitem{DFSLidarNMRImplement}
J.~E. Ollerenshaw, D.~A. Lidar, and L.~E. Kay,
\newblock Phys. Rev. Lett. {\bf 91}, 217904 (2003).

\bibitem{DFSCoryScience}
L.~Viola, E.~M. Fortunato, M.~A. Pravia, E.~Knill, R.~Laflamme, and
D.~G. Cory,
\newblock Science {\bf 293}, 2059 (2001).

\bibitem{KeelerGradients}
J.~Keeler, R.~T. Clowes, A.~L. Davis, and E.~Laue,
\newblock Methods Enzymol. {\bf 239}, 145 (1994).

\bibitem{GradientZQKeeler}
M.~J. Thrippleton and J.~Keeler,
\newblock Angew. Chem. Int. Ed. {\bf 42}, 3938 (2003).

\bibitem{FanoDensityMatrixRMP}
U.~Fano,
\newblock Rev. Mod. Phys. {\bf 29}, 74 (1957).

\bibitem{EraserQuantum}
G.~Teklemariam, E.~M. Fortunato, M.~A. Pravia, Y.~Sharf, T.~F.
Havel, D.~G.
  Cory, A.~Bhattaharyya, and J.~Hou,
\newblock Phys. Rev. A {\bf 66}, 012309 (2002).

\bibitem{MehringNuclearELectronEnt}
M.~Mehring, J.~Mende, and W.~Scherer,
\newblock Phys. Rev. Lett. {\bf 90}, 153001 (2003).

\bibitem{TomographyNMRLong01}
G.~L. Long, H.~Y. Yan, and Y.~Sun,
\newblock J. Opt. B: Quantum Semiclass. Opt. {\bf 3}, 376 (2001).

\bibitem{Sakurai}
J.~J. Sakurai,
\newblock {\em Modern Quantum Mechanics},
\newblock Addison-Wesley Publishing Company, 1994.

\bibitem{CoryPhysComp}
D.~G. Cory, A.~F. Fahmy, and T.~F. Havel,
\newblock in {\em PhysComp '96}, edited by T.~Toffoli, M.~Biafore, and
  J.~Le{\~{a}}o, pages 87--91, New England Complex Systems Institute, 1996.

\bibitem{UnitaryBoundsSorensen}
O.~W. S{\o}rensen,
\newblock Prog. NMR Spectrosc. {\bf 21}, 503 (1989).

\bibitem{KnillChuangEffectivePure}
E.~Knill, I.~Chuang, and R.~Laflamme,
\newblock Phys. Rev. A {\bf 57}, 3348 (1998).

\bibitem{TemporalMarx}
R.~Marx, A.~F. Fahmy, J.~M. Myers, W.~Bermel, and S.~J. Glaser,
\newblock Phys. Rev. A {\bf 62}, 012310 (2000).

\bibitem{Gershenfeld}
N.~A. Gershenfeld and I.~L. Chuang,
\newblock Science {\bf 275}, 350 (1997).

\bibitem{LogicalLabellingVandersypen}
L.~M.~K. Vandersypen, C.~S. Yannoni, M.~H. Sherwood, and I.~L.
Chuang,
\newblock Phys. Rev. Lett. {\bf 83}, 3085 (1999).

\bibitem{KnillCatBenchamrk}
E.~Knill, R.~Laflamme, and C.-H. Tseng,
\newblock Nature {\bf 404}, 368 (2000).

\bibitem{Warren}
W.~S. Warren,
\newblock Science {\bf 277}, 1688 (1997).

\bibitem{MountScalable}
R.~Blume-Kohout, C.~M. Caves, and I.~H. Deutsch,
\newblock Found. Phys. {\bf 32}, 1641 (2002).

\bibitem{SchulmanVazirani}
L.~J. Schulman and U.~Vazirani,
\newblock in {\em {P}roc. 31'st {STOC} ({ACM} {S}ymp. {T}heory {C}omp.)}, ACM
  Press, 1999.

\bibitem{VandersypenSchulman}
D.~E. Chang, L.~M.~K. Vandersypen, and M.~Steffen,
\newblock Chem. Phys. Lett. {\bf 338}, 337 (2001).

\bibitem{AlgCoolProcNatlAcad}
P.~O. Boykin, T.~Mor, V.~Roychowdhury, F.~Vatan, and R.~Vrijen,
\newblock Proc. Natl. Acad. Sci. {\bf 99}, 3388 (2002).

\bibitem{AlgoCoolingJose}
J.~M. Fernandez, S.~Lloyd, T.~Mor, and V.~Rowchowdhury,
\newblock e-print: quant-ph/0401135  (2004).

\bibitem{ErrorCorrectionShor}
P.~Shor,
\newblock Phys. Rev. A {\bf 52}, 2493 (1995).

\bibitem{ErrorCorrectionSteane}
A.~Steane,
\newblock Proc. R. Soc., London A {\bf 452}, 2551 (1996).

\bibitem{ErrorCorrectionVandersypen}
D.~Leung, L.~M.~K. Vandersypen, X.~L. Zhou, M.~Sherwood, C.~Yannoni,
  M.~Kubinec, and I.~Chuang,
\newblock Phys. Rev. A {\bf 60}, 1924 (1999).

\bibitem{ErrorCorrectionNMRCory98}
D.~G. Cory, M.~D. Price, W.~Maas, E.~Knill, R.~Laflamme, W.~H.
Zurek, T.~F.
  Havel, and S.~S. Somaroo,
\newblock Phys. Rev. Lett. {\bf 81}, 2152 (1998).

\bibitem{DetectionVirtualPhotons}
D.~I. Hoult and B.~Bhakar,
\newblock Conc. Magn. Reson. {\bf 9}, 277 (1997).

\bibitem{SpinDetectionMRFM}
D.~Rugar, R.~Budakian, H.~J. Mamin, and B.~W. Chui,
\newblock Nature {\bf 430}, 329 (2004).

\bibitem{SpinDetectionVandersypen}
J.~M. Elzerman, R.~Hanson, L.~H.~W. van Beveren, B.~Witkamp,
L.~M.~K.
  Vandersypen, and L.~P. Kouwenhoven,
\newblock Nature {\bf 430}, 431 (2004).

\bibitem{ODMR1}
J.~Kohler, J.~A.~M. Disselhorst, M.~C.~J.~M. Donkers, E.~J.~J.
Groenen,
  J.~Schmidt, and W.~E. Moerner,
\newblock Nature {\bf 363}, 242 (1993).

\bibitem{ODMR2}
J.~Wratchtrup, C.~von Borczyskowski, J.~Bernhard, M.~Orrit, and
R.~Brown,
\newblock Nature {\bf 363}, 244 (1993).

\bibitem{ODNMRQuantumWell}
M.~Eickhoff and D.~Suter,
\newblock J. Magn. Reson. {\bf 166}, 69 (2004).

\bibitem{Schrodinger}
E.~Schr{\"{o}}dinger,
\newblock Proc. Cambridge Philos. Soc. {\bf 31}, 555 (1935).

\bibitem{TeleportBennettOriginal}
C.~H. Bennett, G.~Brassard, C.~Cr\'{e}peau, R.~Jozsa, A.~Peres, and
W.~K.
  Wooters,
\newblock Phys. Rev. Lett. {\bf 70}, 1895 (1993).

\bibitem{TeleportBeowmeesterExp}
D.~Bowmeester, J.-W. Pan, K.~Mattle, M.~Eibl, H.~Weinfurter, and
A.~Zeilinger,
\newblock Nature {\bf 390}, 575 (1997).

\bibitem{SuperdenseCodingBennettOriginal}
C.~H. Bennett and S.~J. Weisner,
\newblock Phys. Rev. Lett. {\bf 69}, 2881 (1992).

\bibitem{BellInterConversion}
K.~Mattle, H.~Weinfurter, P.~G. Kwiat, and A.~Zeilinger,
\newblock Phys. Rev. Lett. {\bf 76}, 4656 (1996).

\bibitem{BB84}
C.~H. Bennett and G.~Brassard,
\newblock Quantum cryptography: public key distribution and coin tossing,
\newblock in {\em Proceedings of IEEE Conference on Computers, Systems and
  Signal Processing}, page 175, IEEE, New York, 1984.

\bibitem{BBB92}
C.~H. Bennett, F.~Bessette, G.~Brassard, L.~Salvail, and J.~Smolin,
\newblock J. Cryptology {\bf 5}, 3 (1992).

\bibitem{EntanglementNMRLaflamme}
R.~Laflamme, D.~G. Cory, C.~Negreveregne, and L.~Viola,
\newblock Quantum Inform. and Compu. {\bf 1}, 1 (2001).

\bibitem{WernerState}
R.~F. Werner,
\newblock Phys. Rev. A {\bf 40}, 4277 (1989).

\bibitem{Braunstein}
S.~L. Braunstein, C.~M. Caves, R.~Jozsa, N.~Linden, S.~Popescu, and
R.~Schack,
\newblock Phys. Rev. Lett. {\bf 83} (1999).

\bibitem{Pittenger}
A.~O. Pittenger and M.~H. Rubin,
\newblock Opt. Commun. {\bf 179}, 447 (2000).

\bibitem{Gurvits}
L.~Gurvits and H.~Barnum,
\newblock Phys. Rev. A {\bf 68}, 042312 (2003).

\bibitem{CavesSeparableLowerBound}
R.~Schack and C.~M. Caves,
\newblock J. Mod. Optic. {\bf 47}, 387 (2000).

\bibitem{GlaserScience98}
S.~J. Glaser, T.~Schulte-Herbr{\"{u}}ggen, M.~Sieveking,
O.~Schedletzky, N.~C.
  Nielsen, O.~W. S{{\o}}rensen, and C.~Griesinger,
\newblock Science {\bf 280}, 421 (1998).

\bibitem{PeresWernerSep}
A.~Peres,
\newblock Phys. Rev. Lett. {\bf 77}, 1413 (1996).

\bibitem{SchackClassicalNMR}
R.~Schack and C.~M. Caves,
\newblock Phys. Rev. A {\bf 60}, 4354 (1999).

\bibitem{DiscordOllivier}
H.~Ollivier and W.~H. Zurek,
\newblock Phys. Rev. Lett. {\bf 88}, 017901 (2002).

\bibitem{EntanglementNMRBrassard}
E.~Biham, G.~Brassard, D.~Kenigsberg, and T.~Mor,
\newblock Theor. Comp. Sci. {\bf 320}, 15 (2004).

\bibitem{EkertJozsa98}
A.~Ekert and R.~Jozsa,
\newblock Phil. Trans. R. Soc. London Ser. A {\bf 356}, 1769 (1998).

\bibitem{ShorAlgo}
P.~W. Shor,
\newblock Algorithms for quantum computation: discrete logarithm and factoring,
\newblock in {\em Proceedings of the 35th Annual Symposium on Fundamentals of
  Computer Science}, page 124, 1994.

\bibitem{SeparabilityPrimerLewenstein}
M.~Lewenstein, J.~I.~C. D.~Bru{\ss}, B.~Kraus, M.~Kus,
J.~Samsonowicz,
  A.~Sanpera, and R.~Tarrach,
\newblock J. Mod. Optic. {\bf 47}, 2481 (2000).

\bibitem{EntanglementCharacterizeBruss}
D.~Bru{\ss},
\newblock J. Math. Phys. {\bf 43}, 4237 (2002).

\bibitem{HorodeckiSeparability}
M.~Horodecki, P.~Horodecki, and R.~Horodecki,
\newblock Physics Letters A. {\bf 223}, 1 (1996).

\bibitem{BoundEntanglement}
P.~Horodecki,
\newblock Phys. Lett. A {\bf 232}, 333 (1997).

\bibitem{BoundEntanglementUndistillable}
M.~Horodecki, P.~Horodecki, and R.~Horodecki,
\newblock Phys. Rev. Lett. {\bf 80}, 5239 (1998).

\bibitem{MultiPartiteEntDur}
W.~D{\"{u}}r, J.~I. Cirac, and R.~Tarrach,
\newblock Phys. Rev. Lett. {\bf 83}, 3562 (1999).

\bibitem{QuantifyingEntVedral}
V.~Vedral, M.~B. Plenio, M.~A. Rippin, and P.~L. Knight,
\newblock Phys. Rev. Lett. {\bf 78}, 2275 (1997).

\bibitem{BennetMixedEntPRA}
C.~H. Bennett, D.~P. DiVincenzo, J.~A. Smolin, and W.~K. Wootters,
\newblock Phys. Rev. A. {\bf 54}, 3824 (1996).

\bibitem{concurrence}
W.~K. Wootters,
\newblock Phys. Rev. Lett. {\bf 80}, 2245 (1997).

\bibitem{HorodeckiInfoThSep}
R.~Horodecki and M.~Horodecki,
\newblock Phys. Rev. A {\bf 54}, 1838 (1996).

\bibitem{SWAPConcMagRes}
T.~Schulte-Herbr{\"{u}}ggen and O.~W. S{\o}rensen,
\newblock Conc. Magn. Reson. {\bf 12}, 389 (2000).

\bibitem{Overhauser1}
A.~W. Overhauser,
\newblock Phys. Rev. {\bf 91}, 476 (1953).

\bibitem{Overhauser2}
A.~W. Overhauser,
\newblock Phys. Rev. {\bf 92}, 411 (1953).

\bibitem{ENDORRev}
H.~Kurreck, B.~Kirste, and W.~Lubitz,
\newblock Angew. Chem. Int. Ed. Engl. {\bf 23}, 173 (1984).

\bibitem{DNPLiquids}
J.~H. Ardenkj{\ae}r-Larsen, B.~Fridlund, A.~Gram, G.~Hansson,
L.~Hansson, M.~H.
  Lertche, R.~Servin, M.~Thanning, and K.~Golman,
\newblock Proc. Natl. Acad. Sci. {\bf 100}, 10158 (2003).

\bibitem{LaserPolarizedGoodson}
B.~M. Goodson,
\newblock J. Magn. Res. {\bf 155}, 157 (2002).

\bibitem{LaserPolarizedBruner}
E.~Brunner,
\newblock Conc. Magn. Res. {\bf 11}, 313 (1999).

\bibitem{SPINOESciencePines}
G.~Navon, Y.~Q. Song, T.~Room, A.~Appelt, R.~E. Taylor, and
A.~Pines,
\newblock Science {\bf 271}, 1848 (1996).

\bibitem{SPINOEConcMagRes}
Y.~Q. Song,
\newblock Conc. Magn. Reson. {\bf 12}, 6 (1999).

\bibitem{SPINOEPRL}
J.~J. Heckman, M.~P. Ledbetter, and M.~V. Romalis,
\newblock Phys. Rev. Lett. {\bf 91}, 067601 (2003).

\bibitem{ChuangXenon}
A.~S. Verhulst, O.~Liivak, M.~H. Sherwood, H.~Vieth, and I.~L.
Chuang,
\newblock Appl. Phys. Lett. {\bf 79}, 2480 (2001).

\bibitem{PhysicalChemistryAtkins}
P.~W. Atkins,
\newblock {\em Physical Chemistry},
\newblock Oxford University Press, 1994.

\bibitem{SimonPHRev}
S.~B. Duckett and C.~J. Sleigh,
\newblock Prog. NMR Spectrosc {\bf 34}, 71 (1999).

\bibitem{PHImaging}
K.~Golman, O.~Axelsson, H.~J\'{o}hannesson, S.~M{\aa}nsson,
C.~Olofsson, and
  J.~S. Petersson,
\newblock Magn. Reson. Med. {\bf 46}, 1 (2001).

\bibitem{Hubler}
P.~H{\"{u}}bler, J.~Bargon, and S.~J. Glaser,
\newblock J. Chem. Phys. {\bf 113}, 2056 (2000).

\bibitem{AnwarPHReview}
D.~Blazina, S.~B. Duckett, T.~K. Halstead, C.~M. Kozak, R.~J.~K.
Taylor, M.~S.
  Anwar, J.~A. Jones, and H.~A. Carteret,
\newblock Magn. Reson. Chem. (Submitted)  (2004).

\bibitem{AnwarGrover}
M.~S. Anwar, D.~Blazina, H.~A. Carteret, S.~B. Duckett, and J.~A.
Jones,
\newblock Chem. Phys. Lett. (In Press), e-print: quant-ph/0407091  (2004).

\bibitem{WeitekampPersonal}
D.~P. Weitekamp,
\newblock Sensitivity enhancement through spin statistics,
\newblock in {\em Encyclopedia of NMR}, edited by D.~M. Grant and R.~K. Harris,
  page 696, Wiley, New York, 1996.

\bibitem{Bowers}
C.~R. Bowers and D.~P. Weitekamp,
\newblock Phys. Rev. Lett. {\bf 57}, 2645 (1986).

\bibitem{Symmetrization96}
M.~de~Angelis, G.~Gagliardi, L.~Gianfrani, and G.~M. Tino,
\newblock Phys. Rev. Lett. {\bf 76}, 2840 (1996).

\bibitem{Symmetrization98}
G.~Modugno, M.~Inguscio, and G.~M. Tino,
\newblock Phys. Rev. Lett. {\bf 81}, 4790 (1998).

\bibitem{ShankarQM}
R.~Shankar,
\newblock {\em Principles of Quantum Mechanics},
\newblock Kluwer Academic/Plenum Publishers, 1994.

\bibitem{NattererPHRev}
J.~Natterer and J.~Bargon,
\newblock Prog. NMR Spectrosc. {\bf 31} (1997).

\bibitem{BowersAdvMagRes}
C.~R. Bowers, D.~H. Jones, N.~D. Karur, J.~A. Labinger, M.~G.
Pravica, and
  D.~P. Weitekamp,
\newblock Adv. Magn. Reson. {\bf 15}, 269 (1990).

\bibitem{SilveraRMP}
I.~F. Silvera,
\newblock Rev. Mod. Phys. {\bf 52}, 393 (1980).

\bibitem{Herzberg}
G.~Herzberg,
\newblock {\em Molecular Spectra and Molecular Structure}, volume~1,
\newblock Van Nostrand Reinhold Company, 1950.

\bibitem{SpectroscopyHollas}
J.~M. Hollas,
\newblock {\em Modern Spectroscopy},
\newblock John Wiley and Sons, 1987.

\bibitem{SpectroscopyBanwell}
C.~N. Banwell,
\newblock {\em Fundamentals of Molecular Spectroscopy},
\newblock McGraw-Hill Book Company (UK) Limited, 1983.

\bibitem{Arfken}
G.~Arfken,
\newblock {\em Mathematical Methods for Physicists},
\newblock Academic Press, 1985.

\bibitem{OrthoParaConversionPravica98}
M.~G. Pravica and I.~F. Silvera,
\newblock Phys. Rev. Lett. {\bf 81}, 4180 (1998).

\bibitem{PHCatalyticConverter}
A.~M. Juarez, D.~Cubric, and C.~C. King,
\newblock Meas. Sci. Tech. {\bf 13}, N52 (1998).

\bibitem{OrthoParaWater}
V.~I. Tikhonov and A.~A. Volkov,
\newblock Science {\bf 296}, 2363 (2002).

\bibitem{OrthoParaComets}
J.~Crovisier, K.~Leech, D.~Bockel{\'{e}}e-Morvan, T.~Y. Brooke,
M.~S. Hanner,
  B.~Altieri, H.~U. Keller, and E.~Lellouch,
\newblock Science {\bf 275}, 1904 (1997).

\bibitem{OrthoParaTransitions}
A.~Miani and J.~Tennyson,
\newblock J. Chem. Phys. {\bf 120}, 2732 (2004).

\bibitem{LasersChemistry}
D.~L. Andrews,
\newblock {\em Lasers in Chemistry},
\newblock Springer-Verlag, 1986.

\bibitem{GodardPHPhotolysis}
C.~Godard, P.~Callaghan, J.~L. Cunningham, S.~B. Duckett, J.~A.~B.
Lohman, and
  R.~N. Perutz,
\newblock Chem. Comm. {\bf 23}, 2836 (2002).

\bibitem{SchottPHRu}
D.~Schott, C.~J. Sleigh, J.~P. Lowe, S.~B. Duckett, R.~J. Mawby, and
M.~G.
  Partridge,
\newblock Inorg. Chem. {\bf 41}, 2960 (2002).

\bibitem{CroninPHInstant}
L.~Cronin, M.~C. Nicasio, R.~N. Perutz, R.~G. Peters, D.~M. Roddick,
and M.~K.
  Whittlesey,
\newblock J. Am. Chem. Soc. {\bf 117}, 10047 (1995).

\bibitem{BowersThesis}
C.~R. Bowers,
\newblock {\em \textit{Para}-hydrogen and Synthesis Allow Dramatic Enhanced
  Nuclear Alignment},
\newblock PhD thesis, California Institute of Technology, 1991.

\bibitem{Altadena}
M.~G. Pravica and D.~P. Weitekamp,
\newblock Chem. Phys. Lett. {\bf 145}, 255 (1988).

\bibitem{PHIPReactionPathwayJACS}
G.~Buntkowsky, J.~Bargon, and H.~H. Limbach,
\newblock J. Am. Chem. Soc. {\bf 118}, 8677 (1996).

\bibitem{RemoteDetectionPines}
A.~J. Moul\'{e}, M.~M. Spence, S.-I. Han, J.~A. Seeley, K.~L.
Pierce,
  S.~Saxena, and A.~Pines,
\newblock Proc. Natl. Acad. Sci. {\bf 100}, 9122 (2003).

\bibitem{BryndzaCIDNPPH}
P.~F. Siedler, H.~E. Bryndza, J.~E. Frommer, L.~S. Stuhl, and R.~G.
Bergman,
\newblock Organometallics {\bf 2}, 1701 (1983).

\bibitem{GlaserHH}
S.~J. Glaser and J.~J. Quant,
\newblock Adv. Magn. Optic. Reson. {\bf 19}, 59 (1996).

\bibitem{QPTQFT}
Y.~S. Weinstein, T.~F. Havel, J.~Emerson, N.~Boulant, M.~Saraceno,
S.~Lloyd,
  and D.~G. Cory,
\newblock (2004).

\bibitem{Pendulums}
R.~Marx and S.~J. Glaser,
\newblock J. Magn. Reson. {\bf 164}, 338 (2003).

\bibitem{TOCSYOriginalErnst}
L.~Braunschweiler and R.~R. Ernst,
\newblock J. Magn. Reson. {\bf 53}, 521 (1983).

\bibitem{WaughCoherentAveraging}
U.~Haeberlen and J.~S. Waugh,
\newblock Phys. Rev. {\bf 175}, 453 (1968).

\bibitem{MLEV}
M.~H. Levitt, R.~Freeman, and T.~Frenkiel,
\newblock J. Magn. Reson. {\bf 47}, 328 (1982).

\bibitem{DictionaryNMR}
S.~W. Homans,
\newblock {\em A Dictionary of Concepts in NMR},
\newblock Clarendon Press, Oxford, 1989.

\bibitem{Linden99}
N.~Linden, H.~Barjat, E.~Kupce, and R.~Freeman,
\newblock Chem. Phys. Lett. {\bf 296}, 61 (1998).

\bibitem{ShapedFreeman}
R.~Freeman,
\newblock Prog. NMR Spectrosc. {\bf 32}, 59 (1997).

\bibitem{GARPShaka}
A.~J. Shaka, P.~B. Barker, and R.~Freeman,
\newblock J. Magn. Reson. {\bf 64}, 547 (1985).

\bibitem{DataProcessingHochStern}
J.~C. Hoch and A.~S. Stern,
\newblock {\em NMR Data Processing},
\newblock Wiley-Liss, Inc., 1996.

\bibitem{LindonFerridgeDataProcess}
J.~C. Lindon and A.~G. Ferridge,
\newblock Prog. NMR Spectrosc. {\bf 14}, 27 (1980).

\bibitem{AWK}
An {A}wk {T}utorial, http://www.vectorsite.net/tsawk.html.

\bibitem{Grace}
Grace {U}ser's {G}uide (for {G}race-5.1.10),
  http://plasma-gate.weizmann.ac.il/grace/doc/usersguide.html.

\bibitem{FourierBracewell}
R.~N. Bracewell,
\newblock {\em The Fourier Transform and its Applications},
\newblock McGraw-Hill Book Company, 1986.

\bibitem{McIntyreJDbl}
L.~McIntyre and R.~Freeman,
\newblock J. Magn. Reson. {\bf 96}, 425 (1992).

\bibitem{JMatchDblJones}
J.~A. Jones,
\newblock Conc. Magn. Reson. {\bf 8}, 175 (1996).

\bibitem{DeutschJozsaAlgo}
D.~Deutsch and R.~Jozsa,
\newblock Proc. R. Soc. London Ser. A {\bf 439}, 553 (1992).

\bibitem{DeutschRefinedCleveAlgo}
R.~Cleve, A.~Ekert, C.~Macchiavello, and M.~Mosca,
\newblock Proc. R. Soc. London Ser. A {\bf 454}, 339 (1998).

\bibitem{QCircuit}
B.~Eastin and S.~T. Flammia,
\newblock {Q}-circuit tutorial, e-print: quant-ph/0406003  (2004).

\bibitem{GroverAlgo}
L.~K. Grover,
\newblock Phys. Rev. Lett. {\bf 79}, 2645 (1986).

\bibitem{BennetPurificationPRL}
C.~H. Bennett, G.~Brassard, S.~Popescu, B.~Schumacher, J.~A. Smolin,
and W.~K.
  Wootters,
\newblock Phys. Rev. Lett. {\bf 76}, 722 (1996).

\bibitem{Deutsch96}
D.~Deutsch, A.~Ekert, R.~Jozsa, C.~Macchiavello, S.~Popescu, and
A.~Sanpera,
\newblock  {\bf 77}, 2818 (1996).

\bibitem{PetersEntConcentration}
N.~A. Peters, J.~B. Altepeter, D.Branning, E.~R. Jeffrey, T.~C. Wei,
and P.~G.
  Kwiat,
\newblock Phys. Rev. Lett. {\bf 90}, 193601 (2003).

\bibitem{AnwarTwirl}
M.~S. Anwar, L.~Xiao, A.~J. Short, J.~A. Jones, D.~Blazina, S.~B.
Duckett, and
  H.~A. Carteret,
\newblock Phys. Rev. A (Submitted), e-print:quant-ph/0409142  (2004).

\bibitem{TeleportationNielsen}
M.~A. Nielsen, E.~Knill, and R.~Laflamme,
\newblock Nature {\bf 396}, 52 (1998).

\bibitem{AngularMomentumZare}
V.~D. Kleiman, H.~Park, R.~J. Gordon, and R.~N. Zare,
\newblock {\em Companion to Angular Momentum},
\newblock John Wiley and Sons, Inc., London, 1998.

\bibitem{GroupsJones}
H.~F. Jones,
\newblock {\em Groups, Representations and Physics},
\newblock Institute of Physics Publishing, Bristol and Philadelphia, 1996.

\bibitem{MagicHoppingBax}
A.~Bax, N.~M. Szeverenyi, and G.~E. Maciel,
\newblock J. Magn. Reson. {\bf 52}, 147 (1983).

\bibitem{TwirlStellaOctangula}
P.~K. Aravind,
\newblock Phys. Lett. A {\bf 233}, 7 (1997).

\bibitem{OneQubitPowerPRL}
E.~Knill and R.~Laflamme,
\newblock Phys. Rev. Lett. {\bf 81}, 5672 (1998).

\bibitem{OneQubitPowerPoulinPRA}
D.~Poulin, R.~Laflamme, G.~J. Milburn, and J.~P. Paz,
\newblock Phys. Rev. A {\bf 68}, 022302 (2003).

\bibitem{AdiabaticQCVazirani}
W.~V. Dam, M.~Mosca, and U.~Vazirani,
\newblock e-print: quant-ph/0206003  (2003).

\bibitem{AdiabaticCirac2004}
V.~Murg and J.~I. Cirac,
\newblock Phys. Rev. A {\bf 69}, 042320 (2004).

\bibitem{WaughSupercycle82}
J.~S. Waugh,
\newblock J. Magn. Reson. {\bf 50}, 30 (1982).

\bibitem{Convolution}
P.~A. Jansson,
\newblock Convolution and related concepts,
\newblock in {\em Deconvolution: with Applications in Spectroscopy}, edited by
  P.~A. Jansson, pages 1--34, Academic Press, New York, 1984.

\bibitem{ConvolutionNMRBodenhausen}
P.~Huber and G.~Bodenhausen,
\newblock J. Magn. Reson., Ser. A {\bf 102}, 81 (1993).

\bibitem{Kraus}
K.~Kraus,
\newblock {\em State, Effects, and Operations},
\newblock Springer, Berlin, 1983.

\bibitem{OperatorSumSchumacher}
B.~Schumacher,
\newblock Phys. Rev. A {\bf 54}, 2614 (1996).

\bibitem{SolomonRelaxation}
I.~Solomon,
\newblock Phys. Rev. {\bf 99}, 559 (1955).

\end{thebibliography}

\end{document}